\documentclass[aps,prx,reprint,preprintnumbers,showpacs,showkeys,superscriptaddress,longbibliography]{revtex4-1}
\usepackage{amsmath,amssymb,bm,mathrsfs}
\usepackage[colorlinks=true,citecolor=blue,linkcolor=blue]{hyperref}
\usepackage{graphicx}
\usepackage{bm}
\usepackage[Symbol]{upgreek}

\begin{document}
\title{Effective Lagrangian for Nonrelativistic Systems}

\author{Haruki Watanabe}
\email{hwatanabe@berkeley.edu}
\affiliation{Department of Physics, University of California,
  Berkeley, California 94720, USA}
\author{Hitoshi Murayama}
\email{hitoshi@berkeley.edu, hitoshi.murayama@ipmu.jp}
\affiliation{Department of Physics, University of California,
  Berkeley, California 94720, USA} 
\affiliation{Theoretical Physics Group, Lawrence Berkeley National
  Laboratory, Berkeley, California 94720, USA} 
\affiliation{Kavli Institute for the Physics and Mathematics of the
  Universe (WPI), Todai Institutes for Advanced Study, University of Tokyo,
  Kashiwa 277-8583, Japan} 
\begin{abstract}
  The effective Lagrangian for Nambu-Goldstone bosons (NGBs) in
  systems without Lorentz invariance has a novel feature that some of
  the NGBs are canonically conjugate to each other, hence describing
  $1$ dynamical degree of freedom by two NGB fields.  We develop
  explicit forms of their effective Lagrangian up to the quadratic
  order in derivatives.  We clarify the counting rules of NGB degrees
  of freedom and completely classify possibilities of such
  canonically conjugate pairs based on the topology of the coset spaces.
  Its consequence on the dispersion relations of the NGBs is
  clarified.  We also present simple scaling arguments to see whether
  interactions among NGBs are marginal or irrelevant, which justifies a lore in the literature about the possibility of symmetry breaking in $1+1$ dimensions.
\end{abstract}
\preprint{IPMU14-0043,UCB-PTH14/03}
\maketitle
\section{Introduction}
\label{introduction}
In studies of any macroscopic physical systems, the behavior of the system
at low temperatures, small energies, and long distances is determined
predominantly by microscopic excitations with small or zero gap.  It
is, hence, important to develop a general theory to discuss gapless
excitations.  Barring special reasons, however, we generally do not
expect any gapless degrees of freedom in a given system.  The
important exceptions are (1) a Fermi liquid with the Fermi level within
a continuous band, (2) second-order phase transitions with scale (and
often conformal) invariance, (3) states protected by
  topological reasons such as edge states of topological insulators or
  quantum Hall states, and (4) Nambu-Goldstone bosons (NGBs) of
spontaneous symmetry breaking. The first three cases are discussed
extensively in the literature.  We focus on the last case in this
paper because a general theory, so far, has surprisingly been lacking, despite its importance and long history.

Spontaneously broken symmetry is a common theme through all areas of
physics.  The examples are numerous: Bose-Einstein condensates of cold
atoms, superfluids of $^4$He or $^3$He, crystal lattices, neutron stars,
ferromagnets, anti-ferromagnets, liquid crystals, chiral symmetry in
QCD, and cosmic inflation.  The universal feature is that it
guarantees the existence of gapless excitations when the relevant
symmetries are continuous.  Once promoted to gauge symmetries, it is
the basis to discuss superconductivity, the Englert-Bourt-Higgs
mechanism, and cosmic strings.  The crucial question is the following: What is the
general theory that can describe the number of NGB degrees of freedom, and their dispersion relations, their interactions among each other and
to other degrees of freedom?  Ideally, the theory does not depend on
specifics of a given system or perturbation theory but is rather
determined by symmetries alone, so that it is applicable even when the
system is strongly coupled or we lack understanding of the microscopic
description. 

In systems with Lorentz invariance, the general theory has already been
established back in 1960's by the celebrated Nambu-Goldstone
theorem~\cite{Nambu1,Goldstone:1961,Goldstone:1962} and later with
``phenomenological Lagrangians'' by Callan, Coleman, Wess, and
Zumino~\cite{Coleman2,Callan}.  It is important to formulate the
theory using Lagrangians because a Lagrangian is a self-contained
package to describe a system.  It determines the degrees of freedom,
equations of motion, Noether currents for symmetries, and commutation
relations and provides the basis for perturbation theory using Feynman
diagrams and many non-perturbative methods based on path integrals.
In comparison, the Hamiltonian formalism~\cite{HohenbergHalperin,Mazenko} requires additional input: what the degrees of freedom are and what their commutation relations (or Poisson brackets) are.  Especially when at least one of these two is not clear at the beginning of the discussion, which turns out to be the case for our purposes, the Lagrangian formulation is essential.

However, many systems we are interested in are not Lorentz invariant.
A finite temperature violates Lorentz invariance because the Boltzmann
weight depends on the energy, which is the time component of the
energy-momentum four-vector and hence requires a specific choice of
the reference frame.  A chemical potential needed to describe systems
with finite densities couples to the charge density, which is also the
time component of a conserved four-current.  Often, the surrounding
environment violates Lorentz invariance as well.  In all of these
cases, rotational invariance may still be present, while Lorentz
invariance is certainly not there.

{\it It is, therefore, of foremost importance to develop a general
  theory of NGBs based on symmetry principles alone without assuming
  Lorentz invariance.}\/ We develop such a theory in this paper.

NGBs without Lorentz invariance have been discussed for their obvious
importance, as discussed above.  The nonrelativistic~\footnote{When we say “nonrelativistic” in this paper, it just means that the system does not have the Lorentz symmetry to begin with.  The effective Lagrangian for a nonrelativistic system may possess an emergent Lorentz symmetry at the lowest order in the derivative expansion [e.g., $\mathcal{L}_{\text{eff}}=(1/2)\partial_\mu\vec{n}\cdot\partial^\mu\vec{n}$ for antiferromagnets after proper scaling of space and time]. } analog of one
aspect of the NG theorem, that which ensures the appearance of at
least one NGB, was already discussed back in the
1960s~\cite{Lange:1965,Lange:1966,Guralnik:1968,Brauner:2010}.
However, the number and the dispersion of the NGBs have only been
studied on a case-by-case basis until quite recently.

The Nambu-Goldstone theorem says there must be one gapless excitation
for every broken-symmetry generator, assuming Lorentz invariance.
Moreover, Lorentz invariance constrains the dispersion relation for
gapless excitation to be $\omega=c k$, where $c$ is the speed of light.

However, these predictions are known to be false in systems without
Lorentz invariance.  A classic example is a ferromagnet. When spins
line up macroscopically due to the nearest-neighbor interaction, it
spontaneously breaks the $\text{SO}(3)$ spin-rotational symmetry with
three generators down to the unbroken $\text{SO}(2)$ axial symmetry
with only one generator.  Despite the two spontaneously broken
symmetries, the ferromagnet exhibits only one NGB.  Moreover, its
dispersion is quadratic rather than linear.  In contrast, an
antiferromagnet supports two NGBs with a linear dispersion, although
it shows the same symmetry-breaking pattern
$\text{SO}(3)\rightarrow\text{SO}(2)$.

More recent examples appeared in relativistic field theories with
nonzero chemical potentials, where examples of an ``\textit{abnormal}
number of Nambu-Goldstone bosons" are identified in many
contexts~\cite{Miransky,Schafer,Blaschke:2004,lianyi,ebert,ebert2,buchel}.
Also, spinor Bose-Einstein condensates in cold atom systems added a
number of new examples and realized some of them in the actual
experiments~\cite{Ueda,Stamper}.  The dispersion of the softest NGB
immediately modifies the thermodynamic property of the system at a low
temperature. For example, the low-temperature heat capacity behaves as
$C(T)\propto T^{d/z}$ for the NGB with the dispersion $\omega\propto
k^z$ in $d+1$ dimensions.  In general, the low-energy dynamics of
systems with spontaneous symmetry breaking is governed by NGBs, and hence, it is clearly important
to establish a general theorem that predicts the correct number,
dispersion, and interactions of NGBs.

In their pioneering work~\cite{Nielsen:1975}, Nielsen and Chadha
established an inequality that relates the number of NGBs to their
dispersion relations.  In their approach, NGBs are classified as
type-I (type-II) if their dispersion in the long-wavelength limit
behaves as $\omega\propto k^{2n-1}$ ($\omega\propto k^{2n}$). Based on
the analytic property of correlation functions, Nielsen and Chadha
proved that the number of type-I NGBs plus twice the number of type-II
NGBs is greater than or equal to the number of broken-symmetry
generators.  Note that their conclusion is merely an inequality, and
hence, it does not give any lower or upper bound for each type of NGB.  Also, their classification breaks
down when the dispersion is anisotropic, {\it e.g.}\/, $\omega\propto
\sqrt{(k_x)^2+C(k_y)^4}$. (See Sec.~\ref{sec:scaling} for an example.)

In a relatively recent paper, Sch\"afer~\textit{et
  al.}~\cite{Schafer} pointed out the importance of expectation
values of the commutators of the broken generators in reducing the
number of NGBs.  They showed that the number of NGBs must be equal to
the number of broken generators if $\langle [Q_a,Q_b]\rangle=0$ for
all combinations of broken generators.  Although their argument is
physically plausible, it contains a few questionable points.  They
identified the NG state associated with the charge $Q_a$ as
$Q_a|\Psi_0\rangle$ ($|\Psi_0\rangle$ is the quantum many-body ground
state) and discussed the possibility of linear dependence among such
vectors.  However, it is well known that, once symmetries are
spontaneously broken, broken generators themselves are ill defined. We
should rather use commutation relations of generators with other local
quantities.

Nambu~\cite{Nambu:2004, Nambu:2005} was probably the first to obtain
the correct insight into this problem. He observed that the nonzero
expectation value $\langle [Q_a,Q_b]\rangle$ makes zero modes
associated with these generators canonically conjugate to each other,
and hence, the number of NGBs is reduced by 1 per such a
pair. However, he did not prove this claim on general grounds.

With these previous works in mind, the current authors unified all of the above
observations into a simple and well-defined
form by proving them using field theory~\cite{WatanabeMurayama1}:
\begin{eqnarray}
n_{\text{A}}&=&\text{dim}\,G/H-\text{rank}\rho,\label{countingA}\\
n_{\text{B}}&=&\frac{1}{2}\text{rank}\rho,\label{countingB}\\
n_{\text{NGB}}&=&\text{dim}\,G/H-\frac{1}{2}\text{rank}\rho,\label{countingtot}\\
n_{\text{A}}+2n_{\text{B}}&=&\text{dim}\,G/H,\label{countingtot2}\\
i\rho_{ab}&\equiv&\langle[Q_a,j_b^0(0)]\rangle.\label{charge}
\end{eqnarray}
Equation \eqref{countingtot} was conjectured earlier in
Ref.~\cite{WatanabeBrauner1} and was also obtained independently in
Ref.~\cite{Hidaka}.  Here, $n_{\text{A}}$, $n_{\text{B}}$ represent
the numbers of type-A, B NGBs, respectively, and $n_{\text{NGB}}\equiv
n_{\text{A}}+n_{\text{B}}$ is the total number of NGBs.
Equations~\eqref{countingtot} and \eqref{countingtot2} follow from
Eqs.~\eqref{countingA} and \eqref{countingB}.  $j_a^\mu(x)$ is the
conserved current associated with a broken charge
$Q_a=\int\mathrm{d}^dx\,j_a^0(x)$.  The Lie group $G$ represents the
original symmetry of the system, and $H$ is its unbroken subgroup, so
that $\text{dim}\,G/H$ represents the number of broken-symmetry
generators.  Clearly, the symmetry-breaking pattern $G\rightarrow H$
is not sufficient to fix the number of NGBs, and we need additional
information [the matrix $\rho$ in Eq.~\eqref{charge}] about the ground
state.

The definitions of type-A, B NGBs are not based on their dispersion
relations but on their symplectic structure, as we will discuss in
detail later.  For now, we just note that, generically, type-A NGBs have
a linear dispersion and type-B NGBs have a quadratic dispersion, but
there are exceptions. Therefore, Eq.~\eqref{countingtot2} can be
understood as the equality version of the Nielsen-Chadha theorem for
most cases.

The above-explained theorem by Sch\"afer~\textit{et al.} can also be understood as the special case where the matrix $\rho$ vanishes, and hence, $n_{\text{NGB}}=\text{dim}\,G/H$ from Eq.~\eqref{countingtot}.  The matrix $\rho$ must always vanish in the Lorentz-invariant case, because $[Q_a,j_b^0(\vec{x},t)]=if_{ab}^{\phantom{ab}c}j_c^0(\vec{x},t)$ in the absence of central extensions and $j_c^\mu(0)$ is a Lorentz vector, which cannot have an expectation value without breaking the Lorentz symmetry. 

In order to prove the counting rule of NGBs and clarify their
dispersion relations, we develop the nonrelativistic analog of the
``phenomenological Lagrangian'' {\it \`a la}\/ Refs.~\cite{Coleman2,Callan},
following Leutwyler's
works~\cite{Leutwyler:1994nonrel,Leutwyler:1994rel}.  We derive an
explicit expression of the effective Lagrangian for a general symmetry
breaking-pattern $G\rightarrow H$.  In this process, we find a set of
terms that have not been taken into account in the literature.

This fully nonlinear effective Lagrangian contains only a few
parameters that play the role of coupling constants between NGBs.  By
analyzing the scaling law of the dominant interaction, we discuss the
stability of the symmetry-broken ground state.  In sufficiently high
dimensions, the system is essentially free, as expected.  However, it
turns out that, in general, {\it internal symmetries can be
  spontaneously broken even in $1+1$ dimensions}\/.  This is one of
the aspects enriched by the absence of Lorentz invariance --- in a
Lorentz-invariant theory, the well-known Coleman
theorem~\cite{Coleman} prohibits that possibility.

The explicit form of the effective Lagrangian leads to another
nontrivial prediction, that is, a no-go theorem for a certain number
of type-A and type-B NGBs.  One might think that any combination of $n_{\text{A}}$
and $n_{\text{B}}$ subject to Eq.~\eqref{countingtot2} should be possible.
However, for given $G$ and $H$, possibilities are quite restricted, because type-B NGBs are described by symplectic homogeneous
spaces, which are special types of coset spaces that admit the so-called
K\"ahler structure, if $G$ is semisimple.  We will discuss how the
possible numbers for type-A and type-B can be completely enumerated for
any given $G$ and $H$.

This paper is organized as follows.  In Sec.~\ref{sec:effective}, we discuss the most general form of the effective Lagrangian for nonrelativistic systems and derive differential equations for the coefficients appearing in the effective Lagrangians by paying careful attention to the gaugeability of the symmetry $G$.  We present an analytic solution of the differential equations in terms of the Maurer-Cartan form in Sec.~\ref{sec:solution}.  We also clarify the obstacle to gauge Wess-Zumino-Witten terms and algebras with central extensions.  Analyzing the free part of our effective Lagrangian,  we prove the counting rule in Sec.~\ref{sec:number} and derive their dispersion in Sec.~\ref{sec:dispersion}.  We discuss the interaction effect and spontaneous symmetry breaking in $1+1$ dimensions in Sec.~\ref{sec:stability}. 

In Sec.~\ref{sec:topology}, we present the mathematical foundation of the canonically conjugate (presymplectic) structure among some NGBs.  With this preparation, we completely classify the presymplectic structure and prove a no-go theorem that prohibits a certain combination of type-A and type-B NGBs in Sec.~\ref{sec:classification}.  It is followed by concrete demonstration thorough familiar examples in Sec.~\ref{sec:example}.

We will not discuss the counting of NGBs associated to spacetime symmetries.  For those symmetries, the number of NGBs is reduced not only by forming canonically conjugate pairs but also by other mechanisms, {\it e.g.}\/, linear dependence among conserved currents.  Hence the above counting rule does not hold.  See Refs.~\cite{Low,WatanabeMurayama2,Hayata,WatanabeBrauner3} for more details.  Nevertheless, we explain how to impose the Galilean symmetry, if it exists, on the effective Lagrangian in Sec.~\ref{sec:galilei}.

For the reader's convenience, we present a pedagogical introduction to the cohomology of Lie algebra in Appendix~\ref{appendix1}.  We also review how to couple matter fields to NGBs in Appendix~\ref{sec:matter}. Finally, we clarify a confusion in the existing literature on the relation between type-B NGBs and the time-reversal symmetry in Appendix~\ref{sec:discrete}.

\section{Effective Lagrangian for nonrelativistic systems}
\label{sec:effective}

In this section, we describe the general effective Lagrangian for NGBs on the coset space $G/H$.  One way of deriving the effective Lagrangian is to integrate out all high-energy modes from an assumed microscopic model.  However, there is an alternative universal approach, which is more convenient for our general discussion.  Namely, we simply write down the most general Lagrangian that has the assumed symmetry~\cite{Weinberg2}.  Clearly, the Lagrangian derived from the former approach always falls into this general form, and all terms allowed by symmetry should be generated at least in the process of renormalization.

We assume rotational invariance of
space, but no Lorentz invariance.  There are terms that have not been
considered traditionally.  The Lagrangian is considered to be an
expansion in the number of derivatives to study long-range and
low-energy excitations of the system.  We restrict ourselves to terms
up to second order in derivatives because they are sufficient to read
off the number and dispersion relations of NGBs for most purposes.  To work
out symmetry requirements on the functional forms of each term in the
Lagrangian, differential forms turn out to be very useful.

\subsection{Coset space}
Suppose that the symmetry group $G$ of a microscopic Lagrangian is
spontaneously broken down to its subgroup $H$.  The set of the
degenerate ground states forms the coset space $G/H$. The low-energy
effective Lagrangian is the nonlinear sigma model with the target
space $G/H$.  We consider only exact symmetries ({\it i.e.\/}, without
anomalies or explicit breaking).  We also set $\hbar=1$ throughout the
paper.  Except in Sec.~\ref{sec:galilei} and a few examples in \ref{sec:scaling}, we assume
that $G$ and $H$ are compact Lie groups for internal symmetries.

Let $\pi^a$ ($a=1,\ldots,\text{dim}\,G/H$) be a local coordinate of $G/H$.  By definition, the number of \textit{fields} always equals the number of broken generators $\text{dim}\,G/H$.  Every point on this space is equivalent, and we pick the origin $\pi^a=0$ as our ground state.  The NG field $\pi^a(\vec{x},t)$ is a map $\pi:\mathbb{R}^{d+1}\rightarrow G/H$. ($d$ is the spatial dimension.)

$\pi^a$'s form a nonlinear realization of $G$. They transform under $\epsilon^iQ_i$ as
\begin{equation}
\delta_\epsilon \pi^a=\epsilon^ih^a_i(\pi).\label{symmetryh}
\end{equation} 
Generators $h_i^a(\pi)$ can be viewed as vector fields on $G/H$
\begin{equation}
h_i(\pi)= h_i^a(\pi)\partial_a,\quad \partial_a\equiv\frac{\partial}{\partial \pi^a},
\end{equation}
and their Lie bracket is identified with the commutation relation
\begin{equation}
[h_i,h_j]\equiv (h_i^b\partial_b h_j^a-h_j^b\partial_b h_i^a)\partial_a=f_{ij}^{\phantom{ij}k}h_k.
\label{eq:cr}
\end{equation}
Here, $i,j,k,\ldots$ refer to generators of $G$.

In general, we will look for the most general Lagrangian
$\mathcal{L}_{\text{eff}}(\pi,\dot{\pi},\nabla_r\pi,\ddot{\pi}^a,
\nabla_r\dot{\pi}^a,\nabla_r\nabla_s\pi^a,\ldots)$ that only changes
by total derivatives under the transformation in
Eq.~\eqref{symmetryh}.  A particularly useful choice of the nonlinear
realization is given by the Callan-Coleman-Wess-Zumino coset
construction~\cite{Coleman2,Callan}, which we introduce in
Sec.~\ref{sec:prelim}.

If the symmetry can be gauged, parameters of symmetry transformations
are local $\epsilon^i(x)$, and we may introduce gauge fields that
transform as
\begin{align}
\delta_{\epsilon} A^i_{\mu}(x)&=[\mathcal{D}_{\mu}\epsilon(x)]^i=\nabla_\mu\epsilon^i(x)+f_{jk}^{\phantom{jk}i}A^j_\mu\epsilon^k(x),\label{symmetryA}
\end{align}
where $A_\mu^i=(A_t^i,\vec{A}^i)$ and
$\nabla_\mu=(\nabla_t,\vec{\nabla})$.  However, not all symmetries can
be gauged.  Such examples are discussed in Sec.~\ref{sec:central}.  In
order to keep the full generality, we first proceed without gauging
the symmetry.  We will then discuss the local symmetry and clarify the
obstruction.

\subsection{Derivative expansion and symmetry requirements}
We postulate the locality of the microscopic Lagrangian; {\it i.e.}, it does not include terms containing fields at two separated points $(\vec{x},t)$ and $(\vec{x}',t')$. Then, the effective Lagrangian obtained by integrating our higher-energy modes should stay local~\footnote{The condition of the locality can be relaxed to an exponential decay $x^{r}e^{-\kappa|\vec{x}-\vec{x}'|}$ ($r\in\mathbb{R}$, $\kappa>0$). This type of term can be well approximated by the derivative expansion in a strictly local Lagrangian.}.

To study the low-energy structure of the effective Lagrangian systematically, we employ the derivative expansion.  Namely, we expand the Lagrangian 
in the power series of the time derivative $\nabla_t$ and the spatial
derivative $\nabla_r$ ($r,s=1,\ldots,d$).  We do not require Lorentz
invariance but we do require spatial rotational symmetry.  Because of the
lack of the Lorentz invariance, the space and time derivatives may
scale differently.  For example, $O(\nabla_t^2)$ and $O(\nabla_r^2)$
may not be of the same order in a derivative expansion.  We also assume
the broken symmetries are internal symmetries, and hence the NG fields
are spacetime scalars.

To avoid possible confusion, we use $\nabla_r$ to represent the spatial derivative and $\nabla_t$ or a``dot'' to represent the time derivative.  $\partial_a\equiv\partial/\partial\pi^a$ ($a=1,\ldots,\text{dim}\,G/H$) refers to the derivatives with respect to internal coordinates of $G/H$.

With these cautions in mind, we find the most general form of the
effective Lagrangian~\cite{Leutwyler:1994nonrel} up to the second
order in derivatives in $3+1$ dimensions and above:
\begin{equation}
\mathcal{L}_{\mathrm{eff}}=c_a(\pi)\dot{\pi}^a
+\frac{1}{2}\bar{g}_{ab}(\pi)\dot{\pi}^a\dot{\pi}^b
-\frac{1}{2}g_{ab}(\pi)\vec{\nabla}\pi^a\cdot\vec{\nabla}\pi^b.
\label{eq:leff}
\end{equation}
In $1+1$ dimensions, there is no spatial rotation, and therefore, we can
add three more terms: 
\begin{eqnarray}
\tilde{c}_a(\pi)\nabla_x\pi^a+\tilde{g}_{ab}(\pi)\dot{\pi}^a\nabla_x\pi^b+\tilde{b}_{ab}(\pi)\dot\pi^a\nabla_x\pi^b.\label{new1}
\end{eqnarray}
Also, in $2+1$ dimensions, there is an invariant antisymmetric tensor
$\epsilon^{rs}$, and therefore,
\begin{eqnarray}
-\frac{1}{2}b_{ab}(\pi)\epsilon^{rs}\nabla_r\pi^a\nabla_s\pi^b\label{new2}
\end{eqnarray}
is allowed.  $g_{ab}$, $\bar{g}_{ab}$, and $\tilde{g}_{ab}$ are
symmetric, and $b_{ab}$ and $\tilde{b}_{ab}$ are antisymmetric with
respect to $a$ and $b$.  Terms that contain $\ddot{\pi}^a$,
$\nabla_r\dot{\pi}^a$, and $\nabla_r\nabla_s\pi^a$ can be brought to the
above form by integration by parts.

We discuss that the $c_a(\pi)$ term can be interpreted as the Berry
phase in Sec.~\ref{sec:berry}.  The terms in Eqs.~\eqref{new1} and
\eqref{new2} have not been taken into account in
Ref.~\cite{Leutwyler:1994nonrel}.  However, they preserve the assumed
rotational invariance in $1+1$ or $2+1$ dimensions and therefore are
allowed, in general.   We present an example of them in Sec.~\ref{sec:bterm}.  

There are two subtleties about the terms $\tilde{c}_a(\pi)$ and $b_{ab}(\pi)$. First, the energy functional derived by the Lagrangian~\eqref{eq:leff} plus the terms in Eq.~\eqref{new1} is
\begin{eqnarray}
\int\mathrm{d}^dx\left[\frac{1}{2}\bar{g}_{ab}\dot{\pi}^a\dot{\pi}^b+\frac{1}{2}g_{ab}\nabla_x\pi^a\nabla_x\pi^b-\tilde{c}_a\nabla_x\pi^a\right],\label{Hamiltonian}
\end{eqnarray}
In the Fourier space, the second term is $O(k_x^2)$ and the last term is $O(k_x)$. Thus, the energy is minimized by a nonzero $k_x$ and the translational symmetry will be spontaneously broken.  Although the $O(k_x)$ term and the $O(k_x^2)$ term balance against each other, this solution may still be consistent with the derivative expansion if the coefficient of the $O(k_x)$ term is somehow small.  Since our main interest is in the situation with unbroken translational symmetry, we will not discuss the consequences of this term any further.

Second, $\tilde{c}_a(\pi)$ and $b_{ab}(\pi)$ cannot be Wess-Zumino-Witten type terms (see Sec.~\ref{sec:central}).  They appear in the energy functional, unlike the terms $c_a(\pi)$ and $\tilde{b}_{ab}(\pi)$, which are linear in the time derivative.  In order for the energy to be well-defined, $\int\mathrm{d}x\, \tilde{c}_a(\pi)\nabla_x\pi^a$ and $\int\mathrm{d}^2x(1/2)b_{ab}(\pi)\epsilon^{rs}\nabla_r\pi^a\nabla_s\pi^b$ cannot possess the ambiguity  of $2\pi k$ ($k\in \mathbb{Z}$).  Another way of putting it is the Wick rotation.  In the case of $c_a(\pi)$ and $\tilde{b}_{ab}(\pi)$,  the factor of $i$ from their time derivative $\partial_t$ and from $\mathrm{d}t$ in the integral measure cancel each other out under the Wick rotation and the ambiguity of the action remains to be an integer multiple of $2\pi i$. However, if either $\tilde{c}_a(\pi)$ or $b_{ab}(\pi)$ were a Wess-Zumino-Witten-type term, the absolute value of the path-integral weight would not be well defined after the Wick rotation due to the lack of a time derivative.

Our task is to determine coefficients $c_a(\pi)$, $\tilde{c}_a(\pi)$, $g_{ab}(\pi)$, $\bar{g}_{ab}(\pi)$, $\tilde{g}_{ab}(\pi)$, $b_{ab}(\pi)$, and $\tilde{b}_{ab}(\pi)$ by imposing the global symmetry $G$.  

Under global transformation~\eqref{symmetryh}, the first term of the Lagrangian~\eqref{eq:leff} transforms as
\begin{equation}
\delta_i(c_a\dot{\pi}^a)=(h_i^b\partial_b c_a+c_b\partial_ah_i^b)\dot{\pi}^a.
\end{equation}
By requiring that this combination is a total derivative $\nabla_t(e_i+c_ah_i^a)$, we find
\begin{equation}
(\partial_b c_a-\partial_a c_b)h_i^b=\partial_ae_i.\label{eq:delta1}
\end{equation}
Similarly, for $\tilde{c}_a(\pi)$, $b(\pi)$, and $\tilde{b}(\pi)$, we have
\begin{eqnarray}
(\partial_b\tilde{c}_a-\partial_a\tilde{c}_b)h_i^b&=&\partial_a\tilde{e}_i,\label{eq:delta2}\\
(\partial_a b_{bc}+\partial_b b_{ca}+\partial_c b_{ab})h_i^c&=&\partial_a e_{ib}'-\partial_b e_{ia}',\label{eq:delta3}\\
(\partial_a \tilde{b}_{bc}+\partial_b\tilde{b}_{ca}+\partial_c\tilde{b}_{ab})h_i^c&=&\partial_a\tilde{e}_{ib}'-\partial_b\tilde{e}_{ia}'.\label{eq:delta4}
\end{eqnarray}
Here, $\tilde{e}_i(\pi)$, $e_{ia}'(\pi)$, and $\tilde{e}_{ia}'(\pi)$ are also related to the change of the Lagrangian by total derivatives $\nabla_t(\tilde{e}_i+\tilde{c}_ah_i^a)$, $\nabla_r[\epsilon^{rs}(e_{ib}'+b_{ab}h_i^a)\nabla_s\pi^b]$, and $\nabla_t[(\tilde{e}_{ib}'+\tilde{b}_{ab}h_i^a)\nabla_x\pi^b]-\nabla_x[(\tilde{e}_{ib}'+\tilde{b}_{ab}h_i^a)\dot{\pi}^b]$.  

In contrast, the second term of Eq.~\eqref{eq:leff} must be invariant by itself; {\it i.e.}\/, they cannot change by a surface term.
\begin{eqnarray}
&&\delta_i\left(g_{ab}\vec{\nabla}\pi^a\cdot\vec{\nabla}\pi^b\right)\notag\\
&&=(h_i^c\partial_c g_{ab}+g_{c b}\partial_a h_i^c+g_{ac}\partial_b h_i^c)\vec{\nabla}\pi^a\cdot\vec{\nabla}\pi^b=0.\label{changeofg}
\end{eqnarray}
If the left hand side of Eq.~\eqref{changeofg} were a total derivative $\nabla_r\Lambda_i^r$, $\Lambda_i^r$
would take the form $f_{ia}(\pi)\nabla_r\pi^a$.  However,
$\nabla_r\Lambda_i^r$ then contains a term $\nabla_r^2\pi^a$, which
was absent in Eq.~\eqref{changeofg}. Thus, $\Lambda_i^r$ has to be
$0$. Therefore,
\begin{equation}
h_i^c\partial_c g_{ab}+g_{c b}\partial_a h_i^c+g_{ac}\partial_b h_i^c=0.\label{killing11}
\end{equation}
The same equation holds for $\bar{g}_{ab}(\pi)$ and $\tilde{g}_{ab}(\pi)$:
\begin{eqnarray}
h_i^c\partial_c \bar{g}_{ab}+\bar{g}_{c b}\partial_a h_i^c+\bar{g}_{ac}\partial_b h_i^c&=&0,\label{killing12}\\
h_i^c\partial_c \tilde{g}_{ab}+\tilde{g}_{c b}\partial_a h_i^c+\tilde{g}_{ac}\partial_b h_i^c&=&0.\label{killing13}
\end{eqnarray}

In summary, coefficients in the effective Lagrangian must obey the differential equations~\eqref{eq:delta1}--\eqref{eq:delta4} and \eqref{killing11}--\eqref{killing13} in order that the Lagrangian has the symmetry $G$. We also have to derive the differential equations for $e_i(\pi)$, $\tilde{e}_i(\pi)$, $e_{ia}'(\pi)$, and $\tilde{e}_{ia}'(\pi)$ and it can easily be done by using the mathematical technique we introduce in the next section.

\subsection{Geometric derivation}
\subsubsection{Equations on $c(\pi)$'s and $g(\pi)$'s}
Here, we rederive the above differential equations by using differential geometry, to set up notations and introduce useful mathematical tools for later calculation.  The terms in the effective Lagrangian can be viewed as one-forms
\begin{eqnarray}
c(\pi)&=&c_a(\pi)\mathrm{d}\pi^a,\\
\tilde{c}(\pi)&=&\tilde{c}_a(\pi)\mathrm{d}\pi^a,
\end{eqnarray}
symmetric tensors
\begin{eqnarray}
g(\pi)=g_{ab}(\pi)\mathrm{d}\pi^a\otimes\mathrm{d}\pi^b,\\
\bar{g}(\pi)=\bar{g}_{ab}(\pi)\mathrm{d}\pi^a\otimes\mathrm{d}\pi^b,\\
\tilde{g}(\pi)=\tilde{g}_{ab}(\pi)\mathrm{d}\pi^a\otimes\mathrm{d}\pi^b,
\end{eqnarray}
and two-forms
\begin{eqnarray}
b(\pi)=b_{ab}(\pi)\mathrm{d}\pi^a\wedge\mathrm{d}\pi^b,\\
\tilde{b}(\pi)=\tilde{b}_{ab}(\pi)\mathrm{d}\pi^a\wedge\mathrm{d}\pi^b
\end{eqnarray}
on the manifold $G/H$.  Note that $c(\pi)$, $\tilde{c}(\pi)$, $b(\pi)$, and $\tilde{b}(\pi)$ do not necessarily exist globally.

In the following, we use Cartan's magic formula that relates the Lie
derivative $\mathscr{L}_{X}$, the exterior derivative $\mathrm{d}$,
and the interior product $\mathrm{i}_X$:
\begin{equation}
\mathscr{L}_X\omega=(\mathrm{d}\, \mathrm{i}_X+\mathrm{i}_X\mathrm{d})\omega.\label{magicformula}
\end{equation}
Equation \eqref{magicformula} is true for arbitrary forms $\omega$ and vector fields
$X$~\cite{Nakahara,Eguchi:1980jx}.

We require the Lie derivative of the effective Lagrangian along a vector $h_i$ to be a total derivative,
\begin{equation}
\mathscr{L}_{h_i}\mathcal{L}_{\mathrm{eff}}=\mathrm{d}\Lambda_i.
\end{equation}

Let us first focus on the one-form $c$.  To fulfill the symmetry requirement
\begin{equation}
\mathscr{L}_{h_i}c=\mathrm{d}(\mathrm{i}_{h_i}c)+\mathrm{i}_{h_i}\mathrm{d}c=\mathrm{d}(e_i+\mathrm{i}_{h_i}c),
\label{hic}
\end{equation}
we need
\begin{equation}
\mathrm{i}_{h_i}\mathrm{d}c=\mathrm{d}e_i.\label{eq:idc}
\end{equation}
Equation \eqref{eq:idc} is nothing but Eq.~\eqref{eq:delta1}.  In the same way, one can obtain 
\begin{eqnarray}
\mathrm{i}_{h_i}\mathrm{d}\tilde{c}=\mathrm{d}\tilde{e}_i,\quad\mathrm{i}_{h_i}\mathrm{d}b=\mathrm{d}e_i',\quad\mathrm{i}_{h_i}\mathrm{d}\tilde{b}=\mathrm{d}\tilde{e}_i',\label{eq:idb}
\end{eqnarray}
which correspond to Eqs.~\eqref{eq:delta2}--\eqref{eq:delta4}.  Note that the definitions of $e_i$, $\tilde{e}_i$, $e_i'$, and $\tilde{e}_i'$
in Eqs.~\eqref{eq:idc} and \eqref{eq:idb} fix them only up to a
constant or a closed one-form.  We will come back to this
ambiguity shortly.

Finally, Eqs.~\eqref{killing11}--\eqref{killing13} are nothing but the Killing equation for $G$-invariant metrics
\begin{equation}
\mathscr{L}_{h_i}g=0,\quad \mathscr{L}_{h_i}\bar{g}=0,\quad \mathscr{L}_{h_i}\tilde{g}=0.
\label{killing}
\end{equation}
If $\pi^a$ transforms irreducibly under the unbroken symmetry $H$, the invariant metric on $G/H$ is unique and $g$, $\bar{g}$, and $\tilde{g}$ may differ only by an overall factor.  In general, they may differ by overall factors for each irreducible representation [see Eq.~\eqref{conditiong}]. 

\subsubsection{Equations on $e_i(\pi)$'s and $e_i'(\pi)$'s}
In order to solve Eqs.~\eqref{eq:delta1}--\eqref{eq:delta4}, we have to specify the functions $e_i(\pi)$ and $\tilde{e}_i(\pi)$ and one-forms $e_i'(\pi)=e_{ia}'(\pi)\mathrm{d}\pi^a$ and $\tilde{e}_i'(\pi)=\tilde{e}_{ia}'(\pi)\mathrm{d}\pi^a$.  We show that they obey the differential equations
\begin{eqnarray}
\mathscr{L}_{h_i}e_j&=&f_{ij}^{\phantom{ij}k}e_k+z_{ij},\label{eq:e}\\
\mathscr{L}_{h_i}\tilde{e}_j&=&f_{ij}^{\phantom{ij}k}\tilde{e}_k+\tilde{z}_{ij},\label{eq:eb}\\
\mathscr{L}_{h_i}e_j'&=&f_{ij}^{\phantom{ij}k}e_k'+\mathrm{d}z_{ij}',\label{eq:ep}\\
\mathscr{L}_{h_i}\tilde{e}_j'&=&f_{ij}^{\phantom{ij}k}\tilde{e}_k'+\mathrm{d}\tilde{z}_{ij}',\label{eq:epb}
\end{eqnarray}
where $z_{ij}$ and $\tilde{z}_{ij}$ are constants and $z_{ij}'(\pi)$ and $\tilde{z}_{ij}'(\pi)$ are functions.  For example, given the initial condition $e_i(0)$ and the constants $z_{ij}$, we can solve Eq.~\eqref{eq:e} to find $e_i(\pi)$.

If possible, we always remove $z_{ij}$, $\tilde{z}_{ij}$,
$z_{ij}'(\pi)$, and $\tilde{z}_{ij}'(\pi)$ from
Eqs.~\eqref{eq:e}--\eqref{eq:epb} by shifting $e_i(\pi)$ and $\tilde{e}_i(\pi)$ by constants and $e_i'(\pi)$ and $\tilde{e}_i'(\pi)$ by
closed one-forms using the above-mentioned ambiguity.  However, they
cannot always be completely removed.  For example, $z_{ij}$ cannot be
eliminated when the second cohomology of the Lie algebra
$H^2(\mathfrak{g})$ is nontrivial.  (See Appendix~\ref{appendix1}
for a brief review of this subject.)  In Sec.~\ref{sec:central}, we
show that the nontrivial $z_{ij}$ corresponds to a central extension
of the Lie algebra.

To derive Eq.~\eqref{eq:e}, we first note that the Lie derivative of the two-form $\mathrm{d}c$ vanishes,
\begin{equation}
\mathscr{L}_{h_i}\mathrm{d}c=\mathrm{d}^2e_i+\mathrm{i}_{h_i}\mathrm{d}^2c=0.\label{eq:lhidc}
\end{equation}
We also use the commutativity $\mathscr{L}_{h_i}\mathrm{d}=\mathrm{d}\mathscr{L}_{h_i}$ and a property of the interior product,
\begin{equation}
\mathscr{L}_{h_i}\mathrm{i}_{h_j}=f_{ij}^{\phantom{ij} k}\mathrm{i}_{h_k}+\mathrm{i}_{h_j}\mathscr{L}_{h_i}.\label{eq:lhidc22}\\
\end{equation}
Combining Eqs.~\eqref{eq:lhidc} and \eqref{eq:lhidc22} with Eqs.~\eqref{eq:idc}, we obtain
\begin{eqnarray}
\mathrm{d}(\mathscr{L}_{h_i}e_j)&=&\mathscr{L}_{h_i}(\mathrm{d}e_j)=\mathscr{L}_{h_i}(\mathrm{i}_{h_j}\mathrm{d}c)\notag\\
&=&f_{ij}^{\phantom{ij} k}(\mathrm{i}_{h_k}\mathrm{d}c)+\mathrm{i}_{h_j}(\mathscr{L}_{h_i}\mathrm{d}c)\notag\\
&=&\mathrm{d}(f_{ij}^{\phantom{ij} k}e_k),
\end{eqnarray}
which proves Eq.~\eqref{eq:e}.  Exactly the same derivation applies to Eqs.~\eqref{eq:eb}--\eqref{eq:epb}.

\subsection{Local symmetry}
\label{local}
Here we discuss the case where the symmetry $G$ can be gauged.  Since
gauge fields appear in covariant derivatives, it is natural to assume
that $A_\mu^i=(A_t^i,\vec{A}^i)$ is of the same order as
$\nabla_\mu=(\nabla_t,\vec{\nabla})$ in derivative expansion.
Equation~\eqref{eq:leff} is then replaced by the sum of the following
terms~\cite{Leutwyler:1994nonrel,Leutwyler:1994rel}:
\begin{eqnarray}
\mathcal{L}^{(0,1)}_{\mathrm{eff}}&=&c_a(\pi)\dot{\pi}^a+e_i(\pi)A^i_t,\\
\mathcal{L}^{(0,2)}_{\mathrm{eff}}&=&\frac{1}{2}\bar{g}_{ab}(\pi)\dot{\pi}^a\dot{\pi}^b\notag\\
&&\quad-\bar{h}_{ia}(\pi)A^i_t\dot{\pi}^a+\frac{1}{2}\bar{k}_{ij}(\pi)A^i_tA^j_t,\\
\mathcal{L}^{(2,0)}_{\mathrm{eff}}&=&-\frac{1}{2}g_{ab}(\pi)\vec{\nabla}\pi^a\cdot\vec{\nabla}\pi^b\notag\\
&&\quad+h_{ia}(\pi)\vec{A}^i\cdot\vec{\nabla}\pi^a-\frac{1}{2}k_{ij}(\pi)\vec{A}^i\cdot\vec{A}^j.
\end{eqnarray}
Here, $k_{ij}(\pi)$ and $\bar{k}_{ij}(\pi)$ are symmetric with respect to $i$ and $j$

As discussed before, one can add 
\begin{eqnarray}
{\mathcal{L}^{(0,1)}_{\mathrm{eff}}}'&=&\tilde{c}_a(\pi)\nabla_x\pi^a+\tilde{e}_i(\pi)A_x^i,\\
{\mathcal{L}^{(1,1)}_{\mathrm{eff}}}'&=&\tilde{g}_{ab}(\pi)\dot{\pi}^a\nabla_x\pi^b-\tilde{h}_{ia}(\pi)\left(A_t^i\nabla_x\pi^a+A_x^i\dot{\pi}^a\right)\notag\\
&&\quad+\tilde{k}_{ij}(\pi)A_t^iA_x^j,\\
{\mathcal{L}^{(1,1)}_{\mathrm{eff}}}''&=&
\tilde{b}_{ab}(\pi)\dot\pi^a\nabla_x\pi^b+\tilde{e}_{ia}'(\pi)(A^i_t\nabla_x\pi^a-A^i_x\dot\pi^a)\notag\\
&&\quad+\tilde{a}_{ij}(\pi)A^i_tA^j_x\label{eq:btx}
\end{eqnarray}
in $1+1$ dimensions, and 
\begin{eqnarray}
{\mathcal{L}^{(2,0)}_{\mathrm{eff}}}'&=&
-\frac{1}{2}b_{ab}(\pi)\epsilon^{rs}\nabla_r\pi^a\nabla_s\pi^b-e_{ia}'(\pi)\epsilon^{rs}A^i_r\nabla_s\pi^a\notag\\
&&\quad-\frac{1}{2}a_{ij}(\pi)\epsilon^{rs}A^i_rA^j_s\label{eq:b}
\end{eqnarray}
in $2+1$ dimensions.  Here, $\tilde{k}_{ij}(\pi)$ is symmetric and $a_{ij}(\pi)$ and  $\tilde{a}_{ij}(\pi)$ are antisymmetric.

We require that the action $S_{\text{eff}}[\pi,A]=\int\mathrm{d}^dx\mathrm{d}t\,\mathcal{L}_{\text{eff}}$ is invariant under the local transformations $\pi'(x)=\pi(x)+\delta_\epsilon\pi(x)$ and $A'(x)=A(x)+\delta_\epsilon A(x)$, where $\delta_\epsilon \pi^a$ and $\delta_\epsilon A(x)$ are defined in Eqs.~\eqref{symmetryh} and \eqref{symmetryA}.  Here, we assume that the infinitesimal parameters $\epsilon^i(x)$ vanish as $|x|\rightarrow 0$.
The invariance of the action can be reexpressed as
\begin{eqnarray}
0&=&\delta_\epsilon S_{\text{eff}}[\pi,A]\notag\\
&=&\int\mathrm{d}^dx\mathrm{d}t\left[\frac{\delta S_{\text{eff}}}{\delta\pi^a}\delta_\epsilon\pi^a+\frac{\delta S_{\text{eff}}}{\delta A_\mu^i}\delta_\epsilon A_\mu^i\right]\notag\\
&=&\int\mathrm{d}^dx\mathrm{d}t\,\epsilon^i(x)\left[\frac{\delta S_{\text{eff}}}{\delta\pi^a}h_i^a-(\mathcal{D}_\mu)_{i}^{\phantom{i}j}\frac{\delta S_{\text{eff}}}{\delta A_\mu^j}\right],
\end{eqnarray}
where $(\mathcal{D}_\mu)_{i}^{\phantom{i}j}=\delta_i^{\phantom{i}j}\nabla_\mu+f_{ik}^{\phantom{ki}j}A_\mu^k$.
Therefore, the effective Lagrangian must satisfy
\begin{equation}
h_i^a(\pi)\frac{\delta S_{\text{eff}}}{\delta\pi^a}=(\mathcal{D}_\mu)_{i}^{\phantom{i}j}\frac{\delta S_{\text{eff}}}{\delta A_\mu^j}.\label{symmetry}
\end{equation}
This condition leads to the differential equations we have derived above.  For example, Eq.~\eqref{symmetry} for $\mathcal{L}^{(0,1)}_{\mathrm{eff}}$ is
\begin{eqnarray}
0&=&\dot{\pi}^b[h_i^a(\partial_ac_b-\partial_bc_a)-\partial_be_i]\notag\\
&&+A_t^j[h_i^a\partial_a e_j-f_{ij}^{\phantom{ji}k}e_k],
\end{eqnarray}
which leads to the differential equations for $c_a(\pi)$ and $e_i(\pi)$:
\begin{gather}
h_i^a(\partial_a c_b-\partial_bc_a)=\partial_be_i,\label{eq1}\\
h_i^a\partial_a e_j=f_{ij}^{\phantom{ji}k}e_k.\label{eq2}
\end{gather}
Similarly, for ${\mathcal{L}^{(0,1)}_{\mathrm{eff}}}'$,
\begin{gather}
h_i^a(\partial_a \tilde{c}_b-\partial_b\tilde{c}_a)=\partial_b\tilde{e}_i,\label{eq3}\\
h_i^a\partial_a \tilde{e}_j=f_{ij}^{\phantom{ji}k}\tilde{e}_k.\label{eq4}
\end{gather}
We can easily work out all the other terms in the effective Lagrangian in the same way.   

Symmetric terms $\mathcal{L}^{(0,2)}_{\mathrm{eff}}$, $\mathcal{L}^{(2,0)}_{\mathrm{eff}}$, and ${\mathcal{L}^{(1,1)}_{\mathrm{eff}}}'$ can be compactly expressed as
\begin{eqnarray}
\mathcal{L}^{(0,2)}_{\mathrm{eff}}&=&\frac{1}{2}\bar{g}_{ab}(\pi)\mathcal{D}_t\pi^a\mathcal{D}_t\pi^b,\label{gD1}\\
\mathcal{L}^{(2,0)}_{\mathrm{eff}}&=&-\frac{1}{2}g_{ab}(\pi)\vec{\mathcal{D}}\pi^a\cdot\vec{\mathcal{D}}\pi^b,\\
{\mathcal{L}^{(1,1)}_{\mathrm{eff}}}'&=&\tilde{g}_{ab}(\pi)\mathcal{D}_t\pi^a\mathcal{D}_x\pi^b.\label{gD3}
\end{eqnarray}
Here $\mathcal{D}_\mu \pi^a=\nabla_\mu \pi^a-h_i^aA_\mu^i$ is the covariant derivative and $g_{ab}(\pi)$, $\bar{g}_{ab}(\pi)$, and $\tilde{g}_{ab}(\pi)$ are $G$-invariant metrics of $G/H$, obeying the Killing equation~\eqref{killing}.  To verify Eqs.~\eqref{gD1}--\eqref{gD3}, one has to use the Lie bracket Eq.~\eqref{eq:cr} several times.

Similarly, antisymmetric terms ${\mathcal{L}^{(2,0)}_{\mathrm{eff}}}'$ and ${\mathcal{L}^{(1,1)}_{\mathrm{eff}}}''$ can also be written by the covariant derivative:
\begin{eqnarray}
{\mathcal{L}^{(2,0)}_{\mathrm{eff}}}'&=&-\frac{1}{2}b_{ab}(\pi)\epsilon^{rs}\mathcal{D}_r\pi^a\mathcal{D}_s\pi^b,\label{bgauged1}\\
{\mathcal{L}^{(1,1)}_{\mathrm{eff}}}''&=&\tilde{b}_{ab}(\pi)\mathcal{D}_t\pi^a\mathcal{D}_x\pi^b.\label{bgauged2}
\end{eqnarray}
In addition, the two-form $b(\pi)$ obeys the following equations:
\begin{gather}
\mathrm{i}_{h_i}\mathrm{d}b=\mathrm{d}e_i',\\
\mathscr{L}_{h_i}e_j'=f_{ij}^{\phantom{ij}k}e_k',\\
\mathrm{i}_{h_i}e_j'+\mathrm{i}_{h_j}e_i'=0,\label{additional1}
\end{gather}
and $\tilde{b}(\pi)$ obeys
\begin{gather}
\mathrm{i}_{h_i}\mathrm{d}\tilde{b}=\mathrm{d}\tilde{e}_i',\\
\mathscr{L}_{h_i}\tilde{e}_j'=f_{ij}^{\phantom{ij}k}\tilde{e}_k',\\
\mathrm{i}_{h_i}\tilde{e}_j'+\mathrm{i}_{h_j}\tilde{e}_i'=0.\label{additional2}\
\end{gather}

These differential equations are almost identical to those we derived before, except for the following two constraints.
\begin{enumerate}
\item $z_{ij}$, $\tilde{z}_{ij}$, $z'_{ij}(\pi)$, and $\tilde{z}'_{ij}(\pi)$ in Eqs.~\eqref{eq:e}--\eqref{eq:epb} have to vanish.
\item Additional constraints [Eqs.~\eqref{additional1} and \eqref{additional2}] must be satisfied.
\end{enumerate}
Thus, the requirement of the local invariance is stronger than the
global symmetry.  If these additional constraints are not fulfilled,
the symmetry cannot be gauged.  See Sec.~\ref{sec:central} for
a detailed discussion on examples that violate at least one of these
conditions.

\section{Solution with Maurer-Cartan form}
\label{sec:solution}
In this section, we present the exact analytic solutions to the
differential equations derived in the previous section.  We initially
assume the two conditions listed in Sec.~\ref{local}, namely, when the
symmetry is gaugeable.  Since the end result can be understood without
technical details, readers without interest in the derivation can
directly go to Sec.~\ref{ss:summary}, where we summarize our result.
We obtain the same result using an alternative formalism of gauging
the right translation by $H$ in Sec.~\ref{sec:rightH}.  Finally, in
Sec.~\ref{sec:central}, we discuss the additional terms allowed when
the symmetry is not gaugeable.

\subsection{Preliminaries}
\label{sec:prelim}
The Callan-Coleman-Wess-Zumino coset construction is a famous and useful formalism to achieve a nonlinear realization and building blocks of the effective Lagrangian~\cite{Coleman2, Callan}.   

The coset space $G/H$ can be parametrized as $U(\pi)=e^{i \Pi}$ with $\Pi=\pi^a T_a$.  Here, $T_i$ is a faithful representation of the Lie algebra $\mathfrak{g}$.  Throughout this paper, we use the following notation.
\begin{itemize}
\item $i,j,k,\ldots$ refer to generators $\mathfrak{g}$, including both broken and unbroken ones.
\item $a,b,c,\ldots$ refer to broken generators $\mathfrak{g}/\mathfrak{h}$.
\item $\rho, \sigma, \lambda,\ldots$ refer to unbroken generators $\mathfrak{h}$.
\end{itemize}
If $G$ is compact, we can always find a unitary representation of $G$
such that $T_i$'s are Hermitian and orthogonal $\mathrm{tr}(T_i
T_j)=\lambda \delta_{ij}$.  As a result, the structure constants
become fully antisymmetric; {\it i.e.}\/,
$f_{ij}^{\phantom{ij}k}=-f_{ik}^{\phantom{ik}j}=0$. However, it is not
always convenient to work in this orthogonal basis, especially when
$G$ is not semisimple, and in this section, we only use
$f_{ij}^{\phantom{ij}k}=-f_{ji}^{\phantom{ij}k}$, which follows just
by the antisymmetric property of commutators.

The transformation law of NG fields under the action of $g\in G$ is defined through the decomposition of the product $g U(\pi)$ into the form 
\begin{equation}
g U(\pi) = U(\pi'(\pi,g))h_g(\pi),\quad h_g(\pi)\in H.\label{defh}
\end{equation}
  
Now we define an important $\mathfrak{g}$-valued one-form on $G/H$, the so-called Maurer-Cartan one-form:
\begin{eqnarray}
\omega(\pi)&\equiv&-i U(\pi)^{\dagger}\mathrm{d}U(\pi)\notag\\
&=&\sum_{n=0}^\infty \frac{(-i)^n}{(n+1)!}[\underbrace{\Pi,[\Pi,\ldots,[\Pi}_n,\mathrm{d}\Pi]\ldots ]].\label{omega}
\end{eqnarray}
In the following, we use the notation $\omega(\pi)=\omega_a(\pi)\mathrm{d}\pi^a=\omega^i(\pi)T_i=\omega_a^i(\pi)\mathrm{d}\pi^aT_i$ and $A=A^iT_i=A_\mu^iT_i\mathrm{d}x^\mu$.  

Infinitesimal transformation $h_i^a(\pi)$ is defined by ${\pi'}^a=\pi^a+\epsilon^ih_i^a(\pi)+O(\epsilon^2)$ for $g=e^{i\epsilon^iT_i}$.  To find their explicit expression, we compare the order-$\epsilon$ terms in Eq.~\eqref{defh}: 
\begin{equation}
\mathrm{i}_{h_i}\omega\equiv h_i^a(\pi)\omega_a(\pi)=\nu_i^{\phantom{i}j}(\pi)T_j-T_\rho k_i^\rho(\pi),\label{h}
\end{equation} 
where $k_i^\rho(\pi,g)$ is defined by $h_g(\pi)=e^{i \epsilon^i k_i^\rho(\pi,g) T_\rho}$ and
\begin{eqnarray}
\nu_i^{\phantom{i}j}(\pi)T_j&\equiv&U(\pi)^{\dagger}T_iU(\pi)\notag\\
&=&\sum_{n=0}^\infty \frac{(-i)^n}{n!}[\underbrace{\Pi,[\Pi,\ldots,[\Pi}_n,T_i]\ldots]].\label{nu}
\end{eqnarray}
By solving Eq.~\eqref{h}, we can compute $h_i^a(\pi)$ around the origin as
\begin{eqnarray}
h_\rho^a(\pi)&=&\pi^bf_{b \rho}^{\phantom{b\rho}a}+\frac{1}{2}\pi^b\pi^cf_{b\rho}^{\phantom{b\rho}\sigma}f_{c\sigma}^{\phantom{c\sigma}a}+O(\pi^3),\label{ubh}\\
h_b^a(\pi)&=&\delta_b^a+\frac{1}{2}\pi^cf_{c b}^{\phantom{b c}a}+O(\pi^2).\label{bh}
\end{eqnarray}
Note, in particular, that $h_b^a(0)=\delta_b^a$ and $h_{\rho}^a(0)=0$ at $\pi=0$, meaning that the broken generator $h_a$ shifts $\pi^a$ and that the unbroken generator $h_\rho$ does not change the ground state.

The transformation law of the Maurer-Cartan form follows from the definition \eqref{defh}:
\begin{eqnarray}
\omega(\pi')=-i (h_gU^\dagger g^\dagger)\mathrm{d}(gUh_g^\dagger)\notag\\
=h_g\omega(\pi)h_g^\dagger-ih_g\mathrm{d}h_g^\dagger.
\end{eqnarray}
It is convenient to decompose the Maurer-Cartan forms $\omega = \omega_\perp+\omega_{\parallel}$, where $\omega_\perp=\omega^aT_a$ are in $\mathfrak{g}/\mathfrak{h}$, while $\omega_\parallel=\omega^\rho T_\rho$ are in $\mathfrak{h}$.  Since $h_g\mathrm{d}h_g^\dagger\in\mathfrak{h}$, we have
\begin{eqnarray}
  \omega_\perp(\pi')&=& h_g \omega_\perp(\pi)h_g^\dagger, \label{transomegab}\\
  \omega_\parallel(\pi')&=& h_g\omega_{\parallel}(\pi)h_g^\dagger-ih_g\mathrm{d}h_g^\dagger.\label{transomegau}
\end{eqnarray}
Their infinitesimal versions are
\begin{eqnarray}
\mathscr{L}_{h_i}\omega^a(\pi)&=&-f_{\rho b}^{\phantom{\rho b}a}k_i^\rho(\pi)\omega^b(\pi),\label{homegab}\\
\mathscr{L}_{h_i}\omega^\rho(\pi)&=&- f_{\lambda\sigma}^{\phantom{\lambda\sigma}\rho}k_i^\lambda(\pi)\omega^\sigma(\pi)-\mathrm{d}k_i^\rho(\pi).\label{homegau}
\end{eqnarray}

When we gauge the symmetry $G$ by introducing gauge fields that obey the transformation rule in Eq.~\eqref{symmetryA}, the Maurer-Cartan form no longer transforms covariantly, {\it i.e.}\/, does not obey Eq.~\eqref{transomegab} for local transformation.  Instead, the combination
\begin{eqnarray}
(\omega_\perp)_a\mathcal{D}\pi^a&\equiv&(\omega_\perp)_a(\mathrm{d}\pi^a-h_i^aA^i)\notag\\
&=&[-iU^\dagger(\mathrm{d}-iA)U]_\perp\label{leftlocal}
\end{eqnarray}
transforms covariantly.

It is also straightforward to verify the following useful relations:
\begin{eqnarray}
\mathrm{d}\omega^k(\pi)&=&\frac{1}{2}f_{ij}^{\phantom{ij}k}\omega^i(\pi)\wedge\omega^j(\pi),\label{dw}\\
\mathscr{L}_{h_i}\nu_j^{\phantom{j}k}(\pi)&=&f_{ij}^{\phantom{ij}l}\nu_l^{\phantom{l}k}(\pi)-f_{\rho l}^{\phantom{l\rho}k}k_i^\rho(\pi) \nu_j^{\phantom{j}l}(\pi),\label{hiv}\\
\mathrm{d}\nu_i^{\phantom{i}k}(\pi)&=&f_{jl}^{\phantom{jl}k}\omega^j(\pi)\nu_i^{\phantom{i}l}(\pi).\label{dv}
\end{eqnarray}

Finally, we note that the last line of Eqs.~\eqref{omega} and \eqref{nu} is written in terms of commutation relations. Therefore, the Maurer-Cartan form $\omega(\pi)$ and generators $h_i(\pi)$ \textit{do not} fundamentally depend on a specific choice of the representation of $T_i$.

With these preparations, we now present our analytic solutions to the
differential equations derived in Sec.~\ref{sec:effective} one by one.

\subsection{Explicit solutions}
\subsubsection{$g(\pi)$'s}
As the first example, here we show that
\begin{eqnarray}
g(\pi)&=&g_{ab}(0)\omega^a(\pi)\otimes\omega^b(\pi)\label{solg}
\end{eqnarray}
is the solution to the Killing equation \eqref{killing}.  If NGBs transform irreducibly under the unbroken subgroup $H$, constants $g_{cd}(0)$ must be proportional to $\delta_{cd}$. In the most general case, $g_{cd}(0)$ has to be invariant under unbroken symmetries; namely,
\begin{eqnarray}
f_{\rho a}^{\phantom{\rho a}c}g_{c b}(0)+f_{\rho b}^{\phantom{\rho b}c}g_{ac}(0)=0,\label{conditiong}
\end{eqnarray}
which can be derived from the Killing equation \eqref{killing} at the origin $\pi=0$ with the help of Eq.~\eqref{ubh}.  

To see that $g(\pi)$ in Eq.~\eqref{solg} is the solution of Eq.~\eqref{killing}, we use Eq.~\eqref{homegab}:
\begin{eqnarray}
\mathscr{L}_{h_i}g&=&\mathscr{L}_{h_i}[g_{cd}(0)\omega^c(\pi)\otimes\omega^d(\pi)]\notag\\
&=&g_{cd}(0)[(\mathscr{L}_{h_i}\omega^c)\otimes\omega^d+\omega^c\otimes(\mathscr{L}_{h_i}\omega^d)]\notag\\
&=&-k_i^\rho[g_{ed}(0)f_{\rho c}^{\phantom{\rho c}e}+g_{ce}(0)f_{\rho d}^{\phantom{\rho d}e}]\omega^c\otimes\omega^d.
\end{eqnarray}
The combination in the square brackets vanishes thanks to Eq.~\eqref{conditiong}.  Solution~\eqref{solg} also respects the initial value since $\omega^a=\mathrm{d}\pi^a$ at $\pi=0$. Hence, Eq.~\eqref{solg} is the unique solution of Eq.~\eqref{killing}.  

The same is true for $\bar{g}_{ab}(\pi)$ and $\tilde{g}_{ab}(\pi)$; {\it i.e.}\/,
\begin{eqnarray}
\bar{g}(\pi)=\bar{g}_{ab}(0)\omega^a(\pi)\otimes\omega^b(\pi),\\
\tilde{g}(\pi)=\tilde{g}_{ab}(0)\omega^a(\pi)\otimes\omega^b(\pi)
\end{eqnarray}
with
\begin{eqnarray}
f_{\rho a}^{\phantom{\rho a}c}\bar{g}_{c b}(0)+f_{\rho b}^{\phantom{\rho b}c}\bar{g}_{ac}(0)=0,\label{conditiongb}\\
f_{\rho a}^{\phantom{\rho a}c}\tilde{g}_{c b}(0)+f_{\rho b}^{\phantom{\rho b}c}\tilde{g}_{ac}(0)=0.\label{conditiongt}
\end{eqnarray}

\subsubsection{$e_i(\pi)$'s}
We now prove that
\begin{eqnarray}
e_i(\pi)=\nu_i^{\phantom{i}j}(\pi)e_j(0)\label{sole}
\end{eqnarray}
is the solution of Eq.~\eqref{eq:e} when $z_{ij}=0$.  By multiplying $e_k(0)$ to Eq.~\eqref{hiv}, we get
\begin{eqnarray}
&&\mathscr{L}_{h_i}[\nu_j^{\phantom{j}k}(\pi)e_k(0)]\notag\\
&&=f_{ij}^{\phantom{ij}l}[\nu_l^{\phantom{l}k}(\pi)e_k(0)]-[f_{\rho l}^{\phantom{l\rho}k}e_k(0)]k_i^\rho(\pi) \nu_j^{\phantom{j}l}(\pi).
\end{eqnarray}
The second term vanishes because Eq.~\eqref{eq:e} at $\pi=0$ implies
\begin{equation}
f_{\rho i}^{\phantom{\rho i}k}e_k(0)=0.\label{conditione}
\end{equation}
Therefore, Eq.~\eqref{sole} satisfies the differential equation~\eqref{eq:e}.  Combined with $\nu_i^{\phantom{i}j}(0)=\delta_i^j$ [see Eq.~\eqref{nu}], we conclude that this is the unique solution that is consistent with the initial value.  

Similarly, 
\begin{equation}
\tilde{e}_i(\pi)=\nu_i^{\phantom{i}j}(\pi)\tilde{e}_j(0),\quad f_{\rho i}^{\phantom{\rho i}k}\tilde{e}_k(0)=0\label{conditionet}
\end{equation}
is the solution of Eq.~\eqref{eq:eb}.

\subsubsection{$c(\pi)$'s}
\label{sec:solution3}
Next, we claim that
\begin{equation}
c(\pi)=-\omega^i(\pi)e_i(0)+\mathrm{d}\chi
\label{solc}
\end{equation}
is a solution of Eq.~\eqref{eq:idc}, where $\chi$ a smooth function.  First, we multiply $e_k(0)$ to Eqs.~\eqref{dw} and \eqref{dv} to get
\begin{eqnarray}
&\mathrm{d}[\omega^k(\pi)e_k(0)]=\frac{1}{2}f_{lj}^{\phantom{lj}k}\omega^l(\pi)\wedge\omega^j(\pi)e_k(0),&\\
&\mathrm{d}e_i(\pi)=\mathrm{d}[\nu_i^{\phantom{i}j}(\pi)e_j(0)]=f_{jl}^{\phantom{jl}k}\omega^j(\pi)\nu_i^{\phantom{i}l}(\pi)e_k(0).&\label{de2}
\end{eqnarray}
Further operating $\mathrm{i}_{h_i}$ to the former equation, we have
\begin{eqnarray}
&&\mathrm{i}_{h_i}\mathrm{d}c(\pi)\notag\\
&&=\mathrm{i}_{h_i}\mathrm{d}[-\omega^k(\pi)e_k(0)]\notag\\
&&=-f_{lj}^{\phantom{ij}k}[\mathrm{i}_{h_i}\omega^l(\pi)]\omega^j(\pi)e_k(0),\notag\\
&&=f_{jl}^{\phantom{ij}k}\nu_i^{\phantom{i}l}(\pi)\omega^j(\pi)e_k(0)-[f_{j\rho}^{\phantom{ij}k}e_k(0)]k_i^\rho(\pi)\omega^j(\pi)\notag\\
&&=\mathrm{d}e_i(\pi).
\end{eqnarray}
In the derivation, we use Eqs.~\eqref{h}, \eqref{conditione}, and \eqref{de2}.  Therefore, $c(\pi)$ in Eq.~\eqref{solc} indeed obeys the differential equation. The undetermined part $\mathrm{d}\chi$ is a total derivative term in the Lagrangian.

Similarly, $\tilde{c}(\pi)=-\omega^i(\pi)\tilde{e}_i(0)$ up to a closed one-form.

\subsubsection{$e_i'(\pi)$'s and $b(\pi)$'s}
\label{sec:bterm}
In the same way, it is not difficult to verify that
\begin{eqnarray}
e_i'(\pi)&=&e_{bc}'(0)\nu_i^b(\pi)\omega^c(\pi)\label{eq:solep},\\
\tilde{e}'_i(\pi)&=&\tilde{e}_{bc}'(0)\nu_i^b(\pi)\omega^c(\pi)
\end{eqnarray}
are the solutions of Eqs.~\eqref{eq:ep} and \eqref{eq:epb} and that
\begin{eqnarray}
b(\pi)&=&-e_{cd}'(0)\omega^c(\pi)\wedge\omega^d(\pi)+\mathrm{d}\chi',\label{eq:solb}\\
\tilde{b}(\pi)&=&-\tilde{e}_{cd}'(0)\omega^c(\pi)\wedge\omega^d(\pi)+\mathrm{d}\tilde{\chi}'
\end{eqnarray}
are the solutions of Eq.~\eqref{eq:idb}.  Constants $e_{i a}'(0)$ and $\tilde{e}_{i a}'(0)$ have to satisfy
\begin{eqnarray}
&e_{\rho a}'(0)=0,\quad e_{ab}'(0)+e_{ba}'(0)=0,&\notag\\
&f_{\rho a}^{\phantom{a\rho}c}e_{cb}'(0)+f_{\rho b}^{\phantom{\rho d}c}e_{ac}'(0)=0\label{conditionep}&
\end{eqnarray}
and
\begin{eqnarray}
&\tilde{e}_{\rho a}'(0)=0,\quad \tilde{e}_{ab}'(0)+\tilde{e}_{ba}'(0)=0,&\notag\\
&f_{\rho a}^{\phantom{a\rho}c}\tilde{e}_{cb}'(0)+f_{\rho b}^{\phantom{\rho d}c}\tilde{e}_{ac}'(0)=0.\label{conditionept}&
\end{eqnarray}

One can see that a condition for the gaugeability \eqref{additional1} is indeed fulfilled since
\begin{equation}
\mathrm{i}_{h_i}e_j'(\pi)=e_{bc}'(0)\nu_j^b(\pi)[\mathrm{i}_{h_i}\omega^c(\pi)]=e_{bc}'(0)\nu_j^b(\pi)\nu_i^c
\end{equation}
is antisymmetric with respect to $i$ and $j$, thanks to the second relation of Eq.~\eqref{conditionep}.

Among constants $e_{ia}'(0)$ that satisfy the above conditions, those which can be written as 
\begin{equation}
e_{ia}'(0)=f_{ia}^{\phantom{ia}k}C_k,\quad f_{\rho i}^{\phantom{\rho i}k}C_k=0
\end{equation}
give only a total derivative term in the Lagrangian. Indeed, from Eq.~\eqref{dw} and $f_{\rho i}^{\phantom{\rho i}k}C_k=0$, it follows that
\begin{equation}
\mathrm{d}[C_k\omega^k]=f_{ab}^{\phantom{ia}k}C_k\omega^a\wedge\omega^b.
\end{equation}
For example, for $G/H=\text{SO}(3)/\text{SO}(2)=S^2$, the choice $e_{ab}'(0)=\epsilon_{ab}$ satisfies all conditions in Eq.~\eqref{conditionep}. In this case, $\epsilon_{ab}\omega^a\wedge\omega^b$ is nothing but the $\theta$ term:
\begin{equation}
\frac{\theta}{4\pi}\vec{n}\cdot\nabla_x\vec{n}\times\nabla_y\vec{n}
\end{equation}
up to an overall factor, which is expected since $\epsilon_{ab}$ can be written as $f_{ab}^{\phantom{ab}z}=\epsilon_{abz}$ ($C_z=1$ and $C_x=C_y=0$).  

An example of $b_{ab}(\pi)$ terms that are not a total derivative is given by the coset $\text{SU}(3)/\text{U}(1)\times\text{U}(1)$.  We use the standard notation of Gell-Mann matrices $\lambda_i$ ($i=1,\ldots,8$) and set $T_i=\lambda_i/2$.  In this case,
\begin{equation}
\omega^1\wedge\omega^2,\quad \omega^4\wedge\omega^5,\quad \omega^6\wedge\omega^7\label{exampleofb}
\end{equation}
are candidates for $b_{ab}(\pi)\mathrm{d}\pi^a\wedge\mathrm{d}\pi^b$, but we have to pay attention to
\begin{eqnarray}
\mathrm{d}\omega^3&=&\omega^1\wedge\omega^2+\frac{1}{2}(\omega^4\wedge\omega^5-\omega^6\wedge\omega^7),
\label{eq:omega3}\\
\mathrm{d}\omega^8&=&\frac{\sqrt{3}}{2}(\omega^4\wedge\omega^5+\omega^6\wedge\omega^7).\label{eq:omega8}
\end{eqnarray}
Therefore, only one of the three in Eq.~\eqref{exampleofb} is not a
total derivative and affects the equation of motion.  

\subsection{Summary of the Lagrangian}
\label{ss:summary}
Let us summarize what we have shown above.  We found explicit analytic solutions for differential equations derived in Sec.~\ref{sec:effective} under the assumptions that the symmetries can be gauged. (See conditions discussed in Sec.~\ref{local}.)

In $3+1$ dimensions, the most general effective Lagrangian that has the internal symmetry $\delta_\epsilon \pi^a=\epsilon^i(x)h^a_i(\pi)$ and $\delta_{\epsilon} A^i_{\mu}=\nabla_\mu\epsilon^i(x)+f_{jk}^{\phantom{jk}i}A^j_\mu\epsilon^k(x)$ as well as the spatial rotation is given by
\begin{eqnarray}
\mathcal{L}_{\text{eff}}&=&c_a(\pi)\dot{\pi}^a+e_i(\pi)A_t^i\notag\\
&+&\frac{1}{2}\bar{g}_{ab}(\pi)\mathcal{D}_t\pi^a\mathcal{D}_t\pi^b-\frac{1}{2}g_{ab}(\pi)\vec{\mathcal{D}}\pi^a\cdot\vec{\mathcal{D}}\pi^b\label{summary}
\end{eqnarray}
to the quadratic order in derivatives. Here, $\mathcal{D}_\mu\pi^a=\nabla_\mu\pi^a-h_i^a(\pi)A_\mu^i$ is the covariant derivative.  The coefficients $c_a(\pi)$, $e_i(\pi)$, $g_{ab}(\pi)$, and $\bar{g}_{ab}(\pi)$ are given by
\begin{eqnarray}
c_a(\pi)&=&-\omega_a^i(\pi)e_i(0),\\
e_i(\pi)&=&\nu_i^{\phantom{i}j}(\pi)e_j(0),\\
g_{ab}(\pi)&=&g_{cd}(0)\omega_a^c(\pi)\omega_b^d(\pi),\\
\bar{g}_{ab}(\pi)&=&\bar{g}_{cd}(0)\omega_a^c(\pi)\omega_b^d(\pi).
\end{eqnarray}
Here, $\omega_a^i(\pi)T_i=-i U(\pi)^{\dagger} \partial_aU(\pi)$ [$U(\pi)=e^{i\pi^aT_a}$] is the Maurer-Cartan form.  The function $\nu_i^{\phantom{i}j}(\pi)$ is defined by $\nu_i^{\phantom{i}j}(\pi)T_j=U(\pi)^{\dagger}T_iU(\pi)$.  The generator $h_i^a(\pi)$ can also be solved from $h_i^a(\pi)\omega_a^b(\pi)=\nu_i^{\phantom{i}b}(\pi)$.

The Lagrangian contains only few parameters (coupling constants)
$e_i(0)$, $g_{ab}(0)$ and $\bar{g}_{ab}(0)$.  They have to be
invariant under unbroken-symmetry transformation; {\it i.e.}\/,
\begin{gather}
f_{\rho i}^{\phantom{\rho i}j}e_j(0)=0, \label{eiHinv}\\
f_{\rho a}^{\phantom{\rho a}c}g_{c b}(0)+f_{\rho b}^{\phantom{\rho
    b}c}g_{ac}(0)=0, \label{gabHinv}\\
f_{\rho a}^{\phantom{\rho a}c}\bar{g}_{c b}(0)+f_{\rho b}^{\phantom{\rho b}c}\bar{g}_{ac}(0)=0.\label{gbarabHinv}
\end{gather}

If we further demand the Lorentz invariance, $\bar{g}_{ab}(0)=c^{-2}g_{ab}(0)$ and $e_i(0)=0$, so that the Lagrangian is reduced to
\begin{eqnarray}
\mathcal{L}_{\text{eff}}=\frac{1}{2}g_{ab}(\pi)\mathcal{D}_\mu\pi^a\mathcal{D}^\mu\pi^b.\label{effectiveLorentz}
\end{eqnarray}
Equation~\eqref{effectiveLorentz} is exactly the leading-order term of the standard chiral perturbation theory.  Therefore, our effective Lagrangian equally applies to Lorentz-invariant systems.

In $2+1$ dimensions, one can add
\begin{equation}
-\frac{1}{2}b_{ab}(\pi)\epsilon^{rs}\mathcal{D}_r\pi^a\mathcal{D}_s\pi^b
\end{equation}
to the effective Lagrangian~\eqref{summary}, where
\begin{eqnarray}
b_{ab}(\pi)&=&-e_{cd}'(0)\omega_a^c(\pi)\omega_b^d(\pi)
\end{eqnarray}
with constraints Eq.~\eqref{conditionep} on $e_{ab}'(0)$.

Similarly, in $1+1$ dimensions, the following terms are allowed:
\begin{gather}
\tilde{c}_a\nabla_x\pi^a+\tilde{e}_iA_x^i+\tilde{g}_{ab}\mathcal{D}_t\pi^a\mathcal{D}_x\pi^b+\tilde{b}_{ab}\mathcal{D}_t\pi^a\mathcal{D}_x\pi^b,
\end{gather}
where
\begin{eqnarray}
\tilde{c}_a(\pi)&=&-\omega_a^i(\pi)\tilde{e}_i(0),\\
\tilde{e}_i(\pi)&=&\nu_i^{\phantom{i}j}(\pi)\tilde{e}_j(0),\\
\tilde{g}_{ab}(\pi)&=&\tilde{g}_{cd}(0)\omega_a^c(\pi)\omega_b^d(\pi),\\
\tilde{b}_{ab}(\pi)&=&-\tilde{e}_{cd}'(0)\omega_a^c(\pi)\omega_b^d(\pi)
\end{eqnarray}
with constraints Eqs.~\eqref{conditiongt}, \eqref{conditionet}, and \eqref{conditionept} on coupling constants.

\subsection{Gauging ${\cal H}$ rather than modding}
\label{sec:rightH}
It is well known (see Ref.~\cite{Bando:1987br} for a review) that the coset
construction on $G/H$ is equivalent to that on $G$ with the
right translation by ${\cal H}$ gauged.  Here we use the notation
${\cal H}$ that commutes with the left translation by $G$, as opposed
to $H \subset G$ that does not commute with $G$. The gauging of the
unbroken ${\cal H}$ symmetry eliminates unwanted NGBs.  Using this
method, it is now somewhat more transparent to derive the action in
the differential-geometric method above because the transformation
laws are linear.

We first consider $U=e^{i \Pi}$ with $\Pi = \pi^a T_a + \pi^\rho T_\rho$ for {\it all}\/ generators of $\mathfrak{g}$.  Namely, $T_\rho \in \mathfrak{h}$ and $T_a \in \mathfrak{g}/\mathfrak{h}$.  Under the global symmetry $G$, $U$ transforms as the left translation
\begin{equation}
  U(\pi) \rightarrow g U(\pi) = U(\pi').
\end{equation}
On the other hand, we require a local symmetry under the {\it
  right translation}\/ by ${\cal H}$
\begin{equation}
  U(\pi) \rightarrow U(\pi) h(x).
\end{equation}
Note that gauging the right translation of $\mathcal{H}$ is {\it different}\/ from the gauging we studied in the
previous sections that corresponds to the {\it left translation}\/.  

The point here is that one can always take the gauge $\pi^\rho =0$.
In order for $U$ to stay in this gauge, the global transformation
needs to be accompanied by a gauge transformation
\begin{equation}
  U(\pi) \rightarrow g U(\pi) h_g^\dagger (\pi) = U(\pi')
\end{equation}
with a suitable choice of $h_g \in {\cal H}$.  The end result is therefore
equivalent to writing the theory on $G/H$.

We introduce a gauge field ${\cal A} ={\cal A}^\rho T_\rho = {\cal A}_\mu \mathrm{d} x^\mu$
for the right translation gauge group ${\cal H}$ so that the
Lagrangian is invariant under both the global $G$ and the local ${\cal
  H}$.
Note that we use a different symbol from the gauge field $A^i$ in the
previous section [see, {\it e.g.}\/, Eq.~\eqref{leftlocal}] for the
left translation under $G$.  The Maurer-Cartan form $\omega = -i
U^\dagger \mathrm{d} U$ is invariant under the global $G$, while it
transforms as
\begin{equation}
  \omega \rightarrow -i h^\dagger U^\dagger \mathrm{d} (U h)
  = h^\dagger \omega h -i h^\dagger \mathrm{d} h.
\end{equation}
On the other hand, the gauge field transforms as usual:
\begin{equation}
  {\cal A} \rightarrow h^\dagger {\cal A} h + i h^\dagger \mathrm{d} h.\label{rightA}
\end{equation}
Then the combination
\begin{equation}
  \omega + {\cal A}
\end{equation}
is gauge covariant.  As before, we decompose the Maurer-Cartan forms $\omega = \omega_\perp + \omega_{\parallel}$, where
$\omega_\perp = \omega^a T_a$ are in $\mathfrak{g}/\mathfrak{h}$, while
$\omega_\parallel = \omega^\rho T_\rho$ are in $\mathfrak{h}$.  Then, the
inhomogeneous transformation occurs only on $\omega_\parallel$,
\begin{eqnarray}
  \omega_\perp &\rightarrow& h^\dagger \omega_\perp h, \\
  \omega_\parallel + {\cal A} &\rightarrow& h^\dagger (\omega_\parallel +{\cal A} ) h.
\end{eqnarray}
Therefore, we can build an invariant Lagrangian just by focusing on {\it
  local}\/ ${\cal H}$ invariance on $\omega_\perp$ and
$\omega_\parallel + {\cal A}$.

We introduce the notation for the pullback of Maurer-Cartan forms to space and time:
\begin{equation}
  \pi^* \omega = \bar{\omega} \mathrm{d} t 
  + \vec{\omega}\cdot\mathrm{d} \vec{x} 
  = - i U^\dagger \partial_i U (\dot{\pi}^i \mathrm{d} t 
  + {\nabla} \pi^i \cdot \mathrm{d} \vec{x}).
\end{equation}
They are decomposed as
\begin{eqnarray}
  & & \bar{\omega} = \bar{\omega}^a T_a 
  + \bar{\omega}^\rho T_\rho, \\
  & & \vec{\omega} = \vec{\omega}^a T_a 
  + \vec{\omega}^\rho T_\rho.
\end{eqnarray}
The general Lagrangian at the second order in the time derivative is
\begin{eqnarray}
  \lefteqn{
    \mathcal{L}_{\text{eff}} = 
    \frac{1}{2} \bar{g}_{ab}(0) \bar{\omega}^a \bar{\omega}^b
    + \frac{1}{2} \bar{g}_{\rho\sigma}(0) 
    (\bar{\omega}^\rho + {\cal A}_t^\rho)
    (\bar{\omega}^\sigma + {\cal A}_t^\sigma) } \nonumber \\
  &-&
  \frac{1}{2} g_{ab}(0) \vec{\omega}^a \cdot \vec{\omega}^b  
  - \frac{1}{2} g_{\rho\sigma}(0) (\vec{\omega}^\rho + \vec{{\cal A}}^\rho)
  \cdot (\vec{\omega}^\sigma + \vec{{\cal A}}^\sigma). \nonumber \\
\end{eqnarray}
$\bar{g}_{ab}(0)$, $\bar{g}_{\rho\sigma}(0)$, $g_{ab}(0)$, and $g_{\rho\sigma}(0)$ are all constants subject to $H$ invariance as in
the previous section [see Eqs.~\eqref{gabHinv} and \eqref{gbarabHinv}].

Because the Lagrangian is quadratic in ${\cal A}$, we can integrate it out and find
\begin{equation}
  {\cal A} = - \omega_\parallel .
\end{equation}
In addition, we can perform a gauge transformation in ${\cal H}$ to remove
all $\pi^\rho$ without a loss of generality.  Then, the Lagrangian
reduces to the form
\begin{equation}
  \mathcal{L}_{\text{eff}} = 
  \frac{1}{2} \bar{g}_{ab}(0) \bar{\omega}_{\perp}^a \bar{\omega}_{\perp}^b
  - \frac{1}{2} g_{ab}(0) \vec{\omega}_{\perp}^a \cdot \vec{\omega}_{\perp}^b  ,
\end{equation}
which can be easily verified to be the same as what we derived in earlier sections.

So far, everything is well known.  Now come the new terms we discussed
in previous sections.

We first discuss terms with a single derivative.  If the generator
$T_a \in \mathfrak{g}/\mathfrak{h}$ commutes with ${\cal H}$,
$\omega^a \rightarrow h^\dagger \omega_\perp^a h = \omega^a$ and hence
is invariant.  Therefore, we can add it to the Lagrangian.  On the
other hand, if the generator $T_\rho \in \mathfrak{h}$ commutes with
${\cal H}$, it generates a U(1) subgroup, and hence,
\begin{equation}
  \omega^\rho 
  \rightarrow h^\dagger \omega^\rho h - i (h^\dagger \mathrm{d}
  h)^\rho
  = \omega^\rho -i \mathrm{d} (\log h)^\rho.
\end{equation}
Namely, the shift is a total derivative.  It is also allowed as a term
of the Lagrangian.  In addition, the combination $(\omega^\rho +
{\cal A}^\rho)$ is invariant.  Therefore, the following terms are allowed:
\begin{equation}
  \mathcal{L}^{(0,1)}_{\mathrm{eff}}
  = - e_a(0) \omega^a - e_\rho(0) \omega^\rho
  - \bar{e}_\rho(0) (\omega^\rho + {\cal A}^\rho).
\end{equation}
The last term is removed after integrating over ${\cal A}^\rho$ together with
the quadratic terms.  Therefore, we only need to consider the first
two terms, which are nothing but
\begin{equation}
  \mathcal{L}^{(0,1)}_{\mathrm{eff}} = - e_i(0) \omega_a^i \dot{\pi}^a,
\end{equation}
which we derived in Eq.~\eqref{solc}.

The antisymmetric tensor can also be included in the same fashion,
\begin{eqnarray}
  {\mathcal{L}^{(2,0)}_{\mathrm{eff}}}'&=&
  -\frac{1}{2}e'_{ab}(0) \vec{\omega}^a \times \vec{\omega}_\perp^b
  \nonumber \\
  & &-\frac{1}{2}e'_{\rho\sigma}(0) (\vec{\omega}^\rho + \vec{{\cal A}}^\rho) 
  \times (\vec{\omega}^\sigma + \vec{{\cal A}}^\sigma).
\end{eqnarray}
$e'_{ab}(0)$ is invariant under ${\cal H}$ [see Eq.~\eqref{conditionep}].
The second line is again eliminated by integrating out
$\vec{{\cal A}}^\rho$, and the first line again can be shown to be the same
as the previous result.

The central extension or Wess-Zumino-Witten terms, however, cannot
be written using Maurer-Cartan forms, because they are not gaugeable
as we discuss in the following section.

The advantage of this formulation is that the only question is to find
${\cal H}$-invariant tensors.  It is, therefore, easier to generalize to
higher-derivative terms than solve the differential equations.  In
that case, integration over the gauge field needs to be done by an
order-by-order basis because the Lagrangian is no longer quadratic in
the gauge field.

Note that we integrate out the gauge fields ${\cal A}$ 
to show the equivalence to the results in the previous 
sections.  However, they can be kept in the Lagrangian as 
nondynamical auxiliary fields.  For some applications, such as 
large-$N$ expansion, it is more convenient to keep them.  

\subsection{Central extensions and Wess-Zumino-Witten term}
\label{sec:central}
We have presented our analytic expressions of the effective Lagrangian in terms of Maurer-Cartan forms assuming that the symmetry $G$ is gaugeable.  The conditions for the gaugeability are summarized in Sec.~\ref{local}.  In this section, we discuss examples in which at least one of these conditions is violated, making it impossible to gauge the symmetry.

\subsubsection{Central extensions}
Let us consider the case $G=\text{U}(1)\times\text{U}(1)$ and $H=\{e\}$. The NG fields $\varphi^a$ ($a=1,2$) independently change by a constant under $G$.  In such a case, the effective Lagrangian may contain
\begin{equation}
c_a(\varphi)\dot{\varphi}^a=\frac{C}{2}\epsilon_{ab}\varphi^a\dot{\varphi}^b.\label{exampleofz}
\end{equation}
with $C$ a constant.  

Here, we explain that the one-form $c=(C/2)\epsilon_{ab}\varphi^a\mathrm{d}\varphi^b$ ends up with nonzero $z_{ij}$'s in Eq.~\eqref{eq:e}.  To that end, we first compute $e_a(\varphi)$ following the definition in Eq.~\eqref{eq:idc}:
\begin{eqnarray}
\mathrm{d}c&=&\frac{C}{2}\epsilon_{ab}\mathrm{d}\varphi^a\wedge\mathrm{d}\varphi^b,\\
\mathrm{i}_{h_a}\mathrm{d}c&=&C\epsilon_{ab}\mathrm{d}\varphi^b=\mathrm{d}e_a,
\end{eqnarray}
where $h_a=\partial_a$.  Therefore, $e_a=C\epsilon_{ab}\varphi^b$ up to a constant. Their Lie derivative is
\begin{equation}
\mathscr{L}_{h_a}e_b=\partial_ae_b=-C\epsilon_{ab}.\label{eq148}
\end{equation}
Comparing Eq.~\eqref{eq148} with Eq.~\eqref{eq:e}, we see $z_{ab}=-C\epsilon_{ab}\neq0$.  Therefore, the symmetry $G$ cannot be gauged. The Lagrangian
\begin{eqnarray}
\mathcal{L}^{(0,1)}_{\mathrm{eff}}=\frac{C}{2}\epsilon_{ab}\varphi^a\dot{\varphi}^b+C\epsilon_{ab}\varphi^bA_0^a
\end{eqnarray}
changes not only by a surface term $\nabla_t(C\epsilon_{ab}\epsilon^a\varphi^b)/2$ but also by $\epsilon^az_{ab}A_0^b=-C\epsilon_{ab}\epsilon^aA_0^b$.

To make a connection to central extensions, we note that conserved charges of the internal symmetry $G$ are dominated by $Q_a=\int\mathrm{d}^dx\,j_a^0=\int\mathrm{d}^dx\,C\epsilon_{ab}\varphi^b$.  Their commutation relation can be computed by using the commutation relation $[\varphi^1(\vec{x},t),\varphi^2(\vec{x}',t)]=-C^{-1}\delta^d(\vec{x}-\vec{x}')$ as
\begin{equation}
[Q_a,Q_b]=-i\epsilon_{ab}C\Omega,\label{eq150}
\end{equation}
where $\Omega$ is the volume of the system. Naively, the shift
symmetries $\varphi^a\rightarrow\varphi^a+\epsilon^a$ for $a=1$ and
$a=2$ commute with each other but Noether charges do not.  The right hand side of Eq.~\eqref{eq150} is
the central extension of the
$\mathfrak{g}=\mathfrak{u}(1)\times\mathfrak{u}(1)$ algebra.

The shift symmetry $\psi'=\psi+c$ ($c\in\mathbb{C}$) of the free-boson Sch\"ordinger field theory \cite{Brauner}
\begin{equation}
\mathcal{L}=\frac{i}{2}(\psi^\dagger\dot{\psi}-\dot{\psi}^\dagger\psi)-\frac{1}{2m}\vec{\nabla}\psi^\dagger\cdot\vec{\nabla}\psi\label{freeSch}
\end{equation}
cannot be gauged due to the same reason, although the phase rotation $\psi\rightarrow\psi e^{i\epsilon}$ can be gauged.

The central extension is possible only when the second cohomology of
the Lie algebra $H^2(\mathfrak{g})$ is nontrivial.  Namely, $G$ must have at least two Abelian generators that commute with all the other
generators.  [See Appendix \ref{appendix1} for a brief review of $H^2(\mathfrak{g})$.]  Therefore, the corresponding terms in the Lagrangian are
always of the form $-(1/2)z_{ab}\varphi^a\dot{\varphi}^b$, where $\varphi^a$
are the NG fields for such Abelian generators, which leads to the extended algebra $[Q_a,Q_b]=iz_{ab}\Omega$.

Note that the coefficient $C$ is quantized when $G/H$ is compact.  See the discussion at the end of Sec.~\ref{sec:quantization}.

\subsubsection{Example of $\tilde{z}_{ij}'(\pi)$}
We now give an example of nonzero $\tilde{z}_{ij}'(\pi)$ in
Eq.~\eqref{eq:epb}.  We take
$G=\text{U}(1)^3=\{(\varphi^1,\varphi^2,\varphi^3)|\varphi^i \in
[0,2\pi)\}$ and $H=\{e\}$.  The effective Lagrangian may contain
\begin{equation}
{\mathcal{L}^{(1,1)}_{\mathrm{eff}}}''=
\tilde{b}_{ab}(\varphi)\dot\varphi^a\nabla_x\varphi^b
=\frac{k}{3(2\pi)^2}\epsilon_{abc}\varphi^c\dot\varphi^a\nabla_x\varphi^b,
\end{equation}
which can be regarded as the two-from
$\tilde{b}=(k/3!)(2\pi)^{-2}\epsilon_{abc}\varphi^c\mathrm{d}\varphi^a\wedge\mathrm{d}\varphi^b$. The
one-form $\tilde{e}_a'(\varphi)$ can be computed as 
\begin{eqnarray}
\mathrm{d}\tilde{b}&=&\frac{k}{3!(2\pi)^2}\epsilon_{abc}\mathrm{d}\varphi^a\wedge\mathrm{d}\varphi^b\wedge\mathrm{d}\varphi^c,\\
\mathrm{i}_{h_a}\mathrm{d}\tilde{b}&=&\mathrm{d}\left(\frac{k}{2(2\pi)^2}\epsilon_{abc}\varphi^b\mathrm{d}\varphi^c\right)=\mathrm{d}\tilde{e}_a'.
\end{eqnarray}
Therefore, $\tilde{e}_a'(\varphi)=(k/2)(2\pi)^{-2}\epsilon_{abc}\varphi^b\mathrm{d}\varphi^c$ up
to a closed one-form.  

Let us check conditions for gaugeability summarized in Sec.~\ref{local} one by one.  First, Eq.~\eqref{additional1} is satisfied since
\begin{equation}
\mathrm{i}_{h_a}\tilde{e}_b'=\frac{k}{2(2\pi)^2}\epsilon_{abc}\varphi^c
\end{equation}
is antisymmetric with respect to $a$ and $b$.  However,
\begin{equation}
\mathscr{L}_{h_a}\tilde{e}_b'=\mathrm{d}\left(-\frac{k}{2(2\pi)^2}\epsilon_{abc}\varphi^c\right)=\mathrm{d}\tilde{z}_{ab}'.
\end{equation}
Hence, $\tilde{z}_{ab}'(\varphi)=-(k/2)(2\pi)^{-2}\epsilon_{abc}\varphi^c\neq0$ up to a constant.  This nonzero $\tilde{z}_{ab}'$ is the obstruction to gauge the symmetry $G$.

Note that the coefficient $k$ must be an integer to
  ensure that the Lagrangian changes only by integer multiples of
  $2\pi$ under the periodic shift $\varphi^a \rightarrow \varphi^a +
  2\pi$, because the integrand $e^{iS}$ in the path integral
  must be single valued even though the action $S$ itself is
  multivalued.  (See the discussion at the end of
  Sec.~\ref{sec:quantization}.)  On the other hand, this type of term
  is not allowed in
  ${\mathcal{L}^{(2,0)}_{\mathrm{eff}}}'=-(1/2)b_{ab}(\pi)\epsilon^{rs}\nabla_r\pi^a\nabla_s\pi^b$
  in Eq.~\eqref{eq:b} because the Hamiltonian must be single valued.  

\subsubsection{Wess-Zumino-Witten term}
\label{sec:WZW}
In general, we can write a similar term whenever
$H_{\text{dR}}^3(G/H)$~\cite{Nakahara,Eguchi:1980jx} is nontrivial.
(Here and below, $H^n_{\text{dR}}$ refers to de Rham cohomology, the
space of closed but not exact $n$-forms.) Then there is a nontrivial
closed three-form $\omega_3$ on $G/H$.  Because $\omega_3$ is locally
exact $\omega_3=\mathrm{d} \tilde{b}$, we can take the $(1+1)$-dimensional spacetime that is Wick-rotated and compactified to Euclidean space
$S^2 = \partial B_3$ as a boundary of a three-ball $B_3$, and we can
have
\begin{equation}
  \int_{B_3} \omega_3 = \int_{S^2} \tilde{b}
\end{equation}
as a part of a Lagrangian or a Hamiltonian.  

Note that there is, in general, more than $1$ $B_3$ in $G/H$ whose boundary is
$S^2 = \partial B_3$.  Therefore, the action is defined only up to an
integral of $\omega_3$ over a closed three-surface in $G/H$.  To
ensure that $e^{iS/\hbar}$ in the path integral is single valued, the
difference may only be integer multiples of $2\pi\hbar$
\cite{Witten:1983}.  It requires a quantization condition on the
coefficient of terms of this type.  The same quantization condition
can be obtained from the requirement of the associativity of the group
elements \cite{Murayama:1987nj}.

An important example is the Wess-Zumino-Witten term
\cite{Witten:1983ar}.  This term exists for any compact simple $G$ and
$H=\{e\}$ because $H^3_{\text{dR}}(G) = {\mathbb R}$.  It is defined with
\begin{equation}
  \omega_3 =  \frac{k}{12\pi}\mathrm{tr}[(U^{-1} \mathrm{d} U)^3]=\frac{k\lambda}{24\pi}f_{abc}\omega^a\wedge\omega^b\wedge\omega^c,
\end{equation}
with $k$ an integer, which is sometimes referred to as the level.
Here, we normalize $T_a$ as $\mathrm{tr}[T_aT_b]=\lambda\delta_{ab}$
so that the structure constant is completely antisymmetric. In order
for the path integral $e^{iS}$ to be single valued, $k$ must be an
integer in $1+1$ dimensions.  (See the discussion at the end of
Sec.~\ref{sec:quantization}.)  Also, because of this ambiguity of $2\pi k$, the Wess-Zumino-Witten term cannot be used to construct a $b(\pi)$ term since it takes part in the energy functional, as noted before.

Consider the transformation $U(\pi) \rightarrow U(\pi') = g U(\pi)$.
Obviously, for a global $g$, $\omega_3$ does not change.  However,
$\tilde{b}$ can change.  To see this possible change in $\tilde{b}$, let us temporarily regard $g=e^{iv}$
as local and consider infinitesimal change up to the linear order
in $i \mathrm{d}v = g^{-1}\mathrm{d}g$,
\begin{eqnarray}
   \frac{12\pi}{k} \delta (\mathrm{d} \tilde{b})
   &=&\mathrm{tr}[(U^{-1} \mathrm{d} U + U^{-1} (g^{-1}\mathrm{d}g) U)^3-(U^{-1} \mathrm{d} U)^3]\nonumber \\
  &=&3\mathrm{tr}[g^{-1}\mathrm{d}g (U \mathrm{d} U^{-1})^2]\nonumber \\
  &=&3i\mathrm{d}\,\mathrm{tr}[v(U \mathrm{d} U^{-1})^2],
\end{eqnarray}
and hence,
\begin{equation}
\delta \tilde{b}=\frac{ik}{4\pi}\mathrm{tr}[v (U \mathrm{d} U^{-1})^2].
\end{equation}
Now we can set $v$ to be constant.  Then, we see that
\begin{equation}
  \delta \tilde{b}
  =\frac{ik}{4\pi}\mathrm{tr}[v(U \mathrm{d} U^{-1})^2]
  =-\frac{ik}{4\pi}\mathrm{d}\,\mathrm{tr}[v U \mathrm{d} U^{-1}]
\end{equation}
is indeed a total derivative.

There is no compact way to write $\omega_3 = \mathrm{d} \tilde{b}$, but the following
trick works for a power-series expansion in $\pi$.  By defining $U_\tau =
e^{i \tau \Pi}$ for a real parameter $\tau$, it is easy to show
\begin{equation}
  \frac{\partial}{\partial \tau} U_\tau^{-1} \mathrm{d}U_\tau
  = i U_\tau^{-1} (\mathrm{d} \Pi) U_\tau,
\end{equation}
and therefore,
\begin{equation}
  \frac{\partial}{\partial \tau}\mathrm{tr}[(U_\tau^{-1} \mathrm{d} U_\tau)^3]
  = -3 i \mathrm{d}\,\mathrm{tr}[\Pi \mathrm{d} U_\tau \wedge \mathrm{d} U_\tau^{-1}].
\end{equation}
We can integrate the both sides and find
\begin{equation}
  \tilde{b} =-\frac{ik}{4\pi} \int_0^1\mathrm{d}\tau\,
 \mathrm{tr}[\Pi \mathrm{d} U_\tau \wedge \mathrm{d} U_\tau^{-1}]
\end{equation}
to obtain an explicit form in a power-series expansion in $\Pi$.  To the
leading order in $\pi$, we find
\begin{equation}
  \tilde{b} =\frac{k\lambda}{24\pi} f_{abc} \pi^a \mathrm{d}\pi^b \wedge \mathrm{d} \pi^c + O(\pi^4).
\end{equation}
Since $\pi^a$ shifts under the $G$ transformation, we can see that
$\tilde{b}$ changes by a total derivative.  

It is well known that the Wess-Zumino-Witten term cannot be gauged. To clarify the obstruction, we now compute $\tilde{e}_i'(\pi)$:
\begin{eqnarray}
\mathrm{i}_{h_d}\mathrm{d}\tilde{b}
&=&\frac{k\lambda}{4\pi}(\mathrm{i}_{h_d}\omega^a)\left(\frac{1}{2}f_{abc}\omega^b\wedge\omega^c\right)\notag\\
&=&\frac{k\lambda}{4\pi}\nu_d^a\mathrm{d}\omega^a=-\frac{k\lambda}{4\pi}\mathrm{d}(\nu_d^a\omega^a),\label{WZWe}
\end{eqnarray}
where we use Eqs.~\eqref{h} and \eqref{dw}.  (Since we assume all generators are broken, terms with indices $\rho,\sigma,\ldots$ should be neglected.)  The last equality can be shown backward:
\begin{eqnarray}
\mathrm{d}(\nu_d^a\omega^a)&=&(\mathrm{d}\nu_d^a)\wedge\omega^a+\nu_d^a\mathrm{d}\omega^a\notag\\
&=&(f_{abc}\omega^b\nu_d^c)\wedge\omega^a+\nu_d^a\mathrm{d}\omega^a\notag\\
&=&-\nu_d^cf_{abc}\omega^a\wedge\omega^b+\nu_d^a\mathrm{d}\omega^a\notag\\
&=&-2\nu_d^c\mathrm{d}\omega^c+\nu_d^a\mathrm{d}\omega^a=-\nu_d^c\mathrm{d}\omega^c,
\end{eqnarray}
where we use Eq.~\eqref{dv} in the first line.  Comparing Eq.~\eqref{WZWe} with Eq.~\eqref{eq:idb}, we find
\begin{eqnarray}
\tilde{e}_a'=-\frac{k\lambda}{4\pi}\nu_a^c\omega^c
\end{eqnarray}
up to an exact one-form.  

Having obtained $e_a'(\pi)$, let us now check the gaugeability condition.  First, the Lie derivative of $e_a'(\pi)$ satisfies
\begin{eqnarray}
\mathscr{L}_{h_a}\tilde{e}_b'&=&-\frac{k\lambda}{4\pi}\left[(\mathscr{L}_{h_a}\nu_b^c)\omega^c+\nu_b^c(\mathscr{L}_{h_a}\omega^c)\right]\notag\\
&=&-\frac{k\lambda}{4\pi}(f_{abd}\nu_d^c\omega^c+0)=f_{abc}\tilde{e}_c',\label{WZWg1}
\end{eqnarray}
meaning that $z_{ab}'(\pi)$ does vanish, according to Eq.~\eqref{eq:ep}.  However, since
\begin{equation}
\mathrm{i}_{h_a}\tilde{e}_b'=-\frac{k\lambda}{4\pi}\nu_b^c(\mathrm{i}_{h_a}\omega^c)=-\frac{k\lambda}{4\pi}\nu_b^c\nu_a^c=-\frac{k}{4\pi}\delta_{ab},\label{WZWg2}
\end{equation}
$\mathrm{i}_{h_a}\tilde{e}_b'$ is symmetric, rather than antisymmetric, with respect to $a$ and $b$, and therefore does not satisfy Eq.~\eqref{additional2}. Therefore, the Wess-Zumino-Witten term cannot be made gauge invariant.  In the derivation of Eqs.~\eqref{WZWg1} and \eqref{WZWg2}, we use Eqs.~\eqref{h}, \eqref{nu}, \eqref{homegab}, and \eqref{hiv}.

Another example of this type is $G/H = \text{U}(1) \times
\text{SO}(3) / \text{SO}(2) = S^1 \times S^2$ with $H^3_{\text{dR}}
(G/H) = H^1_{\text{dR}}(S^1) \times H^2_{\text{dR}}(S^2) = \mathbb R$.
Parametrizing the coset space with $\varphi$ for $S^1$ and the unit
vector $\vec{n}$ for $S^2$, we can write
\begin{equation}
  {\mathcal{L}^{(1,1)}_{\mathrm{eff}}}'
  = k \frac{\varphi}{4\pi} 
  \vec{n}\cdot(\dot{\vec{n}}\times\nabla_x\vec{n}).
\end{equation}
Under a constant shift of $\varphi$ by $2\pi$, the change is a total
derivative in space, and hence, the Lagrangian is $\text{U}(1)$
invariant.  However for a local shift of $\varphi$, it changes the
Lagrangian and hence is not an invariance.

Note that the shift of $\varphi$ by $2\pi$ does not change $e^{iS}$ in
the path integral because
\begin{equation}
  \int \mathrm{d}t\mathrm{d}x \frac{1}{4\pi} 
  \vec{n}\cdot(\dot{\vec{n}}\times\nabla_x\vec{n}) \in {\mathbb Z}
\end{equation}
is the winding number of $S^2 \rightarrow S^2$, as long as
$k\in{\mathbb Z}$.

\subsection{Berry's phase}
\label{sec:berry}
Finally, we discuss the interpretation of the linear time-derivative term of the effective Lagrangian as the Berry phase. Terms of our interest are
\begin{eqnarray}
\mathcal{L}&=&c_a(\pi)\dot{\pi}^a+e_i(\pi)A_t^i\notag\\
&=&-\omega_a^i(\pi)\dot{\pi}^ae_i(0)+e_i(0)\nu^i_j(\pi)A_t^j.
\end{eqnarray}
We apply a set of infinitesimal external fields $A_t^i=\mu^i(t)$ that slowly depend on time.  NG fields $\pi^a$ condense in such a way that $\{\pi^a\}_{a=1}^{\mathrm{dim}\,G/H}$ minimize the potential
\begin{equation}
V(t)\equiv-e_i(0)\nu^i_j(\pi)\mu^j(t)
\end{equation}
at each time.  Now, we consider a closed path $\mu^i(t)$ in the
parameter space $\{\mu^i\}_{i=1}^{\mathrm{dim}G}$.   NG fields
adiabatically depend on time through external fields, {\it i.e.}\/, $\pi^a=\pi^a(\mu(t))$. Under this process, the ground state $|\Psi_0\rangle$ evolves as
\begin{equation}
|\Psi(t)\rangle=e^{i\pi^a(\mu(t))Q_a}|\Psi_0\rangle,
\end{equation}
where $Q_a=\int\mathrm{d}^dx\,j_a^0(\vec{x},t)$ are broken generators.  Note that $\pi^a$ here is a $c$-number, not an operator, that is fixed by $\mu^i(t)$.

The Berry phase acquired under this cyclic process is
\begin{eqnarray}
\Theta_{\text{BP}}&=&\int\mathrm{d}t\,i\langle\Psi(t)|\frac{\mathrm{d}}{\mathrm{d}t}|\Psi(t)\rangle\notag\\
&=&-\int\mathrm{d}t\,\omega_a^i(\pi(\mu(t)))\frac{\mathrm{d}\pi^a(\mu(t))}{\mathrm{d}t}\langle\Psi_0|Q_i|\Psi_0\rangle\notag\\
&=&-\int\mathrm{d}t\mathrm{d}^dx\,\omega_a^i(\pi)\dot{\pi}^ae_i(0),\label{eq:berry}
\end{eqnarray}
where $e_i(0)=\langle\Psi_0|Q_i|\Psi_0\rangle/\Omega=\langle\Psi_0|j_i^0(\vec{x},t)|\Psi_0\rangle$ due to the translational invariance of the ground state.  Again, we have used the fact that the Maurer-Cartan form $\omega_a^i(\pi)$ only depends on the commutation relation and not on the specific representation.  Equation~\eqref{eq:berry} reproduces the $c_a(\pi)\dot{\pi}^a$ term of the effective Lagrangian, except for the $\vec{x}$ dependence of $\pi^a$. 

To treat the coordinate dependence properly, we introduce external fields $\mu^i=\mu^i(\vec{x},t)$ that are slowly varying over both space and time.  In this case, the ground state is given by
\begin{eqnarray}
|\Psi(t)\rangle&=&e^{i\Pi(t)}|\Psi_0\rangle,\\
\Pi(t)&=&\int\mathrm{d}t\mathrm{d}^dx\,j_a^0(\vec{x},t)\pi^a(\mu(\vec{x},t)).
\end{eqnarray}
To compute the Berry phase, we have to evaluate commutation relations
\begin{eqnarray}
&&\langle\Psi_0|[\underbrace{\Pi,[\ldots,[\Pi}_n,\nabla_t\Pi]\ldots ]]|\Psi_0\rangle\notag\\
&=&\int\mathrm{d}t\mathrm{d}^dx\Big[\dot{\pi}^a\langle\Psi_0|[\underbrace{\Pi,[\ldots,[\Pi}_n,j_a^0(\vec{x},t)]\ldots ]]|\Psi_0\rangle\notag\\
&&-\pi^a\vec{\nabla}\cdot\langle\Psi_0|[\underbrace{\Pi,[\ldots,[\Pi}_n,\vec{j}_a(\vec{x},t)]\ldots ]]|\Psi_0\rangle\Big].
\end{eqnarray}
Here, we use the current conservation $\nabla_tj^0(\vec{x},t)+\vec{\nabla}\cdot\vec{j}_a(\vec{x},t)=0$.  The second line vanishes since we assume the rotational symmetry of the ground state. Also, due to the translational symmetry of the ground state,
the expectation value of the commutator in the first line does not actually depend on $\vec{x}$ or $t$. Using the current algebra $[j_i^0(\vec{x},t),j_j^0(\vec{x}',t)]=if_{i j}^{\phantom{ij}k}j_k^0(\vec{x})\delta^d(\vec{x}-\vec{x}')$, one can easily show Eq.~\eqref{eq:berry} with the proper coordinate dependence of $\pi^a$.

\section{Number of Nambu-Goldstone bosons}
\label{sec:number}
In the next two sections, we will make use of the effective Lagrangian developed in the previous section to derive several rigorous results on the number of NGBs.  To be consistent with the assumed broken symmetries, in this section and the next sections we assume $2+1$ or higher dimensions.  

In order to discuss the number and the dispersion relation of NGBs, we
focus on the free part of the Lagrangian.  We will justify ignoring
the interaction terms in Sec.~\ref{sec:scaling}.  Keeping only the
quadratic terms in $\pi$ in Eq.~\eqref{summary} and setting
$A_\mu^i=0$, we find
\begin{eqnarray}
\mathcal{L}_{\text{eff}}&=&\frac{1}{2}f_{ab}^{\phantom{ab}k}e_k(0)\dot{\pi}^a\pi^b\notag\\
&+&\frac{1}{2}\bar{g}_{ab}(0)\dot{\pi}^a\dot{\pi}^b-\frac{1}{2}g_{ab}(0)\vec{\nabla}\pi^a\cdot\vec{\nabla}\pi^b.
\end{eqnarray}
Note that the $b(\pi)$ term does not contribute to the free part.

When $z_{ij}$ in Eq.~\eqref{eq:e} does not vanish, $e_i(\pi)$ and $c_i(\pi)$ receive a contribution from $z_{ij}$:
\begin{eqnarray}
e_i(\pi)&=&e_i(0)+\pi^b[f_{bi}^{\phantom{bi}k}e_k(0)+z_{bi}]+O(\pi^2),\\
c_a(\pi)&=&c_a(0)+\frac{1}{2}\pi^b[f_{ab}^{\phantom{ab}k}e_k(0)+z_{ab}]+O(\pi^2).
\end{eqnarray}
[The condition Eq.~\eqref{conditione} should also be replaced by $f_{\rho i}^{\phantom{\rho i}b}e_b(0)+z_{\rho i}=0$.]  Including this contribution, we have
\begin{eqnarray}
\mathcal{L}_{\text{eff}}&=&\frac{1}{2}\sigma_{ab}\dot{\pi}^a\pi^b\notag\\
&+&\frac{1}{2}\bar{g}_{ab}(0)\dot{\pi}^a\dot{\pi}^b-\frac{1}{2}g_{ab}(0)\vec{\nabla}\pi^a\cdot\vec{\nabla}\pi^b.\label{free}
\end{eqnarray}
where $\sigma_{ab}\equiv f_{ab}^{\phantom{ab}k}e_k(0)+z_{ab}$.

\subsection{Derivation 1}
\label{sec:number1}
The parameter $e_i(0)$ is related to the expectation value of the conserved charge density.  From theorem, the conserved current associated with $\delta_i\pi^a=h_i^a$ can be derived as
\begin{equation}
j_i^0(x)=e_i(\pi)-\bar{g}_{ab}(\pi)h_i^a(\pi)\dot{\pi}^b.\label{noether}
\end{equation}
Note that the conserved-current operators are free of anomalous
dimensions even in the presence of interactions because $j_i^0
\rightarrow Z j_i^0$ would violate the commutation relations
$[j_i^0(x), j_k^0(y)] = i f_{ik}{}^l j_l^0(x) \delta (x-y)$.  The absence of the anomalous dimensions is the nonrenormalization theorem of conserved currents.  Therefore, its
expectation value is that of the origin
\begin{equation}
\langle j_i^0(x)\rangle=e_i(0).
\end{equation}
We present explicit calculations in Sec.~\ref{sec:fluctuation} and an
alternative argument in Sec.~\ref{sec:SSB} to support this point.

Now, let us define a real and antisymmetric matrix $\rho$ by
\begin{equation}
i\rho_{ab}=\langle[Q_a,j_b^0(x)]\rangle.\label{rho}
\end{equation}
Assuming the translational invariance of the ground state, $\rho_{ab}$
is independent of $x$.  We see that $\rho_{ab}$ is related to the
first term in the effective Lagrangian:
\begin{eqnarray}
\rho_{ab}&=&-i\langle[Q_a,j_b^0(x)]\rangle\notag\\
&=&f_{ab}^{\phantom{ab}i}\langle j_i^0(x)\rangle+z_{ab}=\sigma_{ab}.
\end{eqnarray}
One can always block diagonalize $\rho$ by an orthogonal matrix as
\begin{gather}
\rho=
\begin{pmatrix}
i\sigma_y\lambda_1&&&\\
&\ddots&&&\\
&&i\sigma_y\lambda_m&\\
&&&O\\
\end{pmatrix},\,\,\lambda_\alpha\neq0\,(\alpha=1,\ldots,m).
\end{gather}
Here, $\sigma_y$ is the Pauli matrix and $m=(1/2)\text{rank}\rho$.  On this basis, the first term of the effective Lagrangian becomes
\begin{equation}
\sum_{\alpha=1}^m\lambda_\alpha\pi^{2\alpha}\dot{\pi}^{2\alpha-1}=\lambda_1\pi^2\dot{\pi}^1+\ldots+\lambda_m\pi^{2m}\dot{\pi}^{2m-1}.\label{pq}
\end{equation}
In the presence of these single time-derivative terms, one can neglect $O(\nabla_t^2)$ terms at a sufficiently low energy. Therefore, $\lambda_\alpha\pi^{2\alpha}$ (no sum) is, in fact, a \textit{canonically conjugate} valuable to $\pi^{2\alpha-1}$.  They together represent $1$ low-energy degree of freedom, rather than $2$.  We call those NGBs that are generated by a pair of canonically conjugate generators type-B, while the rest type-A.  By definition, the number of type-A and type-B NGBs are given by
\begin{eqnarray}
n_{\text{A}}=\text{dim}\,G/H-\text{rank}\rho,\quad n_{\text{B}}=\frac{1}{2}\text{rank}\rho.~\label{eq190}
\end{eqnarray}
Equation~\eqref{eq190} proves the counting rules in Eqs.~\eqref{countingA} and \eqref{countingB}.  As a corollary, the number of NGBs always falls into the range,
\begin{eqnarray}
\frac{1}{2}\text{dim}\,G/H\leq n_{\text{NGB}}\leq \text{dim}\,G/H.~\label{eq191}
\end{eqnarray}
Equation~\eqref{eq191} is obvious since $0\leq\text{rank}\,\rho\leq\text{dim}\,G/H$.

Note that our definition of type-A, B NGBs is not based on the
dispersion relation. They are instead classified based on the
structure of time derivatives that defines the {\it presymplectic
  structure} (see, {\it e.g.}\/, Ref.~\cite{Woodhouse:1992de}), as we
discuss in Sec.~\ref{sec:topology}.  These canonically
  conjugate relations among fields are the close analogs of Poisson
  brackets in the Hamiltonian
  formalism~\cite{HohenbergHalperin,Mazenko}. Note,
  however, that they had to provide the Poisson brackets in order to
  reproduce the microscopic theory, while in our case, we derive the
  commutation relations from the first principles for each possibility
  we can classify.

\subsection{Derivation 2}
Another way of deriving the same result is to make use of the canonical commutation relation.  Let us go back to the first term of the Lagrangian $\frac{1}{2}\sigma_{ab}\pi^b\dot{\pi}^a$.  Here, we assume that $\sigma$ is block diagonalized as
\begin{gather}
\sigma=
\begin{pmatrix}
i\sigma_y\lambda_1'&&&\\
&\ddots&&&\\
&&i\sigma_y\lambda_m'&\\
&&&O\\
\end{pmatrix},\,\,\lambda_\alpha'\neq0\,(\alpha=1,\ldots,m).
\end{gather}
We denote by $\sigma'$ the $2m\times 2m$ upper left part of the matrix $\sigma$, which has the full rank.

When we neglect the $O(\nabla_t^2)$ term of the effective Lagrangian, there are $m$ constraints of the second class in the system.  By following Dirac's quantization procedure, one can derive the equal-time commutation relation
\begin{equation}
[\pi^a(\vec{x},t),\pi^b(\vec{x}',t)]=i(\sigma'^{-1})^{ba}\delta^d(\vec{x}-\vec{x}').
\end{equation}
for $0\leq a,b\leq 2m$.  By definition, $n_{\text{A}}=\text{dim}\,G/H-\text{rank}\sigma'$ and $n_{\text{B}}=(1/2)\text{rank}\sigma'$. In this approach, we have to prove that $\text{rank}\sigma'=\text{rank}\rho$.

The Noether current in Eq.~\eqref{noether} can be expanded around the origin as 
\begin{equation}
j_a^0(x)=e_a(0)+\sigma_{ab}\pi^b(x)+O(\pi^2).
\end{equation}
By neglecting the contribution from higher-order terms,
\begin{eqnarray}
\rho_{ab}&\equiv&-i\langle[Q_a,j_b^0(\vec{x},t)]\rangle\notag\\
&=&-i\int\mathrm{d}^dx'\langle[j_a^0(\vec{x}',t),j_b^0(\vec{x},t)]\rangle\notag\\
&=&\sigma'_{ac}({\sigma'}^{-1})^{cd}\sigma'_{db}=\sigma'_{ab}.
\end{eqnarray}
Therefore, $\text{rank}\sigma'=\text{rank}\rho$.

Finally, let us comment on the locality of the effective Lagrangian.  Even if the microscopic model does not have long-range interactions, long-range interactions among NGBs may be mediated by other gapless degrees of freedom in the system.  When the effective Lagrangian fails to be local, would-be NGBs may acquire a gap and the counting rule may not hold. (See Ref.~\cite{fractional} for more details.)

Moreover, if we allow nonlocal effective Lagrangians, the classification of type-A and B becomes ambiguous. As an example, let us take a free theory of a type-B NGB described by a local Lagrangian $\mathcal{L}_{\text{eff}}=\sum_{a,b=1,2}(\sigma/2)\epsilon_{ab}\pi^a\dot{\pi}^b-\sum_{a=1,2}(g/2)\vec{\nabla}\pi^a\cdot\vec{\nabla}\pi^a$ in $3+1$ dimensions. After integrating out the field $\pi^2$, one finds a nonlocal effective Lagrangian in terms of $\pi^1$:
\begin{eqnarray}
L_{\text{eff}}&=&\frac{1}{2}\int\mathrm{d}^3x\mathrm{d}^3x'\,\dot{\pi}^1(\vec{x},t)\frac{\sigma^2}{4\pi g|\vec{x}-\vec{x}'|}\dot{\pi}^1(\vec{x}',t)\notag\\
&&-\frac{g}{2}\int\mathrm{d}x\,\vec{\nabla}\pi^1(\vec{x},t)\cdot\vec{\nabla}\pi^1(\vec{x},t).
\end{eqnarray}
This nonlocal Lagrangian can still describe the mode with the same quadratic dispassion $\omega=(g/\sigma)k^2$, but now, it is described by a single field $\pi^1$ and hence may be classified as type-A.  Therefore, the classification of type-A and B makes sense only when we restrict ourselves to local effective Lagrangians.

\section{Dispersion relation}
\label{sec:dispersion}
In this section, we discuss the dispersion relation of NGBs.  In
particular, we show that type-A NGBs generically have linear
dispersions, while type-B NGBs are quadratic.

The linearized effective Lagrangian in Eq.~\eqref{free} leads to the equation of motion $\mathcal{G}_{ab}\pi^b(k,\omega)=0$, where
\begin{align}
\mathcal{G}=i\sigma\omega+\bar{g}(0)\omega^2-g(0)k^2.
\end{align}
The dispersion relations of NGBs are determined by solving
$\mathrm{det}\mathcal{G}=0$.  If type-A and type-B NGBs do not coexist,
the situation is pretty simple.  When $\sigma=0$ (only type-A), the
dispersion is always linear since $\omega^2$ has to balance with
$k^2$. In contrast, when $\sigma$ has the full rank (only type-B), we
can ignore $\bar{g}(0)\omega^2 \ll i\sigma\omega$ in the low-energy
limit, and the dispersion is quadratic by the same argument.

Note that $g(0)$ must always be full rank as long as we consider an internal symmetry group $G$, because the field-transformation rule in Eq.~\eqref{symmetryh} does not explicitly depend on coordinates, and thus, there are no symmetries that prohibit the appearance of the $O(k^2)$ term.  In Sec.~\ref{sec:scaling}, we explain examples of NGBs associated with spacetime symmetries that lack the $O(k^2)$ term, but for now, let us focus on internal symmetries.

When type-A and type-B NGBs do coexist, and especially when there are NGBs of the same representation under $H$, the metrics $g(0)$ and $\bar{g}$ may mix them and the discussion of the dispersion becomes complicated.  To discuss the dispersion even in such a general situation, here, we develop a perturbation theory for small $\omega$.

Assuming that $g(0)$ is positive and nonsingular, we can always write it as $g(0)=Z^2$, 
with $Z$ a symmetric, positive, and nonsingular matrix. Substituting this expression into $\mathcal{G}$, we have
\begin{align}
\mathcal{G}'\equiv Z^{-1}\mathcal{G}Z^{-1}=i\Sigma\omega+Z^{-1}\bar{g}(0)Z^{-1}\omega^2-k^2,
\end{align}
where $\Sigma=Z^{-1}\sigma Z^{-1}$.  Because $\Sigma$ is still real and antisymmetric, one can always find an orthogonal matrix $O$ such that
\begin{gather}
\Sigma=O\Lambda O^T,\quad \Lambda=
\begin{pmatrix}
i\sigma_y\lambda_1''&&&\\
&\ddots&&&\\
&&i\sigma_y\lambda_m''&\\
&&&O\\
\end{pmatrix}.
\end{gather}
Here, $\lambda''_\alpha>0$ for
$\alpha=1,\ldots,m=(1/2)\mathrm{rank}\,\rho$.  Now, $\mathrm{det}\mathcal{G}=0$ is equivalent to $\mathrm{det}\mathcal{G}''=0$, where
\begin{align}
\mathcal{G}''\equiv O^T\mathcal{G}'O=i\Lambda \omega+\bar{G}\omega^2-k^2
\label{eq:dispersion}
\end{align}
and $\bar{G}=O^TZ^{-1}\bar{g}(0)Z^{-1}O$.

We regard $O(\omega^2)$ terms as a small perturbation.  Following the standard procedure for the degenerate perturbation theory, we diagonalize the bottom right $n\times n$ ($n\equiv\mathrm{dim}G-2m$) block of $\bar{G}$:
\begin{equation}
\bar{G}=
\begin{pmatrix}
*&\cdots&*&*&\cdots&*\\
\vdots&*&\vdots&\vdots&*&\vdots\\
*&\cdots&*&*&\cdots&*\\
*&\cdots&*&s_1&&0\\
\vdots&*&\vdots&&\ddots&\\
*&\cdots&*&0&&s_n\\
\end{pmatrix}
\end{equation}
Asterisks stand for unknown elements. This diagonalization is compatible with the above transformation of $\Sigma$, since all relevant components of $\Lambda$ vanish.  

The upper left $2m\times 2m$ block has a nonzero unperturbed term that reads
\begin{align}
\lambda''_\alpha\omega\sigma_{y}+k^2\,\sigma_{0}=0\,\,&\Leftrightarrow\,\,\omega_\alpha(k)=\frac{k^2}{\lambda_\alpha''}
\end{align}
for $\alpha=1,\ldots,m$.  The off-diagonal component
$\omega\sigma_{y}$ is reminiscent of the presymplectic structure in
Eq.~\eqref{pq}. Therefore, these modes with quadratic dispersion may
still be called type-B NGBs, although, strictly speaking, fields
describing these modes are, in general, a mixture of type-A and type-B NG
fields, according to the definition in Sec.~\ref{sec:number1}.
 
On the other hand, in the bottom right $n\times n$ block, where the zeroth-oder term vanishes,  the linear order correction gives
\begin{align}
s_\xi\omega^2-k^2=0\,\,&\Leftrightarrow\,\,\omega_\xi(k)=\pm\frac{k}{\sqrt{s_\xi}}
\end{align}
for $\xi=1,\ldots,n$. Because there is no presymplectic structure in
this block, these linear dispersions can be regarded as type-A NGBs.
Our ground state is stable only when all of $s_\xi>0$.
Note that the mixing between upper and lower blocks
  induces only negligible corrections of $O(\omega^3)$.

We have shown here that generically type-A NGBs have a linear dispersion and type-B NGBs have a quadratic dispersion.   
Therefore, the equality version of the Nielsen-Chadha theorem is now proven. 

When the $O(\nabla^2)$ term of the effective Lagrangian is somehow absent, type-A NGBs may have a quadratic dispersion and type-B NGBs may have a quartic dispersion. As explained above, that never happens for internal symmetries, but there are examples of NGBs originated from spacetime symmetries that lack the $O(\nabla^2)$ term. See Sec.~\ref{sec:scaling} for more details.

\section{Stability of the symmetry breaking ground state}
\label{sec:stability}

In identifying the degrees of freedom and reading off their dispersion
relations, in previous sections, we used the perturbation theory and studied the quadratic
part of the effective Lagrangian.  One may be concerned that the
interactions may upset the conclusion.  Namely, the question is whether
the cubic and higher terms can modify the dynamics at long distances,
which is equivalent to the question about the stability of a
long-range order.

\subsection{Scaling of interactions among NGBs}
\label{sec:scaling}
Here we examine the scaling law of the most relevant interactions among NGBs to see the stability of the symmetry-breaking ground state.  

We start with the situation when there are only type-A NGBs.  In order to keep the free action 
\begin{equation}
\int\mathrm{d}^dx\mathrm{d}t\left(\frac{\bar{g}_{ab}(0)}{2}\dot{\pi}^a\dot{\pi}^b-\frac{g_{ab}(0)}{2}\vec{\nabla}\pi^a\cdot\vec{\nabla}\pi^b\right)
\end{equation}
invariant, NG fields $\pi^a$ should transform as ${\pi'}^a(\alpha\vec{x},\alpha t)=\alpha^{(1-d)/2}\pi^a(\vec{x},t)$.  In $1+1$ dimensions, we should include $\tilde{g}_{ab}(0)\dot{\pi}^a\nabla_x\pi^b$ in the free action, but it does not change the scaling law. Note again that the $b(\pi)$ and $\tilde{b}(\pi)$ terms do not have the free part, and the $\tilde{c}_a(\pi)$ term causes an instability to a translational symmetry-broken phase as discussed before, and hence, we do not consider them here.
The most relevant interactions $\mathrm{d}^dx\mathrm{d}t\nabla_t^2\pi^3$ and $\mathrm{d}^dx\mathrm{d}t\nabla_r^{2}\pi^3$ then scale with $\alpha^{(1-d)/2}$.  Therefore, if the spatial dimension $d$ is greater than one, all interactions are irrelevant and the system flows into the free fixed point.  In this case, the symmetry-breaking ground state is stable and $1$ can understand the property of the system via the standard perturbation theory.  On the other hand, when $d=1$, the interaction is marginal, so that broken symmetries are restored and the low-energy spectrum may get gapped.

This result is consistent with the Coleman theorem that guarantees the absence of continuous symmetry breaking in $1+1$ dimensions for the Lorentz-invariant case $g_{ab}=\bar{g}_{ab}$~\cite{Coleman}.  Superfluids in $1+1$ dimensions are in the Kosterlitz-Thouless phase, which possesses only a quasi-long-range order (power-law decay) and has a gapless density wave.  The $S=1/2$ antiferromagnetic chain also shows a quasi-long-range order and supports gapless excitations called des Cloizeaux-Pearson modes.  These gapless excitations are qualitatively different from free NGBs; rather, they can be understood as Tomonaga-Luttinger liquids~\cite{Giamarchi}.  In contrast, the $S=1$ antiferromagnetic chain is believed to be in the Haldane phase and to be gapped.

We can easily extend our analysis for other types of dispersion. Although spacetime symmetries are not the main focus of the current paper, type-A NGBs that originated from spontaneously-broken spacetime symmetries sometimes have weird dispersions. In such a case, the criteria we have derived for internal symmetries may be violated.  For example, in a rotating superfluid in $2+1$ dimensions, a vortex lattice breaks the {\it magnetic}\/ translation. The NG bosons, the so-called Tkachenko mode, are described by the effective Lagrangian
\begin{equation}
\int\mathrm{d}^2x\mathrm{d}t\left[\frac{A}{2}\dot{\varphi}^2-\frac{B}{2}(\nabla^2\varphi)^2\right].
\end{equation}
Note that the term $(\nabla\varphi)^2$ is prohibited by symmetry transformation $\delta\varphi\propto\vec{x}$~\cite{WatanabeMurayama2}. In this case, it is easy to see that the dominant interaction is marginal, which destroys the long-range phase correlation even at $T=0$~\cite{MacDonald}.  This conclusion makes contrast with the usual superfluids or crystals in $2+1$ dimensions, which are stable at $T=0$.  Another example is a helical magnet. Because of the spin-orbit coupling, the spin rotation must be accompanied by the spatial one. The helical (spiral) order breaks some combination of the rotation and translation. It turns out that there is only one gapless mode~\cite{Radzihovsky}, which is described by
\begin{equation}
\int\mathrm{d}^3x\mathrm{d}t\left[\frac{A}{2}\dot{\varphi}^2-\frac{B}{2}(\nabla_z\varphi)^2-\frac{C}{2}[(\nabla_x^2+\nabla_y^2)\varphi]^2\right].
\end{equation}
Again, the terms $(\nabla_x\varphi)^2$ and $(\nabla_y\varphi)^2$ are prohibited by symmetry.  As a result, the dispersion of the NGB is anisotropic $\omega=\sqrt{(B/A)k_z^2+(C/A)(k_x^2+k_y^2)^2}$, which is an example of NGBs that cannot be classified as either type-I nor type-II, although it can be unambiguously classified as type-A.  All interactions are irrelevant at $T=0$, but there are marginal interactions at a finite temperature, despite the fact that usually broken symmetries are stable at a finite temperature in three dimensions.

Let us go back to the usual case $z=1$ and instead consider a finite
temperature. When $T>0$, all imaginary-time dependences drop out at a
sufficiently long-distance and low-energy scale, leaving only the $n=0$
component of the Matsubara frequency. Then, the free part of the action
is just $-T\int\mathrm{d}^dx[g_{ab}(0)/2]\vec{\nabla}\pi^a\cdot\vec{\nabla}\pi^b$
and fields transform as
${\pi'}^a(\alpha\vec{x})=\alpha^{(2-d)/2}\pi^a(\vec{x})$. The
most relevant interaction $\mathrm{d}^dx\nabla_r^{2z}\pi^3$ scales as
$\alpha^{(2-d)/2}$, so that the stability condition is given by
$d>2$, which is nothing but the Mermin-Wagner theorem.

Next, we discuss the case where only type-B NGBs are present.  To keep the free action
\begin{equation}
\int\mathrm{d}^dx\mathrm{d}t\left(\frac{\sigma_{ab}}{2}\dot{\pi}^a\pi^b+\frac{g_{ab}(0)}{2}\nabla\pi^a\cdot\nabla\pi^b\right)
\end{equation}
invariant, NG fields should obey the scaling law ${\pi'}^a(\alpha\vec{x},\alpha^2 t)=\alpha^{-d/2}\pi^a(\vec{x},t)$.  
We could add $\tilde{g}_{ab}(0)\dot{\pi}^a\nabla_x\pi^b$ in $1+1$ dimensions, but it is clearly higher order in derivatives. In this case, the most relevant interactions $\mathrm{d}^dx\mathrm{d}t\nabla_t\pi^3$ and $\mathrm{d}^dx\mathrm{d}t\nabla_r^2\pi^3$ scale as $\alpha^{-d/2}$.  Therefore, the theory is essentially free in all dimensions, and hence, broken symmetries can never be restored.  This conclusion might sound surprising for high-energy theorists, but actually, it is a well-known fact in condensed-matter physics~\cite{Nagaosa}.  We will come back to this point in Sec.~\ref{sec:SSB}.

Type-A NGBs with a quadratic dispersion $\omega\propto k^2$ ($z=2$) and type-B NGBs with the same dispersion have a completely different effect on broken symmetries. The former destroys the order parameter if $d\leq 2$, while the latter does not do anything if $d>0$.

The discussion for a finite temperature for type-B NGBs is identical to the type-A
case, since all imaginary-time dependences drop out.  We summarize our result in Table~\ref{tab1}.

\begin{table}
  \begin{center}
    \caption{The stability condition for the symmetry-breaking ground state in $d$ spatial dimensions, obtained by evaluating the scaling law of interactions and the infrared divergence for NGBs.}
       \label{tab1}
    \begin{tabular}{ccc} \hline\hline
    		&\hspace{10pt}$T=0$\hspace{10pt}&\hspace{10pt}$T>0$\\ \hline
Only type-A NGBs\hspace{10pt}&\hspace{10pt}$d>1$\hspace{10pt}&\hspace{10pt}$d>2$\\ 
Only type-B NGBs\hspace{10pt}&\hspace{10pt}$d>0$\hspace{10pt}&\hspace{10pt}$d>2$\\ \hline\hline
    \end{tabular}
  \end{center}
\end{table}

\subsection{Fluctuation of order parameters}
\label{sec:fluctuation}
The stability of the symmetry-breaking ground state can also be discussed by evaluating the quantum correction to the expectation value of order parameters.  The infrared divergence originated from gapless NGBs tends to destroy the symmetry-breaking order parameters in lower dimensions.  

Again, assuming that the free theory is a good starting point, we express the expectation value of order parameters in terms of the free Green functions $\mathcal{G}^{ab}(x-y)=\langle T\pi^a(x)\pi^b(y)\rangle$.  For example, the Noether charge density $j_i^0(\vec{x},t)$ plays the role of the order parameter for charges $Q_a$ for which $\langle[Q_a,j_b^0(x)]\rangle\neq0$ for some $b$.  The current density $j_i^0(\vec{x},t)$ in Eq.~\eqref{noether} can be expanded in terms of NG fields as
\begin{eqnarray}
j_i^0&=&e_k(0)\left[\delta_i^k+\pi^bf_{bi}^{\phantom{bi}k}+\frac{1}{2}f_{ai}^{\phantom{ai}j}f_{jb}^{\phantom{jb}k}\pi^a\pi^b+O(\pi^3)\right]\notag\\
&&-\bar{g}_{ab}(0)\left[\dot{\pi}^a\delta_i^b+C_{icd}^{ab}\pi^c\dot{\pi}^d+O(\nabla_t\pi^3)\right],\label{current}
\end{eqnarray}
where $C_{\rho cd}^{ab}=f_{c\rho}^{\phantom{c\rho}a}\delta_d^b$ for unbroken currents ($i=\rho$) and $C_{ecd}^{ab}=f_{ce}^{\phantom{ce}a}\delta_d^b+(1/2)\delta_e^af_{cd}^{\phantom{cd}b}$ for broken currents ($i=e$).  Therefore, the dominant contribution to the expectation value is given by
\begin{eqnarray}
\langle j_i^0\rangle&\simeq&e_k(0)\left[\delta_i^k+\frac{1}{2}f_{ai}^{\phantom{ai}j}f_{jb}^{\phantom{jb}k}\mathcal{G}^{ab}(0)+\cdots\right].\label{nrt1}
\end{eqnarray}

For superfluids, $\psi(\vec{x},t)\simeq\sqrt{n_0}\, e^{i\theta(\vec{x},t)}$ is the order parameter, and its expectation value with quantum fluctuation is
\begin{equation}
\langle e^{i\theta(\vec{x},t)}\rangle=e^{-\frac{1}{2}\langle\theta(\vec{x},t)^2\rangle}=e^{-\frac{1}{2}\mathcal{G}(0)}.
\end{equation}
(Note that $\theta$ itself is not a good quantity to look at since it
does not have the assumed periodicity of $2\pi$.)  As one can see, we
need $|\mathcal{G}_0^{ab}(\vec{x}=0,t=0)|\ll1$ in order for the
quantum correction to be small compared to the classical value.

We can easily evaluate $\mathcal{G}^{ab}(0)$ by scaling.  When only type-A NGBs appear, $i(\mathcal{G}_0^{-1})_{ab}(\vec{k},\omega)=\bar{g}_{ab}\omega^2-g_{ab}k^2$ and
\begin{eqnarray}
&\int\mathrm{d}^dk\mathrm{d}\omega\,\mathcal{G}^{ab}(\vec{k},\omega)\propto\int_0^{\Lambda}\mathrm{d}k\,k^{d-2},\\
&T\sum_n\int\mathrm{d}^dk\,\mathcal{G}^{ab}(\vec{k},i\omega_n)\propto T\int_0^{\Lambda}\mathrm{d}k\,k^{d-3},
\end{eqnarray}
for $T=0$ and $T>0$, respectively.  We have introduced the ultraviolet cutoff $\Lambda$.  Therefore, for the convergence of the infrared contribution, we need $d>1$ at zero temperature and $d>2$ at a finite temperature.
Similarly, when only type-B NGBs appear, $i(\mathcal{G}_0^{-1})_{ab}(\vec{k},\omega)=-i\sigma_{ab}\omega-g_{ab}k^2$ and 
\begin{eqnarray}
&\int\mathrm{d}^dk\mathrm{d}\omega\,\mathcal{G}^{ab}(\vec{k},\omega)\propto\int_0^{\Lambda}\mathrm{d}k\,k^{d-1},\label{nrt2}\\
&T\sum_n\int\mathrm{d}^dk\,\mathcal{G}^{ab}(\vec{k},i\omega_n)\propto T\int_0^{\Lambda}\mathrm{d}k\,k^{d-3}.
\end{eqnarray}
Therefore, there is no infrared divergence, even at $1+1$ dimensions at zero temperature.  These results are consistent with those summarized in Table~\ref{tab1}.

In Sec.~\ref{sec:number1}, we discussed the nonrenormalization
theorem of $\langle j_i^0(0)\rangle$. However, Eqs.~\eqref{nrt1} and
\eqref{nrt2} may appear to indicate that $\langle j_i^0(0)\rangle$
receives a finite correction due to quantum fluctuations. Now, we show that
it is not the case by explicitly evaluating the magnetization of
ferromagnets at the one-loop level.  The effective Lagrangian
\eqref{free} for the coset $G/H=\text{SO}(3)/\text{SO}(2)$ reads
\begin{equation}
\mathcal{L}=\frac{i}{2}e_0\left(\bar{z}\dot{z}-\dot{\bar{z}}z\right)+\bar{g}_0\dot{\bar{z}}\dot{z}-g_0\vec{\nabla}\bar{z}\cdot\vec{\nabla}z
\end{equation}
to the quadratic order in $z=(\pi^1+i\pi^2)/\sqrt{2}$.  According to
Eq.~\eqref{current}, the magnetization including the fluctuation is
$j_z^0=e_0-e_0\bar{z}z-i\bar{g}_0(\bar{z}\dot{z}-\dot{\bar{z}}z)$.
Therefore,
\begin{equation}
\langle j_z^0\rangle=e_0-\sum_{n}\int\frac{\mathrm{d}^dk}{(2\pi)^d}\frac{e_0+2\bar{g}_0i\omega_n}{-e_0i\omega_n+\bar{g}_0\omega_n^2+g_0k^2}
\end{equation}
We can perform the Matsubara summation using the standard trick and find 
\begin{equation}
\langle j_z^0\rangle=e_0-n(\omega)+n(\omega'),\label{magnetization}
\end{equation}
where $n(\epsilon)=(e^{\beta\epsilon}-1)^{-1}$ is the Bose distribution function,
\begin{equation}
\omega=\frac{\sqrt{e_0^2+4g_0\bar{g}_0k^2}-e_0}{2\bar{g}_0}=\frac{g_0}{e_0}k^2+O(k^4)
\end{equation}
is the dispersion of the gapless Goldstone mode (magnon), and 
\begin{equation}
\omega'=\frac{\sqrt{e_0^2+4g_0\bar{g}_0k^2}+e_0}{2\bar{g}_0}=\frac{e_0}{\bar{g}_0}+O(k^2)
\end{equation}
is the dispersion of the gapped mode. [The existence of the gapped
mode is questionable since this solution balances the $O(\nabla_t)$
term and the $O(\nabla_t^2)$ term of the effective Lagrangian.  It is
easily eliminated from calculation by taking the limit $\bar{g}
\rightarrow 0$.] Since $n(\omega)=n(\omega')=0$ at $T=0$, the one-loop
correction to the expectation value of the magnetization vanishes in
the ground state.  Clearly, the finite-temperature correction is
dominated by magnons and is proportional to $T^{d/2}$ at
low temperature, which is known as Bloch's law~\cite{ashcroft}.

So far, we have only considered the case where only one type of NGB appears, since both of our above arguments are essentially based on scaling.  However, in general, type-A and type-B NGBs can coexist.  In such a case, there is no field transformation that keeps all of the free parts invariant unless type-A and type B NGBs are somehow completely decoupled.  When they interact, we have no choice but to respect the scaling rule of the softer modes (type-B NGBs). Then the free Lagrangian of type-A NGBs are not kept invariant and their velocity diverge in the infrared limit.

In the next section, we present some arguments that can be used in type-A and type-B coexisting cases.

\subsection{Spontaneous symmetry breaking in 1+1 dimensions}
\label{sec:SSB}
The usual argument for ferromagnets in $1+1$ dimensions is as follows~\cite{Nagaosa}.  As the ferromagnetic order parameter $S_z$ commutes with the Hamiltonian $H$, one can simultaneously diagonalize $H$ and $S_z$ and obtain quantum many-body eigenstates $|\Psi_{E,M}\rangle$ labeled by the eigenvalue of $H$ and $S_z$.  Since $|\Psi_{E,M}\rangle$ is an eigenstate, there is no quantum fluctuation of order parameter $\langle\Psi_{E,M}|S_z^2|\Psi_{E,M}\rangle=\langle\Psi_{E,M}|S_z|\Psi_{E,M}\rangle^2$.  From the translational invariance of the ground state, it follows that $\langle\Psi_{E,M}|[S_x,j_y^0(\vec{x},t)]|\Psi_{E,M}\rangle=iM/\Omega$, where $\Omega$ is the volume of the system.  As usual, applying the magnetic field $-B_zS_z$ to pick up a particular state, taking the large volume limit first, and then switching off the field, one finds the definition of symmetry breaking of $S_x$ [$\langle[S_x,j_y^0(\vec{x},t)]\rangle=im\neq0$], with $m$ the magnetization density. 

This argument can be easily extended to a more general case, as long as Cartan generators are not spontaneously broken.  As
discussed above, only Cartan generators, which commute with each other
by definition, can have nonzero expectation values.  We can thus
simultaneously diagonalize all of them (except for the Abelian invariant
algebra of $G$ that never plays the role of an order parameter) and
the Hamiltonian.  This argument is an alternative proof of the
nonrenormalization theorem of the expectation value of the current
operator at $T=0$, discussed in Sec.~\ref{sec:number1}.  (At a finite
temperature, we no longer use a pure quantum eigenstate but take an
ensemble over all states, and the expectation value gets a finite
temperature correction.)

However, the simultaneous eigenstate of the Hamiltonian and Cartan generators can never break those symmetries generated by the Cartan generators themselves. Therefore, this argument has to be modified when applied to, for instance, a magnetic order that completely breaks the $\text{SU}(2)$ symmetry and has a ferromagnetic order $\langle S_z\rangle$, an example of which in $1+1$ dimensions is recently discussed in Ref.~\cite{Furuya}~\footnote{In this case, $S_z$ is not truly broken due to strong quantum fluctuations in $1+1$ dimensions.}.  Even for this case, we can still argue that the ferromagnetic long-range order will not be completely destroyed by quantum fluctuations. In order to break $S_z$, one has to take a superposition of some simultaneous eigenstates with different eigenvalues of $S_z$.  In this superposition, we do not have to include those with positive and negative eigenvalues of $S_z$ with the equal amplitude.  Therefore, the expectation value is generically nonzero, unless dictated by the unbroken time-reversal symmetry etc.

In Ref.~\cite{Momoi}, it has been proved that continuous symmetry breaking in $1+1$ dimensions is possible only when uniform susceptibilities of broken charges diverge.  Indeed, we can show the divergence of uniform susceptibility whenever type-B NGBs appear.  Equation~\eqref{current} tells us that the current-current correlation function of charges associated with type-B NGBs is dominated by
\begin{equation}
\langle \delta j_a^0(\vec{x},t)\delta j_b^0(0)\rangle
=e_k(0)e_\ell(0)f_{ca}^{\phantom{ca}k}f_{db}^{\phantom{db}\ell}\langle\pi^c(\vec{x},t)\pi^d(0)\rangle.
\end{equation}
Therefore, the uniform susceptibility
\begin{equation}
\chi_{ab}=\lim_{|\vec{k}|\rightarrow 0}\left[\langle\delta j_a^0(\vec{k},i\omega_n)\delta j_b^0(-\vec{k},-i\omega_n)\rangle_{\omega_n=0}\right]
\end{equation}
diverges due to poles of Green's functions corresponding to type-B NGBs.

In contrast, when type-B NGBs do not exist, all $e_i(0)'s$s in Eq.~\eqref{current} vanish and the correlation function is dominated by
\begin{eqnarray}
\langle \delta j_a^0(\vec{x},t)\delta j_b^0(0)\rangle=\bar{g}_{ac}(0)\bar{g}_{bd}(0)\langle\dot{\pi}^c(\vec{x},t)\dot{\pi}^d(0)\rangle.
\end{eqnarray}
Additional time derivatives cancel the divergence, and the uniform susceptibility converges.

An example of continuous symmetry breaking at $1+1$ dimensions, which
supports both a linear and a quadratic dispersion, is given by spinor Bose-Einstein Condenates~\cite{Eisenberg,Fuchs,Giamarchi2,Furusaki,Kamenev}.  The model is defined by
\begin{equation}
\mathcal{L}=\frac{i}{2}(\psi^\dagger\dot{\psi}-\text{c.c.})-\frac{\nabla\psi^\dagger\cdot\nabla\psi}{2m}-\frac{g}{2}(\psi^\dagger\psi-n_0)^2.
\end{equation}
Here, $\psi=(\psi_1,\psi_2)^{T}$ is a two-component complex scaler field and $n_0=N/L$. ($N$ is the number of bosons. and $L$ is the system size.) The dimensionless coupling constant is given by $\gamma=mg/n_0$. 

At the tree level (mean-field approximation), the system exhibits a long-range order $\langle\psi\rangle=(0,v)^{T}$ and then the $\text{U}(2)$ symmetry (generated by $S_{x,y,z}$ and $Q$) is spontaneously broken into a $\text{U}(1)$ symmetry (generated by $S_z+Q$).  There are two NGBs, a type-A NGB (sound wave) with a linear dispersion $\omega_{\text{ph}}(k)=(n_0/m)\sqrt{\gamma}\,k$ and a type-B NGB (spin wave) with a quadratic dispersion $\omega_{\text{sw}}(k)=k^2/2m$ as $|\vec{k}|\rightarrow0$.  However, the strong fluctuation caused by the linear dispersion invalidates this simple analysis.  

The ground state in this case cannot be an eigenstate of $S_z$, because $S_z$ is also broken. Instead, the ground state $|0\rangle$ can be taken as an eigenstate of $S_z+Q$.  From the tree level result, it is natural to take the simultaneous eigenstate with $(S_z+Q)|0\rangle=0$.  Then, in particular, $\langle S_z+Q\rangle=0$ and $\langle[S_x,S_y]\rangle=i\langle S_z\rangle=-i\langle Q\rangle\neq0$, which imply the spontaneous breaking of $S_x$ and $S_y$.

Surprisingly, there exists an exact solution of this model based on the Bethe-anzatz~\cite{Fuchs}.  The solution exhibits the ferromagnetic long-range order, showing the spontaneous breaking of spin rotation.  Correspondingly, there is a well-defined spin-wave excitation with the dispersion $\omega_{\text{SW}}(k)=\left[1-(2\sqrt{\gamma}/3\pi)+\cdots\right](k^2/2m)$ in the week-coupling limit $\gamma\ll1$ and $\omega_{\text{SW}}(k)=\left[(2\pi^2/3\gamma)+\cdots\right](k^2/2m)$ in the strong-coupling limit $\gamma\gg1$~\cite{Fuchs}.

On the other hand, the phase-phase correlation is not truly long ranged.  As a result, the sound wave should be understood as a Tomonaga-Luttinger liquid rather than as a type-A NGB~\cite{Giamarchi2,Furusaki,Kamenev}.

\section{Topology}
\label{sec:topology}

In this section, we discuss the geometry behind the type-B NGBs that
do not appear in Lorentz-invariant theories.  There is an underlying
geometrical foundation called a presymplectic structure.  Understanding the 
geometry of NGBs turns out to be important for classifying a possible
division between type-A and type-B NGBs in the next section.

\subsection{Presymplectic structure}

We have seen that the one-form $c=c_a\mathrm{d}\pi^a$ on the cotangent
space $T^*(G/H)$ is in general not invariant under $G$, while the 
two-form $\omega = \mathrm{d} c$ is [see Eq.\eqref{eq:lhidc}].  Therefore,
we should focus on $\omega$, which is a closed and $G$-invariant
two-form on $G/H$.  If the antisymmetric matrix $\omega =
\omega_{ab}(\pi) \mathrm{d}\pi^a \wedge \mathrm{d}\pi^b$ has a
nonzero determinant $\text{det}\,\omega_{ab}(\pi) \neq 0$, it defines a
{\it symplectic structure}\/ on $G/H$.  The combination of a manifold
and a nondegenerate closed two-form $(M,\omega)$ is called a
symplectic manifold.  In physics terminology, it is nothing but a
phase space of a dynamical system with well-defined canonical
commutation relations among its coordinates given by $[\pi^a,\pi^b] =
i (\omega^{-1})^{ab}$.  It is obvious that it requires $G/H$ to be
even dimensional.  If $G/H$ is compact, its second cohomology
$H^2(G/H)$ must be nontrivial.  Note that many coset spaces do not
satisfy these requirements.

If $\omega$ is degenerate, namely, if $\text{det}\,\omega_{ab} = 0$, it
is called a {\it presymplectic structure}\/, or partially symplectic,
because only a subset of the coordinates $\pi^a$ participates in the
matrix $\omega_{ab}$.  Recall that a symplectic structure on a
manifold is what defines the canonical commutation relation on a phase
space $[\pi^a, \pi^b] = i(\omega^{-1})^{ab}$.  If it is only partially
symplectic, $\omega^{-1}$ is singular.  Then, the coset space $G/H$ is
partially a phase space and partially a coordinate space.  Only a subset of
the coordinates participates in the canonical conjugate pairs, while
the remainder does not.  The former corresponds to type-B NGBs, while
the latter corresponds to type-A.

\begin{figure}[t]
  \centering
  \includegraphics[clip,width=0.5\columnwidth]{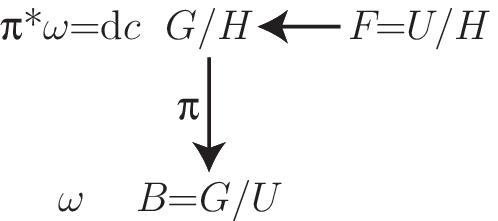}
  \caption{Fibration responsible for the presymplectic structure.  $U
    \subseteq G$ is the subgroup that commutes with all Cartan
    generators $T_i$ with nonvanishing $e_i(0)$.  The base manifold
    $B=G/U$is symplectic, which describes the type-B NGBs, while the
    fiber $F=U/H$ describes the type-A NGBs.  The symplectic form
    $\omega$ on $B$ is pulled back to $\uppi^* \omega = \mathrm{d} c$
    on $G/H$.}
  \label{fig:fiber}
\end{figure}

One crucial theorem from mathematics on the presymplectic structure
was proven by Chu \cite{Chu:1974}:
\begin{quote}
\textit{If the second dimension cohomology group $H^2({\mathfrak g})$ of the Lie algebra ${\mathfrak g}$ for a connected Lie group $G$ is trivial, then every left-invariant closed 2-form on $G$ induces a symplectic homogeneous space.}
\end{quote}
In our case, we have a presymplectic form on $G/H$ that can be pulled
back to $G$.  If $G$ is semisimple, $H^2({\mathfrak g})$ is trivial
(see Appendix \ref{appendix1}).  Then, the theorem states that $G$ can
be projected down to a symplectic homogeneous space $G/U$. Namely,
there is the structure of fibration, as shown in Fig.~\ref{fig:fiber}.
For a nonsemisimple case, however, there is a possibility of central
extension that we will discuss in Sec.~\ref{sec:central2}.

\subsection{Compact semisimple case}
It is important to ask the following question: What kinds of coset spaces support
a presymplectic structure?  We have a definite answer to this question
when $G$ is compact semisimple. 

As we have seen, $c(\pi) = e_i(0) \omega^i$ is completely specified in
terms of constants $e_i(0)$, where the generator $T_i$ commutes with
the entire $H$ [see Eq.~\eqref{conditione}].  Therefore, we can
enlarge $H$ to include all generators that commute with $T_i$ to
define the subgroup $U$ such that $U^\dagger e_i(0) T_i U = e_i(0)
T_i$ in $G$.  Mathematically, $e_i(0) T_i$ generates an Abelian group $T$,
which is called a {\it torus}\/.  Then, $U$ is called a {\it
  centralizer}\/ of the torus $T$ in $G$.  The following theorem
 proven by Borel~\cite{Borel:1954} is then useful:
\begin{quote}
  \textit{Let $G$ be compact semisimple and $U$ be the centralizer of a
  torus. Then, $G/U$ is homogeneous K\"ahlerian and algebraic.}
\end{quote}
A torus $T$ in this context means an Abelian subgroup of $G$.
Now, here is a new theorem of our own that follows from Eq.~\eqref{solc}:
  \begin{quote}
    \textit{The presymplectic structure is determined uniquely with a
      Cartan element of the Lie algebra.}
  \end{quote}
  Namely, once $e_i(0)$ is specified, we know the symplectic
  structure.  And, $e_i(0)$ generates a torus.  For
instance, an $\mathrm{SU}(N)$ group is simple and has many possible
Abelian subgroups $T=\mathrm{U}(1),\mathrm{U}(1)^2,\ldots\mathrm{U}(1)^{N-1}$. In
general, a simple group admits a torus up to $T_{\text{max}} =
\mathrm{U}(1)^r$, where $r$ is the rank of its Lie algebra, called the
maximal torus $T_{\text{max}}$. An Abelian subgroup is called a torus
because it is a manifold of coordinates with periodic boundary
conditions for each, just like the surface of a doughnut (a
two-torus). A centralizer $U$ of a torus $T$ is defined by the
collection of elements in $G$ that commute with every element of $T$,
\textit{i.e.}\/, $U=\{u\in G|utu^{-1}=t,\,\,\forall t\in T\}$. For
instance, for
\begin{equation*}
T=\{e^{ i\,\mathrm{diag}(
\overbrace{\alpha_1,\ldots,\alpha_1}^{n_1},\,
\overbrace{\alpha_2,\ldots,\alpha_2}^{n_2},\,
\cdots,
\overbrace{\alpha_k,\ldots,\alpha_k}^{n_k}\,)}\},
\end{equation*}
where $\sum_{i=1}^kn_i=N$ and $\sum_{i=1}^kn_i\alpha_i=0$ (traceless),
$T=\mathrm{U}(1)^{k-1}\subset\mathrm{SU}(N)$, and its centralizer is
$U=\mathrm{U}(1)^{k-1} \times \prod_{i=1}^k\mathrm{SU}(n_i)$. Borel's theorem then states then $G/U=\mathrm{SU}(N)/[\mathrm{U}(1)^{k-1}
\times \prod_{i=1}^k\mathrm{SU}(n_i)]$ is K\"ahler.  A K\"ahler
manifold always allows for a symplectic structure.

Therefore, this kind of a partially symplectic structure is possible
on the coset space by considering the following fiber bundle $F\hookrightarrow G/H\stackrel{\uppi}{\rightarrow}B$, where the base
space $B = G/U$ is symplectic.  (Note that we use the boldface $\uppi$
here to avoid a possible confusion with the NG field $\pi$.)  The
fiber is $F = U/H$. The symplectic structure $\omega$ on $B$ is pulled
back by the projection $\uppi$ as $\uppi^* \omega$ on the entire coset
space $G/H$. Since the closedness $\mathrm{d}\omega=0$ on $B$ implies
the closedness $\mathrm{d}(\uppi^*\omega) = 0$ on $G/H$, we can always
find a one-form $c$ such that $\mathrm{d}c = \uppi^*\omega$ locally on
$G/H$. Therefore, what we see in the Lagrangian at the first order in the time derivative is this pullback $\uppi^* \omega$ (further pulled back
to spacetime by $\pi$).

\begin{figure}[t]
  \centering
  \includegraphics[clip,width=\columnwidth]{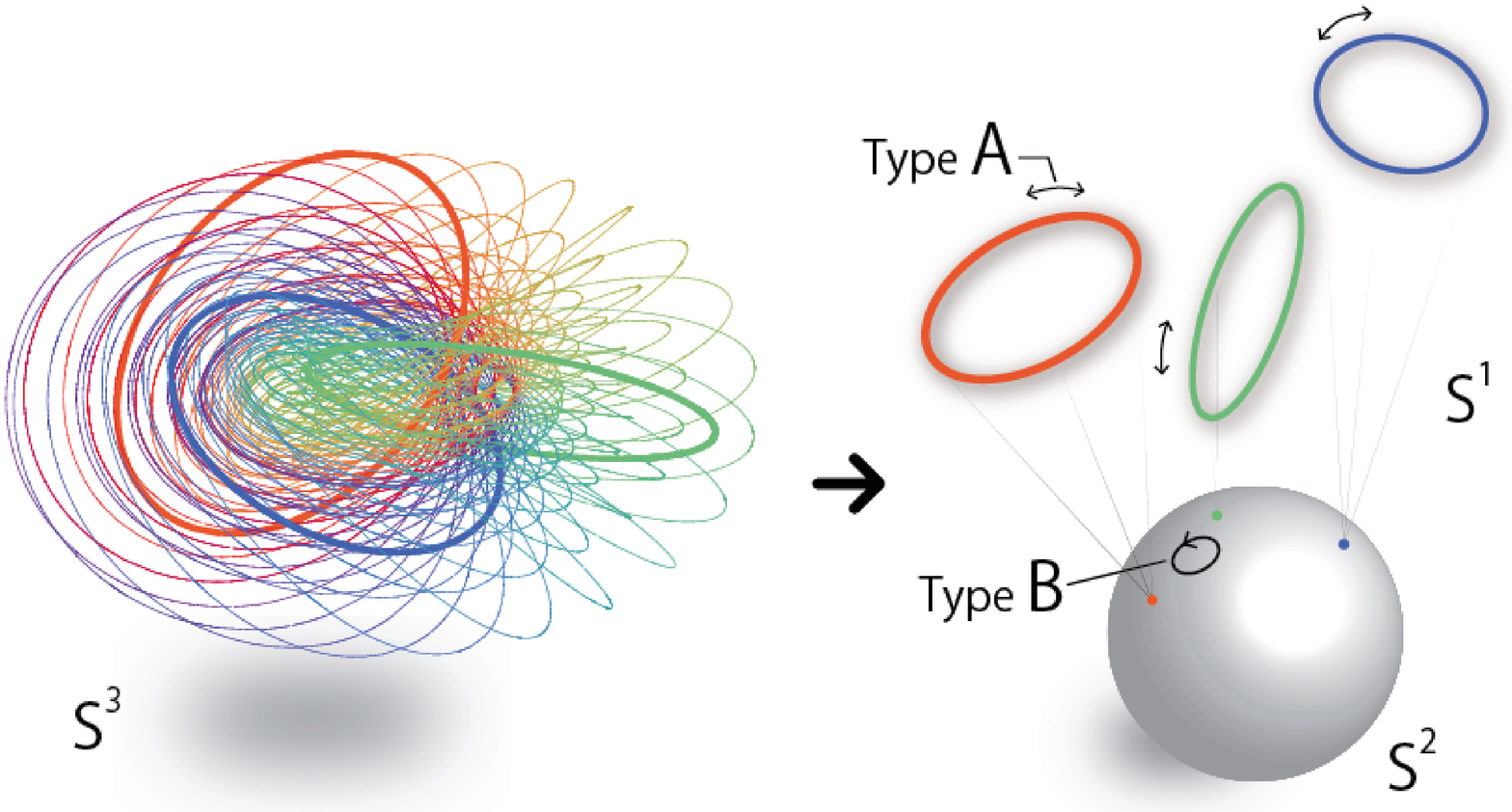}
  \caption{Graphical representation of the fibration $S^1\hookrightarrow S^3\stackrel{\uppi}{\rightarrow}S^2$, where the projection $\uppi$ is the Hopf map.  On each point on $S^2$, there is an $S^1$ fiber where the type-A NGB can fluctuate.  The fiber on each point is shown with different colors.  The type-B NGBs fluctuate on $S^2$.  On the left, the entire $S^3$ is shown using a stereographic projection onto $\mathbb{R}^3$.  $S^1$ Fibers are shown as circles, and the collection of circles form the entire $S^3$.  Note that every circle is intertwined with every other circle.}
  \label{fig:fiberation2}
\end{figure}

The simplest example to see this structure is the $S^3=\text{U}(2)/\text{U}(1)$ as an $S^1$ fibration over $S^2$, as shown in Fig.~\ref{fig:fiberation2}.   In this example, type-B NGBs live on the base space $S^2$, while the type-A NGB fluctuates along the $S^1$ fibers.

The projection on a symplectic manifold makes sense from a physics point
of view.  In the long-distance limit, the modes with quadratic
dispersion (typically, type-B) have much lower energies than those with
linear dispersion (typically, type-A).  Therefore, keeping only the
type-B modes, namely, those with canonically-conjugate pairs, would
make sense in this limit.  It corresponds to the projection on the
symplectic base manifold that describes type-B NGBs while eliminating
the fiber that describes type-A NGBs.

The symplectic structure on $B=G/U$ is specified by parameters
$e_i(0)$.  Going back to the example of
$G/U=\mathrm{SU}(N)/(\mathrm{U}(1)^{k-1} \times
\prod_{i=1}^k\mathrm{SU}(n_i))$, using the exact sequence of the
homotopy groups, it is seen that $\pi_2(G/U) =
\pi_1(\mathrm{U}(1)^{k-1})/\pi_1(G) = {\mathbb Z}^{k-1}$, while the Hurewicz theorem says that $H_2(G/U) = \pi_2(G/U)$ when $\pi_1(G/U)=0$.  In addition, because $G/U$ is compact without a boundary,
$H_{\text{dR}}^2=H_2(G/U,{\mathbb R})$ (de Rham theorem).  Therefore,
there are $k-1$ generators of $H^2_{\text{dR}}(G/U)$, $\omega_1,
\omega_2, \ldots, \omega_{k-1}$, that can be used for the symplectic
form $\omega = \sum_{i=1}^{k-1} a_i \omega_i$ on $G/U$.  These numbers
$a_i (i=1, \ldots, k-1)$ specify $\omega = dc$, and hence, $c=c_a
d\pi^a$ in the Lagrangian.  The number of $a_i$ is precisely the same number of
parameters as $e_i(0)$ for this coset space.

In general, $\text{dim}\, H^2_{\text{dR}}(G/U)$ is the same as the
number of $\text{U}(1)$ factors in $U$ when $G$ is semisimple [{\it i.e.}\/, no $\text{U}(1)$ factors in $G$].  Pulled back to $G/H$, the
possibilities of presymplectic structure correspond to the number
$N_{\text{C}}$ of Cartan generators in $G$ that commute with $H$.  We will use
this fact extensively when we present the classification of possible
presymplectic structures in the next section.

Note, however, that the linear combination $\omega = \sum_{i=1}^{k-1}
a_i \omega_i$ may be degenerate for a certain choice of the parameters
$a_i$.  For instance, $G/H=\text{SU}(3)/\text{U}(1)\times \text{U}(1)$
is K\"ahler, has $H^2(G/H)={\mathbb Z}^2$, and supports a
symplectic structure.  There are two linearly independent closed
invariant two-forms in Eq.~\eqref{exampleofb}: $\text{d}\omega^3$
[\eqref{eq:omega3}] and $\text{d}\omega^8$ [\eqref{eq:omega8}].  Note that
$\omega^3$ and $\omega^8$ are not globally defined, as they transform
inhomogeneously under the group transformations [see
Eq.~\eqref{transomegau}].  Therefore, these two two-forms are closed
but not exact, generate
$H^2_{\text{dR}}[\text{SU}(3)/\text{U}(1)\times \text{U}(1)]$, and are
candidates for the symplectic structure.  Indeed, $\text{d}\omega^3 =
\text{d} \pi^1 \wedge \text{d} \pi^2 + (1/2) (\text{d} \pi^4
\wedge \text{d} \pi^5 - \text{d} \pi^6 \wedge \text{d} \pi^7) +
O(\pi)^3$ and hence is nondegenerate.  On the other hand, if we pick
$\text{d}\omega^8 = (\sqrt{3}/2) (\text{d} \pi^4 \wedge \text{d}
\pi^5 + \text{d} \pi^6 \wedge \text{d} \pi^7) + O(\pi)^3$, it does not
provide a canonical structure between $\pi^1$ and $\pi^2$, and hence, it
is degenerate.  There is actually a larger symmetry that preserves
this choice because the torus is $\text{U}(1)$ generated by $T_8$
and its centralizer is $\text{U}(2)$.  Then, it can be projected down
to $\text{SU}(3)/\text{U}(2)=\mathbb{C}P^2$, where the fiber is
$\text{U}(2)/\text{U}(1)\times \text{U}(1) = S^2$.  This fibration is an example
where the fiber is not a group \footnote{We thank Alan Weinstein for
  this example.}.

\subsection{Case with central extensions}
\label{sec:central2}

So far, we have assumed that $G$ is compact semisimple.  If $G$ is not
semisimple, especially if it has more than one $\text{U}(1)$ factor,
its second cohomology $H^2(\mathfrak{g})$ is nontrivial and it allows
for a central extension.  See Appendix~\ref{appendix1} for more
discussions on the central extension.

In this case, $G/H$ may not necessarily be projected down to a
symplectic manifold.  Considering $G=\text{U}(1)^3$ and $H=\{e\}$, for an
example, parametrized by three angles, $T^3 = G/H=\{\varphi^a \in
[0,2\pi) | a=1,2,3\}$ is a three-torus.  We can introduce a
presymplectic structure \cite{Chu:1974}
\begin{equation}
  \omega = \mathrm{d} \varphi^1 \wedge (\mathrm{d} \varphi^2 + r \mathrm{d} \varphi^3).
\end{equation}
If $r$ is a rational number $r=p/q$ for $p$ and $q$ relatively prime,
the orbit winds around $T^3$ $q$ times and closes on itself.  Then,
there is a well-defined projection down to $T^2$.  On the other hand,
if $r$ is an irrational number, there is no well-defined projection
because the orbit winds around $T^3$ infinite times without closing on
itself.

We suspect that such a pathological case would not arise in physical
systems.  Yet, we do not have a concrete proof of what goes wrong in such
a case.

\subsection{Quantization condition}
\label{sec:quantization}
The normalization of the presymplectic structure may be quantized.
All discussions above are, so far, concerned with the invariance of the
action up to a surface term.  In classical physics, the action itself
does not have a physical meaning while its variation leads to the
equations of motion.  In quantum physics, however, the action itself
goes into the path integrals as $e^{iS/\hbar}$, and hence, its value
matters.  Yet, a change in the action by integer multiples of
$2\pi\hbar$ does not change the path integral.  Recall that we use the
unit $\hbar=1$ in this paper, and henceforth, we drop $\hbar$ in
expressions.

When $\omega = \text{d} c$ is closed but not exact, namely, an element
of $H^2_{\text{dR}}(G/H)$, its coefficient is quantized.  Considering a
time integral to be a periodic loop $L^1$ on $G/H$, the loop can be
viewed as a boundary of a two-disk. [Here, we assume $\pi_1(G/H)=0$, so
that every loop on $G/H$ is contractible to a point.] However,
nontrivial $H^2_{\text{dR}}$ implies nontrivial $H_2$, and hence,
there are noncontractible two-cycles on $G/H$.  Namely, there are
nontrivial closed two-dimensional surfaces $C_2$ in $G/H$.  Then, 
$C_2 = C_2^+ \cup C_2^-$ is a union of two surfaces that share the
same boundary $L^1 = \partial C_2^+ = -\partial C_2^-$.  The simplest
example is $C_2 \simeq S^2$, where $L^1$ is the equator, $C_2^+$ the
northern hemisphere, and $C_2^-$ the southern hemisphere.  For the
action
\begin{equation}
  S \ni \int \text{d}^d x\int_{L_1} c
\end{equation}
to give a single valued  $e^{iS}$, its ambiguity
\begin{equation}
  \Delta S = \int  \mathrm{d}^d x\int_{C_2^+} \mathrm{d} c
  - \int \mathrm{d}^d x\int_{-C_2^-} \mathrm{d} c
  = \int \mathrm{d}^d x\int_{C_2} \mathrm{d} c
\end{equation}
must be quantized in units of $2\pi$.  This discussion is the same as the
one on Wess-Zumino-Witten terms in Sec.~\ref{sec:WZW}.

When the system is finite $\Omega = \int \mathrm{d}^d x < \infty$, the
quantization condition restricts the normalization of $c$.  In other
words, $\Omega \mathrm{d}c$ is an element of $H^2(G/H,{\mathbb Z})$ rather
than $H^2_{\text{dR}}(G/H) = H^2 (G/H, {\mathbb R})$.

The same consideration applies to central extensions.  When the target
space is compact, the (pre)symplectic form is quantized.  For example, 
for $\text{U}(1)^2 = \{(\varphi^1, \varphi^2) | \varphi^i \in
[0,2\pi)\}$, $\Omega \mathrm{d} c = k(2\pi)^{-1} \mathrm{d} \varphi^1
\wedge\mathrm{d} \varphi^2$ with $k \in {\mathbb Z}$.  On the other
hand, when the target space is non-compact, such as ${\mathbb R}^2 =
{\mathbb C}$ in the case of the free Schr\"odinger field mentioned in
Sec.~\ref{sec:central} and Appendix~\ref{sec:discrete}, the coefficient
is not quantized.

\section{Classification of Possible Presymplectic Structures}
\label{sec:classification}
As we have seen in Sec.~\ref{sec:topology}, a presymplectic structure
on a coset space $G/H$ is characterized by its fibration on a
symplectic base space $B=G/U$ with the fiber $F=U/H$, when $G$ and $H$
are compact semisimple.  $U \subset G$ is the subgroup that commutes
with generators with nonzero $e_i(0)$.  Since $e_i(0)$'s need to be
invariant under $H$ [Eq.~\eqref{conditione}], $H \subset U$.  The base
space describes type-B NGBs while the fiber describes type-A NGBs.  In this
section, we show how such structures can be completely classified.

\subsection{Preliminary discussions}

The number of type-A and type-B NGBs is given by the counting rule in Eqs.~\eqref{countingA} and \eqref{countingB}.  
If the rank of $\rho$ explores all the possible integral values in the range
\begin{eqnarray}
0\leq \text{rank}\rho\leq\text{dim}\,G/H,\label{range}
\end{eqnarray}
the number of type-A and type-B NGBs can be any combinations between
$(n_{\text{A}},n_{\text{B}})=(\text{dim}\,G/H,0)$ and
$(0,\frac{1}{2}\text{dim}\,G/H)$.  Indeed, in the case of Heisenberg
magnets $G/H=\text{SO}(3)/\text{SO}(2)$ ($\text{dim}\,G/H=2$),
antiferromagnets and ferromagnets, respectively, realize the case
$\text{rank}\rho=0,1$.  However, in this section, we discuss that, in
general, allowed values of $\text{rank}\rho$ are strongly constrained. 

In general, we can always choose the basis of generators in such a way
that only Cartan generators~\cite{Georgi} of $G$ that commute with all
generators of $H$ may have a nonzero expectation value $\langle
j_i^0(\vec{x},t)\rangle\neq0$~\cite{WatanabeBrauner1}, as we have
discussed in previous sections.  Their expectation values specify
$e_i(0)$, and the corresponding generators generate the torus $T$.
Each nonzero expectation value of conserved charge densities defines a
presymplectic structure on $G/H$ by $c = -e_i(0) \omega^i$
[Eq.~(\ref{solc})]; namely, it makes NG fields associated with broken
generators $Q_a$ and $Q_b$ canonically conjugate to each other, as
discussed in Sec.~\ref{sec:topology}

For a given $G$ and $H$, let $N_{\text{C}}$ be the number of Cartan generators
of $\mathfrak{g}$ that commute with $\mathfrak{h}$.  Based on the
above considerations, we know that these generators are the only ones
that are allowed to have nonvanishing $e_i(0)$.  Therefore, there are
$N_{\text{C}}$ parameters to specify the possible presymplectic structure on
$G/H$.  This counting takes into account only the connected component
$G_0$ of the identity, and the discrete subgroup $G/G_0$ might further
restrict allowed presymplectic structures.

Therefore, we first consider the case when $H$ is generated by Cartan
generators alone, so that all Cartan generators commute with
$\mathfrak{h}$ to maximize $N_{\text{C}}$.

\subsection{Flag manifolds}
To study the case of maximum $N_{\text{C}}$ for a given $G$, let us consider
the flag manifolds $G/\text{U}(1)^{r}$, where $N_{\text{C}} = r\geq 1$ is the
rank of the simple group $G$.  We can systematically enumerate all possibilities of
presymplectic structures for them.  It turns out that
this list allows us to also classify possibilities for other $G/H$ as
well.  In this sense, the discussion here is the basis of all other
cases.  For concreteness, we first discuss $\text{SU}(n+1)/\text{U}(1)^{n}$. 

A flag manifold is K\"ahler, thanks to the Borel
theorem~\cite{Borel:1954}, and is hence symplectic.  Indeed, for
$\text{SU}(n+1)/\text{U}(1)^{n}$, $\text{dim}\,{G/H}=n(n+1)$ is always
even.  Since all Cartan generators of $G$ remain unbroken, $N_{\text{C}}=n$ and
there are many presymplectic structures that can control the number of
type-A and type-B NGBs.  The simplest case of $\text{SU}(3)/\text{U}(1)^2$
with $N_{\text{C}}=2$ is shown in Table~\ref{tab:SU(3)}.

The two limiting cases can easily be understood.  Any symplectic manifold is endowed with an associated symplectic two-form, which always realizes the case $\mathrm{rank}\rho=\text{dim}\,{G/H}$. (Unless discrete the subgroup puts out an obstacle.)  Thus, we know that $(n_{\text{A}},n_{\text{B}})=(0,n(n+1)/2)$ is possible.  Also, by setting all expectation values of charge densities to be $0$, one can realize the case where $\mathrm{rank}\rho=0$, and hence, $(n_{\text{A}},n_{\text{B}})=(n(n+1),0)$.  

The question is whether it is possible to realize combinations of
$(n_{\text{A}},n_{\text{B}})$ between these two limiting cases.
Although there are $N_{\text{C}}=n$ parameters to control, the number of
integers in the range Eq.~\eqref{range} grows as $n^2$, so obviously,
it is not possible to realize all of these values for a large $n$.
For example, there is a minimum value of $\mathrm{rank}\,\rho$ (except
for $0$), which is achieved by the presymplectic structure that
appeared in the above discussion of
$\text{SU}(n+1)/\text{U}(n)=\mathbb{C}P^{n}$ model.  This presymplectic structure gives
$\mathrm{rank}\rho=n$, and $0<\mathrm{rank}\rho<n$ is prohibited.

\begin{table}[t]
  \centering
    \caption{Possible number of type-A and type-B NGBs for
    $\text{SU}(3)/\text{U}(1)\times \text{U}(1)$.\label{tab:SU(3)}}
  \begin{tabular}{cccc}
    \hline\hline
    $n_{\text{A}}$ & $n_{\text{B}}$ & $F=U/H$ & $B=G/U$ \\ \hline
    6 & 0 & $\text{SU}(3)/\text{U}(1)\times \text{U}(1)$ & $\{e\}$ \\
    2 & 2 & $\text{SU}(2)/\text{U}(1)$ & $\text{SU}(3)/\text{SU}(2)\times \text{U}(1)$ \\ 
    0 & 3 & $\{e\}$ & $\text{SU}(3)/\text{U}(1)\times \text{U}(1)$ \\ \hline\hline
  \end{tabular}
\end{table}

The case for simple classical groups is straightforward to work out.
The smallest possible $H$ that makes $G/H$ symplectic is the flag
manifold $H=\text{U}(1)^r$, where $r$ is the rank of $G$.  All Cartan
generators commute with $\text{U}(1)^r$, and hence, $N_{\text{C}}=r$.
Therefore, this case allows for the largest number of possible choices
for $U$.

Because $e_i(0)$ belong to the adjoint representation, the
corresponding generators $T_i$ generate a torus $T$, and its
centralizer $U$ is generated by all generators of $\mathfrak{g}$ that
leave $e_i(0)$ invariant.  Such symmetry-breaking patterns have been
studied extensively in the literature (see, {\it e.g.}\/, Ref.~\cite{Intriligator:1995ax}).  

For $\text{SU}(n)$ groups, the possible form of $e_i(0)T_i$ is
\begin{equation}
  e_i(0) T_i = \mathrm{diag}(
  \overbrace{\alpha_1,\ldots,\alpha_1}^{n_1},\,
  \overbrace{\alpha_2,\ldots,\alpha_2}^{n_2},\,
  \ldots,
  \overbrace{\alpha_k,\ldots,\alpha_k}^{n_k}\,),
\end{equation}
and the corresponding centralizer is
\begin{equation}
  U = \text{U}(1)^{k-1} \times \prod_{k} \text{SU}(n_k), \quad
  n = \sum_k n_k, \quad \sum_k n_k \alpha_k = 0.
\end{equation}
In this expression, $\text{SU}(1)$ counts as a trivial group.

For $\text{SO}(n)$ groups, any element of the adjoint representation
is an antisymmetric matrix that can be skew diagonalized.  Therefore,
the possible form of $e_i(0) T_i$ is
\begin{equation}
  e_i(0) T_i = \mathrm{diag}(
  \overbrace{0,\ldots,0}^{m},\,
  \overbrace{\alpha_1,\ldots,\alpha_1}^{n_1},\,
  \ldots,
  \overbrace{\alpha_k,\ldots,\alpha_k}^{n_k}\,) \otimes i\sigma_2,
\end{equation}
and we find the centralizer
\begin{equation}
  U = \text{SO}(m) \times \prod_k \text{U}(n_k), \qquad
  n = m + 2 \sum_k n_k.
\end{equation}
Finally, for $\text{Sp}(n)$ groups [we use the notation that the rank is $n$
for $\text{Sp}(n)$], every element $g\in \text{Sp}(n)$ preserves
\begin{equation}
  J = \left(
    \begin{array}{c|c}
      0 & -I_n\\ \hline I_n & 0
    \end{array} \right), \qquad
  g J g^T = J.
\end{equation}
Therefore, the adjoint representation is a $2n\times 2n$ matrix of the
form
\begin{equation}
  S = \left(
    \begin{array}{c|c}
      A & B\\ \hline C & -A^T
    \end{array} \right),
  \qquad SJ+JS^T=0.
\end{equation}
Here, $B^T=B$ and $C^T=C$ are symmetric matrices.  The Cartan
generators are given by the diagonal matrices in $A$ with $B=C=0$ and
therefore have the form $S=A_\text{diag} \otimes \sigma_3$.  In general,
\begin{equation}
  e_i(0) T_i = \mathrm{diag}(
  \overbrace{0,\ldots,0}^{m},\,
  \overbrace{\alpha_1,\ldots,\alpha_1}^{n_1},\,
  \ldots,
  \overbrace{\alpha_k,\ldots,\alpha_k}^{n_k}\,) \otimes \sigma_3,
\end{equation}
and we find
\begin{equation}
  U = \text{Sp}(m) \times \prod_k \text{U}(n_k), \qquad
  n = m + \sum_k n_k.
\end{equation}
The problem is basically listing up a partition of integers.  

Once all possibilities $U$ are listed, it is easy to count
$n_{\text{A}} = \text{dim}\,U/H$ and $n_{\text{B}} = \text{dim}\,G/U$.
We present all possible cases for rank-five groups in tables:
$\text{SU}(6)$ (Table~\ref{tab:SU(6)}), $\text{SO}(10)$
(Table~\ref{tab:SO(10)}), and $\text{SO}(11)$ and $\text{Sp}(5)$
(Table~\ref{tab:SO(11)Sp(5)}).

Looking at Table~\ref{tab:SO(10)}, one might think that
$U=\text{SO}(6) \times \text{U}(1)^2$ and $U=\text{U}(4)\times
\text{U}(1)$ are the same because $\mathfrak{so}(6)$ and
$\mathfrak{su}(4)$ are identical Lie algebras.  They are not.  The
spectrum of the 14 type-B NGBs on $\text{SO}(10)/[\text{SO}(6) \times
\text{U}(1)^2]=\text{SO}(10)/[\text{SU}(4)/\mathbb{Z}_2 \times
\text{U}(1)^2]$ consists of $14=6+6+1+1$ under $\text{SO}(6)$, while
those on $\text{SO}(10)/(\text{U}(4) \times
\text{U}(1))=\text{SO}(10)/\{[\text{SU}(4) \times \text{U}(1)]/{\mathbb
  Z}_4\times \text{U}(1)\}$ consist of $14=4+4+6$ under $\text{SU}(4)$.  The
same comment applies to $\text{SO}(4) \times \text{U}(1)^3$ vs
$\text{U}(2)^2 \times \text{U}(1)$ as
$\mathfrak{so}(4)=\mathfrak{su}(2)\oplus \mathfrak{su}(2)$.  On
$\text{SO(10)}/[\text{SO}(4) \times \text{U}(1)^3]$, the type-B spectrum
is $18=4\times 3 + 1 \times 6$ under $\text{SO}(4)$, while for
$\text{SO}(10)/[\text{U}(2)^2 \times \text{U}(1)]$, it is
$18=(2,2)\times 2+(2,1)\times 2+(1,2)\times 2+(1,1)\times 2$ under
$\text{U}(2)\times \text{U}(2)$. Therefore, one has to be careful
about not identifying local isomorphisms among groups.

On the other hand, in the case of $\text{SO}(n)$ with $n$ even, it can
break to $U=\text{SO}(2) \times \prod_k \text{U}(n_k)$.  Turning
$e_i(0)$ for the $\text{SO}(2)$ generator would ``break'' it further
to $\text{U}(1)$ with no difference in the group structure or
representations of NGBs.  Namely, two cases are continuously connected
without an order parameter that distinguishes them.  Therefore, we can
identify $\text{SO}(2)$ and $\text{U}(1)$ and we have eliminated
duplicates from Table~\ref{tab:SO(10)}.

Note that there is a duality between $\text{Sp}(n)$ and
$\text{SO}(2n+1)$ groups in each symmetry-breaking pattern because the
dimensions of the group match: $(1/2)(2n+1)2n=n(2n+1)$ for
$\text{SO}(2n+1)$, and $(1/2)2n(2n+1) = n(2n+1)$ for
$\text{Sp}(n)$.

\begin{table}[tb]
  \centering
    \caption{Possible number of type-A and type-B NGBs for
    $\text{SU}(6)/\text{U}(1)^5$.\label{tab:SU(6)}}
  \begin{tabular}{cccc}
    \hline\hline
    $n_{\text{A}}$ & $n_{\text{B}}$ & $U$ \\ \hline 
    30 & 0 & $\{e\}$ \\ 
    20 & 5 & $\text{SU}(5)\times \text{U}(1)$ \\ 
    14 & 8 & $\text{SU}(4)\times \text{SU}(2) \times \text{U}(1)$ \\
    12 & 9 & $\text{SU}(4)\times \text{U}(1)^2$ \\ 
    12 & 9 & $\text{SU}(3)^2\times \text{U}(1)$ \\ 
    8 & 11 & $\text{SU}(3)\times \text{SU}(2)\times \text{U}(1)^2$ \\
    6 & 12 & $\text{SU}(3)\times \text{U}(1)^3$ \\
    6 & 12 & $\text{SU}(2)^3\times \text{U}(1)^2$ \\ 
    4 & 13 & $\text{SU}(2)^2\times \text{U}(1)^3$ \\ 
    2 & 14 & $\text{SU}(2)\times \text{U}(1)^4$ \\ 
    0 & 15 & $\text{U}(1)^5$ \\ \hline\hline
  \end{tabular}
\end{table}

\begin{table}[tb]
  \centering
    \caption{Possible number of type-A and type-B NGBs for
    $\text{SO}(10)/\text{U}(1)^5$.\label{tab:SO(10)}}
  \begin{tabular}{cccc}
    \hline\hline 
    $n_{\text{A}}$ & $n_{\text{B}}$ & $U$ \\ \hline
    40 & 0 &  $\{e\}$\\ 
    24 & 8 & $\text{SO}(8)\times \text{U}(1)$ \\ 
    20 & 10 & $\text{U}(5)$ \\
    14 & 13 & $\text{SO}(6)\times \text{U}(2)$ \\ 
    12 & 14 & $\text{SO}(6)\times \text{U}(1)^2$ \\ 
    12 & 14 & $\text{U}(4)\times \text{U}(1)$ \\
    10 & 15 & $\text{SO}(4)\times \text{U}(3)$ \\ 
    8 & 16 & $\text{U}(3)\times \text{U}(2)$ \\ 
    6 & 17 & $\text{SO}(4)\times \text{U}(2)\times \text{U}(1)$ \\
    6 & 17 & $\text{U}(3)\times \text{U}(1)^2$ \\ 
    4 & 18 & $\text{SO}(4)\times \text{U}(1)^3$ \\ 
    4 & 18 & $\text{U}(2)^2\times \text{U}(1)$ \\ 
    2 & 19 & $\text{U}(2)\times \text{U}(1)^3$ \\ 
    0 & 30 & $\text{U}(1)^5$ \\ \hline\hline
  \end{tabular}
\end{table}

\begin{table}[tb]
  \centering
  \caption{Possible number of type-A and type-B NGBs for
    $\text{SO}(11)/\text{U}(1)^5$ and $\text{Sp}(5)/\text{U}(1)^5$.\label{tab:SO(11)Sp(5)}}
  \begin{tabular}{cccc}
    \hline\hline
    $n_{\text{A}}$ & $n_{\text{B}}$ & $U\subset \text{SO}(11)$ 
    & $U\subset \text{Sp}(5)$ \\ \hline 
    50 & 0 & $\{e\}$ & $\{e\}$ \\  
    32 & 9 & $\text{SO}(9)\times \text{U}(1)$ 
    & $\text{Sp}(4)\times \text{U}(1)$ \\  
    20 & 15 & $\text{SO}(7)\times \text{U}(2)$ 
    & $\text{Sp}(3)\times \text{U}(2)$ \\  
    20 & 15 & $\text{U}(5)$ & $\text{U}(5)$ \\  
    18 & 16 & $\text{SO}(7)\times \text{U}(1)^2$ 
    & $\text{Sp}(3)\times \text{U}(1)^2$ \\  
    14 & 18 & $\text{SO}(5)\times \text{U}(3)$ 
    & $\text{Sp}(2)\times \text{U}(3)$ \\  
    14 & 18 & $\text{SO}(3)\times \text{U}(4)$ 
    & $\text{Sp}(1)\times \text{U}(4)$ \\ 
    12 & 19 & $\text{U}(4)\times \text{U}(1)$ 
    & $\text{U}(4)\times \text{U}(1)$ \\ 
    10 & 20 & $\text{SO}(5)\times \text{U}(2)\times \text{U}(1)$ 
    & $\text{Sp}(2)\times \text{U}(2)\times \text{U}(1)$ \\ 
    8 & 21 & $\text{SO}(5)\times \text{U}(1)^3$ 
    & $\text{Sp}(2)\times \text{U}(1)^3$ \\ 
    8 & 21 & $\text{SO}(3)\times \text{U}(3)\times \text{U}(1)$ 
    & $\text{Sp}(1)\times \text{U}(3)\times \text{U}(1)$ \\ 
    8 & 21 & $\text{U}(3)\times \text{U}(2)$ 
    & $\text{U}(3)\times \text{U}(2)$ \\ 
    6 & 22 & $\text{SO}(3)\times \text{U}(2)^2$ 
    & $\text{Sp}(1)\times \text{U}(2)^2$ \\ 
    6 & 22 & $\text{U}(3)\times \text{U}(1)^2$ 
    & $\text{U}(3)\times \text{U}(1)^2$ \\ 
    4 & 23 & $\text{SO}(3)\times \text{U}(2)\times \text{U}(1)^2$ 
    & $\text{Sp}(1)\times \text{U}(2)\times \text{U}(1)^2$ \\ 
    4 & 23 & $\text{U}(2)^2\times \text{U}(1)$ 
    & $\text{U}(2)^2\times \text{U}(1)$ \\ 
    2 & 24 & $\text{SO}(3)\times \text{U}(1)^4$ 
    & $\text{Sp}(1)\times \text{U}(1)^4$ \\ 
    2 & 24 & $\text{U}(2)\times \text{U}(1)^3$ 
    & $\text{U}(2)\times \text{U}(1)^3$ \\ 
    0 & 25 & $\text{U}(1)^5$ & $\text{U}(1)^5$ \\ \hline\hline
  \end{tabular}
\end{table}

It should be possible to enumerate possibilities for exceptional
groups $G_2$, $F_4$, and $E_{6,7,8}$ as well, but
we do not attempt it here.

\subsection{General $H$}

For more general $G/H$, we start with the list of possible $U$ for
$G/\text{U}(1)^r$ and remove those that do not commute with $H$.  It
gives all possible presymplectic structures.  The number of type-B
NGBs is given by $n_{\text{B}}= (1/2) \text{dim}\,G/U$, while $n_{\text{A}} = {\rm
  dim}U/H$.  Let us discuss a few examples below.

For instance, one can consider $\text{SU}(6)/\text{SU}(5)$, whose
dimension is $35-24=11$.  Note that
$\text{SU}(6)/\text{SU}(5)=\text{U}(6)/\text{U}(5)=S^{11}$ which is
discussed in Sec.~\ref{subsec:S2n+1}.  Looking at the list in
Table~\ref{tab:SU(6)}, the only $U$ that commutes with $\text{SU}(5)$
is in the top two.  Therefore, there are two types of presymplectic
structures possible on $\text{SU}(6)/\text{SU}(5)$.  If $U$ is
trivial, all 11 are type-A NGBs.  If $U=\text{SU}(5)\times
\text{U}(1)$, $B=\text{SU}(6)/[\text{SU}(5)\times\text{U}(1)]={\mathbb
  C}\text{P}^5$ and there are five type-B NGBs for $(1/2){\rm
  dim}B=5$.  There is only one type-A NGB.

If the same $\text{SU}(6)$ is broken by an order parameter in
a rank-three antisymmetric tensor, the unbroken group is
$H=\text{SU}(3) \times \text{SU}(3)$.  In this case, there is no $U$
that commutes with $H$ except for the trivial one.  Namely, this coset
space allows for no presymplectic structure, and hence, $n_{\text{A}} = 19$ and $n_{\text{B}}=0$.  However, if one of the $\text{SU}(3)$ is further broken
completely by order parameters in fundamental representations (at
least two of them), $H=\text{SU}(3)$ commutes with the first seven choices
of $U$ in Table~\ref{tab:SU(6)}, and there are accordingly seven possibilities of
$(n_{\text{A}}, n_{\text{B}})$. 

This way, one can work out all possibilities of $(n_{\text{A}}, n_{\text{B}})$ for a
given $G$ and $H$ if compact and simple.  Then, we look at discrete
subgroups if $G$ or $H$ has more than one connected component to
further eliminate some possibilities.  It is also straightforward to
study examples with additional $\text{U}(1)$ factors, paying attention to
possible central extensions.

This way, one can enumerate all possible presymplectic structures for
a given $G/H$ and write down the most general effective Lagrangians
using the explicit forms we found in Sec.~\ref{sec:solution}.

\section{Examples}
\label{sec:example}
Having developed a complete classification of presymplectic
structures, we revisit popular examples of coset spaces in the
literature and show what effective Lagrangians are possible for them.

\subsection{$\text{O}(n+1)/\text{O}(n)=S^{n}$}
For $\text{O}(n+1)/\text{O}(n)=S^{n}$,
$\text{SO}(n+1)/\text{SO}(n)=S^{n}$, and
$\text{O}(n+1)/[\text{O}(n)\times\mathbb{Z}_2]=\mathbb{R}P^{n}$,
there is no possible presymplectic structure for $n\geq 3$.  As seen
in Tables~\ref{tab:SO(10)} and \ref{tab:SO(11)Sp(5)}, there is no
nontrivial $U$ that commutes with the $\text{SO}(n)$ subgroup within
$\text{SO}(n+1)$, and hence, $N_{\text{C}}=0$.  Therefore, we can only have $n$
type-A NGBs.  The most general Lagrangian is hence
\begin{equation}
\mathcal{L}_{\text{eff}}= \frac{1}{2} \bar{g}_0\dot{n}_i \dot{n}_i
  - \frac{1}{2} g_0 \vec{\nabla}n_i \cdot \vec{\nabla}n_i 
\end{equation}
up to the second order in derivatives, where $\vec{n}$ is a normalized $(n+1)$-component vector.

When $n=2$, all of these examples have $N_{\text{C}}=1$ and the coset
$\text{SO}(3)/\text{SO}(2)=S^2$ indeed describes both ferro- and
antiferromagnets.  However, for $\text{O}(3)/\text{O}(2)=S^2$, there
is no presymplectic structure that is consistent with the discrete
subgroup $\{+\openone,-\openone\}$, at least when we realize it as an
internal symmetry.  To see this point, let us parametrize the coset $S^2$ by
the spherical coordinate $(\theta,\phi)$.  The candidate of a one-form
that is associated with the would-be symplectic structure is
$\cos\theta\dot{\phi}$, but it changes sign under $-\openone$:
$\theta\rightarrow\pi-\theta$ and $\phi\rightarrow\phi+\pi$ unless the
discrete symmetry incorporates with the time reversal $t\rightarrow
-t$.  The coset
$\text{O}(3)/[\text{O}(2)\times\mathbb{Z}_2]=\mathbb{R}P^2$ can
be discussed in a similar fashion, but since $\mathbb{R}P^2$ is
not even orientable, there is obviously no symplectic structure that
is consistent with the global topology of $G/H$.

\subsection{$\text{SU}(n+1)/\text{U}(n)=\mathbb{C}P^{n}$}
The $\mathbb{C}P^{n}$ ($n\geq 1$) model is a natural
generalization of ferromagnets based on $S^2=\mathbb{C}P^1$.
For $G/H=\text{SU}(n+1)/\text{U}(n)=\mathbb{C}P^{n}$, $N_{\text{C}}=1$
because there is a unique Cartan generator $\text{diag}(n, -1, \ldots,
-1)$ that commutes with $H=\text{U}(n)$.  Therefore, there is a unique
symplectic structure on $G/H$ (up to an overall normalization).  The
effective Lagrangian can be most conveniently expressed in terms of
an $n$-component complex field $z(\vec{x},t)\in\mathbb{C}^n$, and the most
general effective Lagrangian to the quadratic order in derivatives is
given by
\begin{equation}
\mathcal{L}_{\text{eff}}=is_0\frac{z^\dagger\dot{z}-\dot{z}^\dagger z}{1+z^\dagger z}+G_{ab}\left(\bar{g}_0\dot{\bar{z}}^a\dot{z}^b-g_0\vec{\nabla}\bar{z}^a\cdot\vec{\nabla}z^b\right),
\end{equation}
where
\begin{equation}
G_{ab}(\bar{z},z)=\frac{\delta_{ab}(1+\bar{z}z)-\bar{z}^bz^a}{(1+\bar{z}z)^2}
\end{equation}
is the Fubini-Study metric on
$\mathbb{C}P^{n}$~\cite{Nakahara,Eguchi:1980jx}.  In $2+1$
dimensions, we can add a topological term ($\theta$-term)
$(i/2\pi)G_{ab}(\bar{z},z)\epsilon^{ij}\partial_i\bar{z}^a\partial_j
z^b$.  The $n=1$ case is identical to ferromagnets (recall that
$\mathbb{C}P^1=S^2$).  The coefficient of the first term $s_0$
is the charge density of the ground state $\langle
j_{\rho_0}^0(x)\rangle$, where $\rho_0$ is the U(1) part of the
unbroken subgroup $H=\text{U}(n)$.  This term $s_0\Omega$ must be
quantized to a half-integer, where $\Omega$ is the volume of the
system, as discussed in Sec.~\ref{sec:quantization}.  When $s_0\neq0$,
the system resembles ferromagnets: The real and imaginary parts of
$z^a$ become canonically conjugate to each other, and there are $n$
type-B NGBs.  On the other hand, when $s_0=0$, the ground state is
antiferromagnetic and there are $2n$ type-A NGBs.  Other
possibilities $(n_{\text{A}},n_{\text{B}})=(2,n-1),(4,n-2),\ldots,(2n-2,1)$ cannot
be realized.

\subsection{$\text{U}(n+1)/\text{U}(n)=S^{2n+1}$\label{subsec:S2n+1}}
$\text{U}(n+1)/\text{U}(n)=S^{2n+1}$ ($n\geq 1$) is topologically the
same as $\text{SO}(2n+2)/\text{SO}(2n+1)$, yet its field theory is
very different because $N_{\text{C}}=1$ for the generator $\text{diag}(n,-1,
\ldots, -1)$.  It is closely related to the $\mathbb{C}P^{n}$
model since it admits a fibration $S^1\hookrightarrow
S^{2n+1}\stackrel{\uppi}{\rightarrow}\mathbb{C}P^{n}$, where
type-B NGBs live on the base manifold $\mathbb{C}P^{n}$ and a
type-A NGB is in the fiber $S^1$.  Therefore, there are only two
possibilities $(n_{\text{A}},n_{\text{B}})=(1,n)$ and $(2n+1,0)$,
which is expected from $N_{\text{C}}=1$.  The case $n=1$ of this model
describes the physics of Kaon
condensation~\cite{Miransky,Schafer}. The generalization to
$n\geq1$ is discussed in Ref.~\cite{Andersen}.

As a concrete example, let us consider a $\mathrm{U}(n+1)$-symmetric
Schr\"odinger field
\begin{equation}
  \mathcal{L}=i\psi^\dagger\dot{\psi}
  -\frac{1}{2m}\vec{\nabla}\psi^\dagger\cdot\vec{\nabla}\psi
  -\frac{\lambda}{2}(\psi^\dagger\psi-n_0)^2,\label{UV1}
\end{equation}
where $\psi(x)$ is a complex $(n+1)$-dimensional column vector. A
similar model was discussed in
Refs.~\cite{Miransky,Schafer}. At the tree level, it has the
vacuum
\begin{equation}
  \langle\psi\rangle
  =\sqrt{n_0}(1,0,\ldots,0)^T.
\end{equation}
In this case, the original $\mathrm{U}(n+1)$ symmetry is broken to
$\mathrm{U}(n)$ symmetry. The coset space
$\mathrm{U}(n+1)/\mathrm{U}(n)=S^{2n+1}$ does not admit a symplectic
structure.

Therefore, we have to carefully parametrize the coset space.  Since
$\mathrm{U}(n+1)/[\mathrm{U}(n)\times\mathrm{U}(1)]\cong\mathbb{C}P^{n}$,
which does admit a symplectic structure, we view $S^{2n+1}$ as a
$\mathrm{U}(1)$ bundle on $\mathbb{C}P^{n}$. The symplectic two-form
lives on $\mathbb{C}P^{n}$. We parametrize the field $\psi(x)$ as
\begin{equation}
\psi=\sqrt{n}\frac{e^{-i\theta}}{\sqrt{1+z^\dagger z}}
\begin{pmatrix}1\\z\end{pmatrix},
\end{equation}
where $z(x)$ is an $n$-dimensional column vector. Substituting the
above parametrization, we find
\begin{eqnarray}
\mathcal{L}_{\text{eff}}
&=&n_0\left(\dot{\theta}+\frac{i}{2}\frac{z^\dagger\dot{z}-\dot{z}^\dagger z}{1+z^\dagger z}\right)-\frac{1}{2\lambda}\left(\dot{\theta}+\frac{i}{2}\frac{z^\dagger\dot{z}-\dot{z}^\dagger z}{1+z^\dagger z}\right)^2\nonumber \\ 
&&-\frac{n_0}{2m}\left(\vec{\nabla}\theta+\frac{i}{2}\frac{z^\dagger\vec{\nabla}z-\vec{\nabla}z^\dagger z}{1+z^\dagger z}\right)^2\nonumber \\ 
&&-\frac{n_0}{2m}\left(\frac{\vec{\nabla}z^\dagger \vec{\nabla}z}{1+z^\dagger z}-\frac{(\vec{\nabla}z^\dagger z)(z^\dagger\vec{\nabla}z)}{(1+z^\dagger z)^2}\right)+\cdots.\label{UV2}
\end{eqnarray}
The second term arises from integrating out $n$ at the tree level and
looks the same as the terms $O(\vec{\nabla}^2)$, except for the overall
normalization because of the irreducible nature of $\theta$ and $z^i$
under $H=\text{U}(n)$. 

The terms in the last parentheses above are nothing but the
Fubini-Study metric on $\mathbb{C}P^{n}$, which is K\"ahler. On the other hand, the first term defines a one-form
\begin{equation}
c=i\frac{z^\dagger\mathrm{d}z-\mathrm{d}z^\dagger z}{1+z^\dagger z},
\end{equation}
while its exterior derivative
\begin{equation}
\mathrm{d}c=i\frac{(1+z^\dagger z)\mathrm{d}z^\dagger\wedge\mathrm{d}z
-(\mathrm{d}z^\dagger z)\wedge(z^\dagger \mathrm{d}z)}{(1+z^\dagger z)^2}
\end{equation}
is the K\"ahler form on $\mathbb{C}P^{n}$ associated with the Fubini-Study metric.
The coordinate $\theta$ represents the U(1), which is orthogonal to the tangent vectors of $\mathbb{C}P^{n}$.

In Sec.~\ref{sec:galilei}, we derive the effective Lagrangian for $n=1$ based purely on the Galilean symmetry and the $\text{U}(2)$ internal symmetry.  We should be able to rewrite the Lagrangian~\eqref{UV2} in terms of the Galilean-covariant derivatives,
\begin{gather}
\mathscr{D}_t\theta=\dot{\theta}-\frac{(\vec{\nabla}\theta)^2}{2m},\\
\mathscr{D}_tz=\dot{z}-\frac{\vec{\nabla}\theta\cdot\vec{\nabla}z}{m},\quad\vec{\mathscr{D}}z=\vec{\nabla}z,
\end{gather}
neglecting higher-order derivatives. Comparing Eq.~\eqref{UV1} with
Eq.~\eqref{withgalilean}, we notice that the Lagrangian lacks the term
that contains $\mathscr{D}_tz^\dagger\mathscr{D}_tz$.  In general, if
we start from a particular microscopic model and work only at tree
levels, the effective Lagrangian may not include all possible terms
allowed by symmetries. Missing terms are often generated by higher
corrections~\footnote{We have found that an interaction term
  $\left[(i/2)\psi^\dagger\vec{\sigma}\dot{\psi}+\text{c.c.}-(1/2m)\nabla_r\psi^\dagger\vec{\sigma}\nabla_r\psi\right]^2$,
  which respects both Galilean and $\text{U}(2)$ symmetry, contains
  $\mathscr{D}_tz^\dagger\mathscr{D}_tz$.}.

\section{Galilean invariance}
\label{sec:galilei}

So far, our discussions have focused on the spontaneous breaking of internal
symmetries.  However, in many interesting physical systems, spacetime
symmetries are also spontaneously broken.  For the sake of the clarity of
our discussions, we restrict ourselves to translationally and
rotationally invariant systems in this paper.  Therefore, we discuss
spontaneously broken Galilean invariance as an illustrative example in
this section.  We demonstrate how spacetime symmetries can be
discussed within our effective Lagrangian formalism and see how they
provide additional constraints on the parameters in the theory.
The so-called {\it inverse Higgs mechanism}\/ provides a heuristic method
to show how would-be NGB degrees of freedom can be consistently
removed from the physical spectrum in accordance with observations.
This method was discussed mostly in Lorentz-invariant systems, and our
presentation here shows how it can be successfully extended to
Lorentz-noninvariant systems.

It has recently been argued~\cite{Son2} that some classes of Galilean-invariant theories can be promoted to be nonrelativistic general-coordinate invariant, by introducing the spatial metric $g_{ij}(\vec{x},t)$ and the $\text{U}(1)$ gauge field and by assigning their nontrivial transformation rule.  The Galilean symmetry itself is global in the sense that the velocity parameter in $\vec{x}'=\vec{x}+\vec{v}t$ is a constant, but the nonrelativistic general-coordinate invariance allows a more general {\it local} transformation $\vec{x}'(\vec{x},t)$ with arbitrary time dependence (but still, $t'=t$).  Such an extended symmetry strongly restricts the response of the system to external fields.  Our discussion below should be useful to systematically produce general-coordinate-invariant combinations.

\subsection{Coset construction with spacetime symmetries}
In condensed-matter physics, superfluid helium and various types of
Bose-Einstein condensates often spontaneously break the Galilean
symmetry as well as the U(1) phase rotation.  In such a situation, one
has to make sure that the effective Lagrangian has the Galilean
symmetry.
 
Here, we discuss how to incorporate spacetime symmetries in our
effective Lagrangian. Spacetime symmetries are those which change
coordinates $x^\mu=(t,\vec{x})$ in addition to the fields. For
example, the transformation rule of the superfluid phase under the
Galilean transformation is
\begin{eqnarray}
\vec{x}'&=&\vec{x}+\vec{v}_0t,\quad t'=t,\\
\theta'(\vec{x}',t')&=&\theta(\vec{x},t)-m\vec{v}_0\cdot\vec{x}-\frac{mv_0^2}{2}t,
\end{eqnarray}
for a constant vector $\vec{v}_0\in\mathbb{R}^3$.  Since $\vec{x}$ changes, Galilean symmetry is a spacetime symmetry.

For simplicity, here, we discuss the situation where the spacetime translation $P_\mu=(H,-\vec{P})$ is not broken, and unbroken generators $Q_\rho$ are internal symmetries, while broken generators $Q_a$ may contain spacetime symmetries such as the Galilean boost generator.

Following Ref.~\cite{Ivanov:1975zq}, we use
\begin{equation}
U(x,\pi(x))=e^{ix^\mu P_\mu}e^{i\pi^a(x)Q_a}
\end{equation}
to define the Maurer-Cartan form $\omega$:
\begin{eqnarray}
\omega(x,\pi(x))&=&-iU^\dagger\mathrm{d}U\notag\\
&=& e^\mu P_\mu+\omega_\perp+\omega_\parallel.
\end{eqnarray}
Again, $\omega_\perp=\omega^aQ_a$ is the broken part and
$\omega_\parallel=\omega^\rho Q_\rho$ is the unbroken part.  $e^\mu=e^\mu_\nu\mathrm{d}x^\nu$ is called vielbein and $G_{\mu\nu}\equiv\eta_{\rho\sigma}e_\mu^{\rho}e_\nu^{\sigma}$ gives a spacetime metric that transforms nicely.  Especially, the spacetime-invariant volume-form is given by $\mathrm{d}^dx\mathrm{d}t\sqrt{|\mathrm{det}G|}$.

The symmetry transformation of $x$ and $\pi(x)$ under the action of $g$ is defined by [see Eq.~\eqref{defh}]
\begin{equation}
g U(x,\pi(x))= U(x',\pi'(x'))h_g(x,\pi(x)).
\end{equation}
Since $P_\mu$ is unbroken, one may be confused by the $e^{ix^\mu P_\mu}$ factor of $U$, but, thanks to this factor, we can realize the spacetime symmetry in this way.
Analogously to Eqs.~\eqref{transomegab} and \eqref{transomegau}, we have
\begin{eqnarray}
e^\mu(x',\pi'(x'))&=&e^\mu(x,\pi(x)),\label{vb}\\
\omega_\perp(x',\pi'(x'))&=&h_g\omega_\perp(x,\pi(x)) h_g^\dagger,\label{cd}\\
\omega_\parallel(x',\pi'(x'))&=&h_g\omega_\parallel(x,\pi(x))h_g^\dagger-ih_g\mathrm{d}h_g^\dagger.\label{cd2}
\end{eqnarray}
Here, we have used the assumption that unbroken generators are internal.

Let us first discuss the broken part of the Maurer-Cartan form.  We define the spacetime-covariant derivative $\mathscr{D}_\mu\pi^a$ through
\begin{equation}
e^\mu\mathscr{D}_\mu\pi^a=\omega^a.\label{covariant}
\end{equation}
According to Eq~\eqref{cd}, it indeed transforms covariantly:
\begin{equation}
(\mathscr{D}_\mu\pi^a)'Q_a=h_g(\mathscr{D}_\mu\pi^aQ_a)h_g^\dagger,
\end{equation}
thanks to the covariance of the vielbein $e^\mu(x,\pi(x))$ [see Eq.~\eqref{vb}].  If we had defined the covariant derivative by
\begin{equation}
\mathrm{d}x^\mu\tilde{\mathscr{D}}_\mu\pi^a=\omega^a
\end{equation}
instead of Eq.~\eqref{covariant}, $\tilde{\mathscr{D}}_\mu\pi^a$ would not transform covariantly, since $\mathrm{d}x^\mu$ is not covariant; {\it i.e.,}\/ $\mathrm{d}{x'}^\mu\neq\mathrm{d}x^\mu$.

For the same reason, the unbroken part $\partial_\mu\pi^a\omega_a^\rho$ does not transform covariantly.   From Eq~\eqref{cd2}, we have
\begin{equation}
(\partial_\mu{\pi}^a{\omega}_a^\rho)'Q_\rho
=\frac{\partial x^\nu}{\partial {x'}^\mu}\left[h_g(\partial_\nu\pi^a\omega_a^\rho Q_\rho)h_g^\dagger-i h_g\partial_\nu h_g^\dagger\right].\label{transfunbroken}
\end{equation}
If the factor $\partial x^\nu/\partial {x'}^\mu$ were absent, as in the case for internal symmetries, the unbroken part would transform covariantly up to the inhomogeneous term $-ih_g\partial_\nu h_g^\dagger$, which may be just a total derivative. In such a case, the unbroken part can be added to the effective Lagrangian, as discussed in Sec.~\ref{sec:rightH}.  
However, nontrivial $\partial x^\nu/\partial {x'}^\mu$ poses an obstacle, as we shall see shortly.

Covariant derivatives in Eq.~\eqref{covariant} are the building blocks of the effective Lagrangian.  The case considered in Sec.~\ref{sec:solution}, where only internal symmetries are broken, can be understood as the spacial case of $e^\mu(x,\pi(x))=\mathrm{d}x^\mu$.  In the following, we will demonstrate what we have said here using a concrete example.

\subsection{Example}
In this section, we discuss the effective Lagrangian for the microscopic model
\begin{equation}
\mathcal{L}=\frac{i}{2}(\psi^\dagger\dot{\psi}-\text{c.c.})-\frac{\vec{\nabla}\psi^\dagger\cdot\vec{\nabla}\psi}{2m}-\frac{g}{2}(\psi^\dagger\psi-n_0)^2.\label{GP}
\end{equation}
This model can be seen as the nonrelativistic version of the model for the Kaon condensation discussed in Refs.~\cite{Miransky,Schafer}.  Here, $\psi=(\psi_1,\psi_2)^{T}$ is a two-component complex scaler field.  The ground-state expectation value $\langle\psi\rangle=\sqrt{n_0}(0,1)^{T}$ breaks the $\text{U}(2)$ symmetry down to $\text{U}(1)$ symmetry.  Broken-symmetry generators are $\sigma_1$, $\sigma_2$, and $\sigma_3-\sigma_0$, where $\sigma_{1,2,3}$ are Pauli matrices and $\sigma_0$ is the identity matrix.

The Lagrangian~\eqref{GP} possesses the Galilean symmetry
\begin{eqnarray}
\vec{x}'&=&\vec{x}+\vec{v}_0t,\quad t'=t,\\
\psi'(\vec{x}',t')&=&e^{m\vec{v}_0\cdot\vec{x}+\frac{1}{2}mv_0^2t}\psi(\vec{x},t),
\end{eqnarray}
in addition to the internal $\text{U}(2)$ symmetry.  The low-energy effective Lagrangian must respect it.  

Note that our discussion below is solely based on the internal $\text{U}(2)$ symmetry and the Galilean symmetry, so that it applies to any microscopic Lagrangians as long as they respect these symmetries and show the same symmetry-breaking pattern.

\subsubsection{Without Galilean symmetry}
Before going into the detailed discussion on the consequences of Galilean invariance, let us first review what we developed in Sec.~\ref{sec:solution} without paying attention to the Galilean symmetry for comparison.  We parametrize the coset as
\begin{eqnarray}
U=e^{i\pi^aT_a}=e^{i\left[\pi^1\sigma_1+\pi^2\sigma_2+\theta(\sigma_3-\sigma_0)\right]}.
\end{eqnarray}
We compute the Maurer-Cartan form
\begin{eqnarray}
&&\omega\equiv-iU^\dagger\mathrm{d}U\notag\\
&&\equiv\omega^0(\sigma_3+\sigma_0)+\left[\omega^1\sigma_1+\omega^2\sigma_2+\omega^3(\sigma_3-\sigma_0)\right].\label{u2u1}
\end{eqnarray}
Then $\omega_a^i$s defined by $\omega=\mathrm{d}\pi^a\omega_a^iT_i$ are building blocks of the effective Lagrangian, as explained in Sec.~\ref{sec:solution}.  To the quadratic order in derivatives, the most general form of the effective Lagrangian for this symmetry-breaking pattern, is
\begin{eqnarray}
&&\mathcal{L}_{\text{eff}}=-e_3(0)\bar{\omega}^3-e_0(0)\bar{\omega}^0\notag\\
&&+\frac{\bar{g}_{11}(0)}{2}\left(\bar{\omega}^1\bar{\omega}^1+\bar{\omega}^2\bar{\omega}^2\right)-\frac{g_{11}(0)}{2}\left(\vec{\omega}^1\cdot\vec{\omega}^1+\vec{\omega}^2\cdot\vec{\omega}^2\right)\notag\\
&&+\frac{\bar{g}_{33}(0)}{2}\bar{\omega}^3\bar{\omega}^3-\frac{g_{33}(0)}{2}\vec{\omega}^3\cdot\vec{\omega}^3,
\label{withoutG}
\end{eqnarray}
where we use the notation introduced in Sec.~\ref{sec:rightH}; namely, $\bar{\omega}^i=\omega_a^i\dot{\pi}^a$ and $\vec{\omega}^i=\omega_a^i\vec{\nabla}\pi^a$.  \emph{There are six parameters in this Lagrangian.  We will see soon that the Galilean invariance reduce them to four.}

There is a trick to easily compute the Maurer-Cartan form $\omega$ for this example.  We decompose $U$ into the product $U=U_0U_1$, where $U_0=e^{-i\theta\sigma_0}$ and 
\begin{eqnarray}
U_1=e^{i\pi^a\sigma_a}=\sigma_0\cos\rho+\frac{i}{\rho}\pi^a\sigma_a\sin\rho
\end{eqnarray}
with the constraint $\pi^3=\theta$. Here, $a=1,2,3$ and $\rho\equiv\sqrt{\pi^a\pi^a}$ .  Using the property of Pauli matrices, the Maurer-Cartan form for $U_0$ and $U_1$
\begin{eqnarray}
-iU_0^\dagger\mathrm{d}U_0&\equiv&\sigma_0\Omega^0,\\
-iU_1^\dagger\mathrm{d}U_1&\equiv&\sigma_1\Omega^1+\sigma_2\Omega^2+\sigma_3\Omega^3
\end{eqnarray}
can easily be evaluated as
\begin{eqnarray}
\Omega^0&=&-\mathrm{d}\theta,\\
\Omega^a&=&\mathrm{d}\pi^b\left[\left(\delta^{ab}-\frac{\pi^a\pi^b}{\rho^2}\right)\frac{\sin2\rho}{2\rho}\right.\notag\\
&&\quad\quad\quad\left.+\frac{\pi^a\pi^b}{\rho^2}-\epsilon^{abc}\pi^c\left(\frac{\sin\rho}{\rho}\right)^2\right].\label{SU(2)}
\end{eqnarray}

The full Maurer-Cartan form $\omega=-iU_0^\dagger\mathrm{d}U_0-iU_1^\dagger\mathrm{d}U_1$ is given by
\begin{eqnarray}
&\omega^1=\Omega^1,\quad\omega^2=\Omega^2,&\\
&\omega^0=\frac{\Omega^3+\Omega^0}{2},\quad\omega^3=\frac{\Omega^3-\Omega^0}{2}.&\label{omegaOmega}
\end{eqnarray}

\subsubsection{With Galilean symmetry}
To implement the Galilean symmetry, we introduce the boost operator
$\vec{B}$ as well as the spacetime translation $P_\mu=(H,-\vec{P})$.
Their nonzero commutation relations are
$[Q_a,Q_b]=2 i \epsilon_{abc}Q_c$, $[\vec{B},H]=-i\vec{P}$ and
$[B^i,P^j]=-imQ\delta^{ij}$ is centrally extended (see
Appendix~\ref{appendix1}).  $Q_1$, $Q_2$, $Q_3-Q$, and $\vec{B}$ are
spontaneously broken.  The unbroken generator $Q+Q_3$ is internal, so
that the assumption in the previous section is fulfilled.  Therefore,
we use
\begin{eqnarray}
\tilde{U}&=&e^{ix^\mu P_\mu}e^{i\pi^a(\vec{x},t)Q_a-i\theta(\vec{x},t) Q-i\vec{v}(\vec{x},t)\cdot\vec{B}}.
\end{eqnarray}
Here we introduced a new vector field $\vec{v}(\vec{x},t)$ that does not describe any physical modes and will be eliminated later in favor of real NG fields $\pi^1$, $\pi^2$, and $\pi^3\equiv\theta$.

The Maurer-Cartan form $\tilde{\omega}=-i\tilde{U}^\dagger\mathrm{d}\tilde{U}$ is given by
\begin{eqnarray}
\tilde{\omega}&=&\tilde{\omega}^0(Q_3+Q)+\left[\omega^1Q_1+\omega^2Q_2+\tilde{\omega}^3(Q_3-Q)\right]\notag\\
&&+e^\mu P_\mu-\vec{B}\cdot\mathrm{d}\vec{v}\label{QGalilei}
\end{eqnarray}
where $\omega^{0,1,2,3}$ stands for those defined in Eq.~\eqref{u2u1}:
\begin{eqnarray}
\tilde{\omega}^0&=&\omega^0+\frac{1}{2}\left(\frac{mv^2}{2}\mathrm{d}t-m\vec{v}\cdot\mathrm{d}\vec{x}\right),\label{QGalilei2}\\
\tilde{\omega}^3&=&\omega^3-\frac{1}{2}\left(\frac{mv^2}{2}\mathrm{d}t-m\vec{v}\cdot\mathrm{d}\vec{x}\right),\label{QGalilei3}
\end{eqnarray}
and
\begin{eqnarray}
e^0(\vec{x},t)=\mathrm{d}t,\quad\vec{e}(\vec{x},t)=\mathrm{d}\vec{x}-\vec{v}(\vec{x},t)\mathrm{d}t.
\end{eqnarray}
$\vec{e}$ is indeed covariant:
\begin{eqnarray}
\vec{e}'(\vec{x}',t')&=&\mathrm{d}(\vec{x}+\vec{v}_0t)-[\vec{v}(\vec{x},t)+\vec{v}_0]\mathrm{d}t\notag\\
&=&\mathrm{d}\vec{x}-\vec{v}(\vec{x},t)\mathrm{d}t=\vec{e}(\vec{x},t).
\end{eqnarray}
In this case, $\mathrm{det}G$ is trivial and $\mathrm{d}^dx\mathrm{d}t$, by itself, is an invariant volume form.

Following the definition in Eq.~\eqref{covariant}, covariant derivatives are given by
\begin{eqnarray}
\vec{\mathscr{D}}\pi^1&=&\vec{\omega}^1,\\
\vec{\mathscr{D}}\pi^2&=&\vec{\omega}^2,\\
\vec{\mathscr{D}}\pi^3&=&\vec{\omega}^3+\frac{m\vec{v}}{2},\\
\mathscr{D}_t\pi^1&=&\bar{\omega}^1+\vec{v}\cdot\vec{\mathscr{D}}\pi^1,\\
\mathscr{D}_t\pi^2&=&\bar{\omega}^2+\vec{v}\cdot\vec{\mathscr{D}}\pi^2,\\
\mathscr{D}_t\pi^3&=&\bar{\omega}^3-\frac{mv^2}{4}+\vec{v}\cdot\vec{\mathscr{D}}\pi^3.
\end{eqnarray}

Let us now focus on $\vec{\mathscr{D}}\pi^3$.  It contains a linear term of $\vec{v}$ without derivatives.  Thus, we \emph{can} impose a \emph{covariant constraint} $\vec{\mathscr{D}}\pi^3=0$, so called {\it the inverse Higgs constraint}\/~\cite{Ivanov:1975zq}, to eliminate the unphysical field $\vec{v}$ in terms of true NG fields
\begin{equation}
\vec{v}=-\frac{2\vec{\omega}^3}{m}.
\end{equation}
This constraint is a heuristic way to get rid of unphysical fields in the coset construction with spacetime symmetries.  See Refs.~\cite{McArthur, Nicolis, WatanabeBrauner3} for more details.

After imposing this constraint,  covariant derivatives become
\begin{eqnarray}
\vec{\mathscr{D}}\pi^1&=&\vec{\omega}^1,\label{cov1}\\
\vec{\mathscr{D}}\pi^2&=&\vec{\omega}^2,\label{cov2}\\
\mathscr{D}_t\pi^1&=&\bar{\omega}^1-\frac{2}{m}\vec{\omega}^3\cdot\vec{\mathscr{D}}\pi^1,\label{cov3}\\
\mathscr{D}_t\pi^2&=&\bar{\omega}^2-\frac{2}{m}\vec{\omega}^3\cdot\vec{\mathscr{D}}\pi^2,\label{cov4}\\
\mathscr{D}_t\pi^3&=&\bar{\omega}^3-\frac{1}{m}\vec{\omega}^3\cdot\vec{\omega}^3.\label{cov5}
\end{eqnarray}
Combinations in Eqs.~\eqref{cov1}--\eqref{cov5} are the Galilean-covariant building blocks of the effective Lagrangian.  

For the usual superfluid, the inverse Higgs constraint is $\vec{\mathscr{D}}\theta=\vec{\nabla}\theta+m\vec{v}=0$ and the combination in Eq.~\eqref{cov5} corresponds to $\mathscr{D}_t\theta=\dot{\theta}-(\vec{\nabla}\theta)^2/2m$.  Quantities in Eqs.~\eqref{cov3} and \eqref{cov4} correspond to the second term in Eq.~(12) of Ref.~\cite{Son} for supersolids.

According to Eq.~\eqref{transfunbroken}, $\bar{\omega}^0=\omega^0_a\dot{\pi}^a$ transforms as
\begin{eqnarray}
&&(\bar{\omega}^0)'(\vec{x}+\vec{v}_0t,t)\notag\\
&&=\bar{\omega}^0(\vec{x},t)+\vec{v}_0\cdot\vec{\omega}^0(\vec{x},t)+(\nabla_t+\vec{v}_0\cdot\vec{\nabla})\Lambda
\end{eqnarray}
for some $\Lambda$.  Therefore, the change of $\bar{\omega}^0$ is more
than a surface term and it cannot be added to the effective Lagrangian.

In summary, the most general form of the effective Lagrangian that respects the Galilean symmetry is
\begin{eqnarray}
&&\mathcal{L}_{\text{eff}}=-e_3(0)\mathscr{D}_t\pi^3\notag\\
&&+\frac{\bar{g}_{11}(0)}{2}\left[(\mathscr{D}_t\pi^1)^2+(\mathscr{D}_t\pi^2)^2\right]+\frac{\bar{g}_{33}(0)}{2}(\mathscr{D}_t\pi^3)^2\notag\\
&&-\frac{g_{11}(0)}{2}\left(\vec{\mathscr{D}}\pi^1\cdot\vec{\mathscr{D}}\pi^1+\vec{\mathscr{D}}\pi^2\cdot\vec{\mathscr{D}}\pi^2\right),\label{withgalilean}
\end{eqnarray}
which now contains only four parameters.  Compared to Eq.~\eqref{withoutG},  we have two restrictions:
\begin{equation}
e_0(0)=0,\quad g_{33}(0)=-\frac{2e_3(0)}{m}\,\,(>0).
\end{equation}
Since $e_0(0)$ represents the classical expectation value of $(Q_3+Q)/\Omega$, the spin must be fully polarized and $e_3(0)=(Q_3-Q)/\Omega=-2n<0$, where $n$ is the number density of the particles. This conclusion is consistent with the rigorous result in Ref.~\cite{Eisenberg}.  

Galilean-invariant combinations contain mixed powers of derivatives, and one can drop higher-order-derivative terms, as it does not affect the physics to the aimed order of the derivative expansion.

One may think that introducing the unphysical field $\vec{v}(\vec{x},t)$ first and eliminating it by imposing a covariant condition is just a complicated and useless way of deriving the effective Lagrangian. However, as we have demonstrated here, it is actually a convenient way to systematically generate terms with proper spacetime symmetries.

Finally, let us discuss the power counting of the derivative expansion.  In this paper, we assign $\pi^a=O(1)$ so that $\nabla_\mu\pi^a=O(k_\mu)$ and expand the Lagrangian in the series of derivatives.
However, Refs.~\cite{Son,Son2} introduced an alternative way of power counting, which assigns $\nabla_\mu\pi^a=O(1)$, \emph{provided that the Lagrangian does not depend on $\pi^a$ without derivatives}.  In this power-counting method, the lowest-order term is the sum of all invariant combinations with one derivative per a field.  This counting has an advantage that it can deal with the situation with large fluctuation $\pi^a=O(k^{-1})$ from the ground state, but \emph{it works only for Abelian groups $G$}; otherwise the effective Lagrangian depends on fields without derivative, as one can see from the example discussed in this section.

\section{Conclusion}
\label{sec:conclusion}

In this paper, we derived the explicit form of the most general
nonrelativistic Lagrangian of NGBs in terms of Maurer-Cartan form,
which must be quite useful to systematically discuss quantum
corrections.  By using the free part of the effective Lagrangian, we
proved the counting rule of NGBs and clarified the dispersion relation of
NGBs for a general setup.  We also completely classified possible
numbers of type-A and type-B NGBs for a given choice of $G/H$.

To discuss additional constraints on the effective Lagrangian from
spacetime symmetries, we showed explicitly the consequence of Galilean
invariance.  In addition, we presented an intuitive interpretation of
the presymplectic structure as Berry's phase of the ground state. 

Having derived the most general effective Lagrangian, we could develop
simple scaling arguments and show why a long-range order is
stable in $1+1 d$ when only type-B NGBs are present, while the
stability requires $2+1 d$ and above for type-A NGBs.  It remains an
interesting question whether there is a general rule of thumb when
both types of NGBs coexist.

\acknowledgments

We thank Tom\'a\v{s} Brauner, Sergej Moroz, Tsutomu Momoi, Akira Furusaki, and
Yoshimasa Hidaka for fruitful discussions and Aron Beekman for
informing us of the confusion on the time-reversal symmetry.  We are
especially indebted to Alan Weinstein, who helped us understand the
mathematical foundations.  We came up with the interpretation of the
linear derivative term as the Berry phase in the discussion with
Huan-Hang Chi.  We thank Tom\'a\v{s} Brauner for letting us know that the $b$ and $\tilde{b}$ terms can be cast in simple forms in Eqs.~\eqref{bgauged1} and~\eqref{bgauged2}.

H.W. appreciates financial support from the Honjo International
Scholarship Foundation.  The work of H.M. was supported by the U.S. DOE
under Contract No. DE-AC03-76SF00098, by the NSF under Grants No. PHY-1002399
and No. PHY-1316783, by the JSPS Grant No. (C) 23540289, and by WPI, MEXT,
Japan.

\appendix

\section{LIE-ALGEBLA COHOMOLOGY}
\label{appendix1}
The cohomology of Lie algebra was introduced by Chevalley and Eilenberg
\cite{CHevalleyEilenberg} as a way to compute the de Rham cohomology of
compact connected Lie groups using their Lie algebras.  On the other
hand, most physics literature is more familiar with de Rham
cohomology.  We use the work by Chevalley and Eilenberg backward to
describe Lie-algebra cohomology using de Rham cohomology.

The existence of a central extension of a Lie algebra $\mathfrak{g}$ is
determined by its second cohomology $H^2(\mathfrak{g})$.  The question
relevant to us is whether a central extension
\begin{equation}
  [T_i, T_j] = i f_{ij}{}^k T_k + i z_{ij},
\end{equation}
where $z_{ij}$ is the center (an element that commutes with the rest
of $\mathfrak{g}$), is possible for a given Lie algebra.  Then the question
is whether it is consistent with the Jacobi identity
\begin{equation}
  [T_i, [T_j, T_k]] + [T_j, [T_k, T_i]] + [T_k, [T_i, T_j]] = 0.
\end{equation}

A form on a Lie algebra $\omega_k \in \Omega^k(\mathfrak{g})$ is a map
from $\wedge^k \mathfrak{g}$ to $\mathbb R$
\begin{equation}
  \omega_k(\mathfrak{g}_1, \ldots, \mathfrak{g}_k) \in {\mathbb R}
\end{equation}
antisymmetric among arguments, 
\begin{eqnarray}
  \lefteqn{
    \omega_k(\mathfrak{g}_1, \ldots, \mathfrak{g}_i, \ldots, \mathfrak{g}_j,
    \ldots, \mathfrak{g}_k)} \nonumber \\
  &=& - \omega_k(\mathfrak{g}_1, \ldots, \mathfrak{g}_j, \ldots, \mathfrak{g}_i, \ldots, \mathfrak{g}_k).
\end{eqnarray}
A two-form $\omega_2$ is {\it exact}\/ if it can be obtained from a
one-form $\omega_2 = \mathrm{d}\omega_1$, 
\begin{equation}
  \mathrm{d}\omega_1(\mathfrak{g}_1, \mathfrak{g}_2) = \omega_1([\mathfrak{g}_1, \mathfrak{g}_2]).
\end{equation}
On the other hand, it is {\it closed}\/ if
\begin{eqnarray}
  \lefteqn{
    \mathrm{d}\omega_2(\mathfrak{g}_1, \mathfrak{g}_2, \mathfrak{g}_3) } \nonumber \\
  &=& 
  \omega_2(\mathfrak{g}_1, [\mathfrak{g}_2, \mathfrak{g}_3])
  + \omega_2(\mathfrak{g}_1, [\mathfrak{g}_2, \mathfrak{g}_3])
  + \omega_2(\mathfrak{g}_1, [\mathfrak{g}_2, \mathfrak{g}_3]) = 0 \nonumber \\
\end{eqnarray}
for any $\mathfrak{g}_{1,2,3}$.  This condition is called the cocycle condition.
For an exact two-form, it is nothing but the Jacobi identity, and
hence, it is automatically closed.

The possibility of $\omega_2(\mathfrak{g}_1, \mathfrak{g}_2)$ that
cannot be written as the original commutation relation yet satisfies
the Jacobi identity is the central extension and hence can be described
by the second cohomology $H^2(\mathfrak{g})$.

According to the theorem by Chevalley and Eilenberg,
$H^2(\mathfrak{g})=H^2_{\text{dR}}(G)$ if $G$ is the compact connected group
generated by $\mathfrak{g}$.  Since all compact simple Lie groups have
trivial second cohomology, central extensions are not possible for
their Lie algebras.  On the other hand, if there are $U(1)$ factors,
\begin{equation}
  \text{dim}\,H_{\text{dR}}^2 (\text{U}(1)^n) = \frac{n(n-1)}{2}\ ,
\end{equation}
generated by $\mathrm{d}\varphi^a \wedge \mathrm{d}\varphi^b$.  Therefore, the Lie
algebra cohomology $H^2(\mathfrak{u}(1)^n)$ is also nontrivial, and
hence, a central extension is possible.  

Note that the Lie algebra knows only about the local information, and
hence, it makes no distinction between $\mathfrak{u}(1)$ and $\mathbb
R$.  For instance, consider the Galilean group of rotations $M_{ij}$,
translations $P_i$, and Galilean boosts $B_i$:
\begin{eqnarray}
  [M_{ij}, P_k] &=& i (\delta_{ik} P_j - \delta_{ij} P_k), \\
  \left[M_{ij}, B_k\right] &=& i  (\delta_{ik} B_j - \delta_{ij} B_k), \\
  \left[M_{ij}, M_{kl} \right] &=& i  (\delta_{ik} M_{jl} - \delta_{il}
  M_{jk} - \delta_{jk} M_{il} + \delta_{jl} M_{ik}), \nonumber \\ & & \\
  \left[P_i, B_j\right] &=& 0.
\end{eqnarray}
$\vec{P}$ and $\vec{B}$ form ${\mathbb R}^d$ individually, which
allows for a central extension
\begin{equation}
  [P_i, B_j] = i \delta_{ij} M,
\end{equation}
where the eigenvalue of the operator $M$ is the mass of the particle
and a center of the Lie algebra ({\it i.e.}\/, commutes with
everything else).  The rotational invariance restricts the form to be
proportional to $\delta_{ij}$.  The exception is for the $2+1$
dimension, where $\epsilon_{ij}$ allows for alternative extensions
$[P_x, P_y] \propto \epsilon_{xy} = 1$ \cite{Watanabe:2014pea}.

Another example of central extension based on $\mathbb R$ is the shift
symmetry of the Schr\"odinger field mentioned in Sec.~\ref{sec:central}.
It has a
central extension thanks to $H^2({\mathbb R}^2)={\mathbb R} \neq0$.

\section{MATTER FIELDS}
\label{sec:matter}
In this paper, we establish the effective Lagrangian of NGBs for systems
without Lorentz invariance.  The effective Lagrangian can also
describe the situation where other low-energy degrees of freedom
(matter fields) couple to NGBs.  In this Appendix, we review how to
write down such low-energy theory for the reader's convenience.  Such
matter fields are important in many physical systems, {\it e.g.}\/,
fermions coupled to a spin system and nucleons coupled to pions.

\subsection{Approach 1: Modding $H$}
As discussed originally in Ref.~\cite{Callan}, any representation of $H$
$\psi \rightarrow \rho(h)\psi$, where $\rho(h)$ is a representation
matrix, can be promoted to transform under the full $G$ by
\begin{equation}
  \psi \rightarrow \psi' = \rho(h_g(\pi)) \psi.
\end{equation}
Since $h_g(\pi)$ is an element of $H$, the above expression is
well defined.  To see that it is a consistent transformation law, we
perform two successive transformations
\begin{eqnarray}
  U(\pi) \rightarrow g_2 g_1 U(\pi) &=& g_2 U(\pi') h_{g_1}(\pi) \notag\\
  &=& U(\pi'') h_{g_2}(\pi') h_{g_1}(\pi),
\end{eqnarray}
while
\begin{equation}
  \psi \rightarrow \rho(h_{g_2}(\pi') h_{g_1}(\pi)) \psi
  = \rho(h_{g_2}(\pi')) \rho(h_{g_1}(\pi)) \psi,
\end{equation}
given that $\rho$ is a representation of $H$.  

Note that this transformation law is {\it local}\/ in the sense that $\rho(h_g(\pi(\vec{x},t)))$ is position-dependent. As a result, $\mathrm{d}\psi$ does not transform in the same way as $\psi$ does:
\begin{equation}
(\mathrm{d}\psi)'=\mathrm{d}[\rho(h_g)\psi]=\rho(h_g)[\mathrm{d}+\rho(h_g^\dagger\mathrm{d}h_g)]\psi.
\end{equation}
However, the inhomogeneous part can be exactly compensated by the unbroken component of the Maurer-Cartan form [see Eq.~\eqref{homegau}]
\begin{equation}
\rho(\omega_\parallel')=\rho(h_g\omega_{\parallel}h_g^\dagger)-i\rho(h_g\mathrm{d}h_g^\dagger).
\end{equation}
Therefore, the combination
\begin{equation}
D\psi=[\mathrm{d} + i \rho(\omega_\parallel)]\psi
\end{equation}
is covariant. [This fact also means that $D^n\psi$ ($n\geq0$) is covariant.] Then, the question is how to write down $H$-invariant combinations out of these $H$-covariant building blocks. For example, $\psi^\dagger\psi$, $i(\psi^\dagger D_\mu\psi-\text{c.c.})$, and $D_\mu\psi^\dagger D_\nu\psi$ are all invariant combinations.   Since $D$ contains $\omega_\parallel$, they describe interactions between NGBs and matter fields.  We can also multiply invariants such as $g_{ab}(0)\vec{\omega}^a\cdot\vec{\omega}^b$ to them.  Since all Maurer-Cartan forms come with at least $1$ derivative acting on NG fields, all interactions become smaller and smaller in the low-energy limit.

What may be surprising is that the matter fields need to be only in linear representations of $H$, not $G$.  For instance, when electrons are coupled to ferromagnets, $G=\text{SO}(3)$, $H=\text{SO}(2)$, and the electrons need to transform only under $\text{U}(1)$ representation with a particular charge $q$, namely, $\rho(T_z)=q$ and $\psi'=e^{iq\theta}\psi$. Then, the low-energy effective Lagrangian for the interacting system of electrons and magnons (the NGB in ferromagnets) is given by $\mathcal{L}_{\text{eff}}=\mathcal{L}_{\text{mag}}+\mathcal{L}_{\text{el$+$int}}$, where
\begin{eqnarray}
\mathcal{L}_{\text{mag}}&=&-s\bar{\omega}^z-\frac{1}{2}g_0\left[(\vec{\omega}^x)^2+(\vec{\omega}^y)^2\right],\\
\mathcal{L}_{\text{el$+$int}}&=&\frac{i}{2}\left(\psi^\dagger D_t\psi-\text{c.c.}\right)-\mu\psi^\dagger\psi-\frac{\vec{D}\psi^\dagger\cdot\vec{D}\psi}{2m}\notag\\
&&-\lambda\left[(\vec{\omega}^x)^2+(\vec{\omega}^y)^2\right]\psi^\dagger\psi
\label{electronmagnon}
\end{eqnarray}
to the order $O(\nabla_t,\vec{\nabla}^2)$.  Here, $D_t=\nabla_t+iq\bar{\omega}^z$ and $\vec{D}=\vec{\nabla}+iq\vec{\omega}^z$, $s$ is the magnetization density, $m$ is the effective mass, and $\mu$ is the chemical potential of electrons.

The interaction Lagrangian~\eqref{electronmagnon} may be derived from a microscopic model
\begin{eqnarray}
\mathcal{L}&=&\frac{i}{2}(\Psi^\dagger\nabla_t\Psi-\text{c.c.})-\mu\Psi^\dagger\Psi-\frac{\vec{\nabla}\Psi^\dagger\cdot\vec{\nabla}\Psi}{2m}\notag\\
&&-J\vec{n}\cdot\Psi^\dagger\frac{\vec{\sigma}}{2}\Psi,\label{microelectronmagnon}
\end{eqnarray}
where $\Psi(\vec{x},t)$ is a two-component spinor and $\vec{n}(\vec{x},t)$ represents the magnetization of the ferromagnet including the fluctuation.  At this moment, the interaction term $\lambda\vec{n}\cdot\vec{s}$ does not contain any derivatives and the weakness of the interaction at long-distance is less apparent.  To get the effective Lagrangian~\eqref{electronmagnon}, we define locally a unitary transformation $U(\vec{n}(\vec{x},t))$~\cite{Nagaosa,Altland} such that
\begin{equation}
U^\dagger\vec{n}\cdot\vec{\sigma}U=\sigma_z
\end{equation}
and rewrite Eq.~\eqref{microelectronmagnon} in terms of $(\psi,\psi')^T\equiv U\Psi$. Then, $\lambda\vec{n}\cdot\vec{s}$ becomes just a constant $\lambda\sigma_z/2$, giving different chemical potentials to $\psi$ and $\psi'$.  The derivative of $\Psi$ now contains the Maurer-Cartan form
\begin{equation}
\mathrm{d}\Psi=U(\mathrm{d}+i\omega)(\psi,\psi')^T.
\end{equation}
Since $\psi'$ electrons have a gap $J$, we can integrate them out, ending up with Eq.~\eqref{electronmagnon} with $q=1/2$ and $\lambda=1/8m$ to the current order of the derivative expansion.

\subsection{Approach 2: Gauging $H$}
The translation law in the previous section is often awkward to deal
with because it is nonlinear.  Using the formalism to gauge the
right translation of $U$ by $H$, we can identify the above
transformation law as the gauge transformation with the gauge
$\pi^\rho = 0$.  Therefore, we consider $U = e^{i(\pi^a T_a + \pi^\rho
  T_\rho)}$ for the entire $G$ and its global transformation
\begin{equation}
  U(\pi) \rightarrow g U(\pi), \quad \psi \rightarrow \psi,
\end{equation}
while the local $H$ transformation is
\begin{equation}
  U(\pi) \rightarrow U(\pi) h(x), \quad \psi \rightarrow \rho(h^\dagger(x)) \psi.
\end{equation}
Then, we can construct an invariant Lagrangian using $\psi$ and its
covariant derivatives
\begin{equation}
  D \psi = [\mathrm{d} - i \rho(\mathcal{A})] \psi,\label{eqB}
\end{equation}
where $\mathcal{A}={\cal A}^\rho T_\rho$ transforms as in Eq.~\eqref{rightA}.  Equation~\eqref{eqB} is indeed covariant under the right translation:
\begin{eqnarray}
  (D \psi)'
  &=& [\mathrm{d} - i  \rho(h^\dagger {\cal A} h + i h^\dagger \mathrm{d} h)] \rho(h^\dagger)\psi\notag\\
  &=&\rho(h^\dagger) [\mathrm{d}+\rho(h\mathrm{d}h^\dagger) - i  \rho({\cal A} + i (\mathrm{d} h) h^\dagger)] \psi\notag\\
  &=&\rho(h^\dagger) [\mathrm{d}- i  \rho({\cal A})] \psi=\rho(h^\dagger) D \psi.
\end{eqnarray}

At the end of the day, we integrate the gauge fields out and stick to
the gauge $\pi^\rho = 0$.  Within this gauge, $h(x) = h_g^\dagger(\pi)$ and
$\rho(h^\dagger(x)) = \rho(h_g(\pi))$, as desired.

\section{TIME-REVERSAL SYMMETRY}
\label{sec:discrete}
In this Appendix, we clarify a confusion on discrete symmetries in the
existing literature~\cite{Tomas, Anton}.  Contrary to the claim made
in these references, we argue that type-B NGBs can appear without
breaking any discrete symmetries such as the time-reversal symmetry
(TRS).

In the case of ferromagnets, the expectation value
\begin{equation}
\langle [S_x,S_y]\rangle=i\langle S_z\rangle\neq0
\end{equation}
spontaneously breaks not only the spin-rotational symmetry but also TRS, since under the time reversal, the spin operator $\vec{S}$ flips its sign $\vec{S}\rightarrow-\vec{S}$.  However, in general,
\begin{equation}
\langle [Q_a,Q_b]\rangle=if_{ab}^{\phantom{ab}c}\langle Q_c\rangle\neq0\label{TRS1}
\end{equation}
does not necessarily mean that TRS is broken.  In order to respect
TRS, all generators that have a nonzero expectation value $\langle
Q_c\rangle$ have to be even under the time reversal.  Then,
Eq.~\eqref{TRS1} dictates that either the $Q_a$ or $Q_b$ that appears in
the commutator must be even and the other one must be odd, since TRS
is antiunitary and flips the sign of the right-hand side.

The simplest example is again given by the free-boson model in
Eq.~\eqref{freeSch}.  In this model, we identify the free bosons with
the dispersion $\omega=k^2/2m$ as the type-B NGB corresponding
to the spontaneously broken shift symmetry $\psi\rightarrow \psi+c$
($c\in\mathbb{C}$)~\cite{Brauner:2010}.  The Noether charge for
shifting the real and imaginary parts of $\psi$ is given by $Q_R=i\int
\mathrm{d}^dx(\psi-\psi^\dagger)$ and $Q_I=\int
\mathrm{d}^dx(\psi+\psi^\dagger)$, respectively. Because of the
commutation relation
$[\psi(\vec{x},t),\psi^\dagger(\vec{x}',t)]=\delta^d(\vec{x}-\vec{x}')$,
$Q_R$ and $Q_I$ do not commute and $[Q_R,Q_I]=2i \Omega$.  In this
case, the field $\psi$ transforms under TRS as
$\mathcal{T}\psi(\vec{x},t)\mathcal{T}^{-1}=\psi(\vec{x},-t)$, and
hence, $Q_R$ is odd and $Q_I$ is even under TRS.

A more nontrivial example is the model discussed in Sec.~\ref{sec:galilei} that exhibits the symmetry-breaking pattern $\text{U}(2)\rightarrow\text{U}(1)$.  The field $\psi$ transforms as $\psi'=e^{i\epsilon^i\sigma_i}\psi$ under the $\text{SU}(2)$ symmetry, and corresponding conserved charges are given by $Q_i=\int\mathrm{d}^dx\,\psi^\dagger\sigma_i\psi$.  

There are several consistent definitions of the time-reversal symmetries for this model.  If $\psi$ is a scalar, $\mathcal{T}$ acts $\psi$ as
\begin{equation}
\mathcal{T}\psi(\vec{x},t)\mathcal{T}^{-1}=\psi(\vec{x},-t).
\end{equation}
In this case, $Q_1$ and $Q_3$ are even and $Q_2$ is odd since $\sigma_2$ is imaginary.  Thus, $\langle [Q_1,Q_2]\rangle=2i\langle Q_3\rangle\neq0$ does not break this TRS while a type-B NGB appears in this model. $\vec{Q}/2$ represents a pseudospin.   Another way of defining $\mathcal{T}$ symmetry is
\begin{equation}
\mathcal{T}\psi(\vec{x},t)\mathcal{T}^{-1}=i\sigma_2\psi(\vec{x},-t).
\end{equation}
This time, all of the $Q_i$'s are odd under $\mathcal{T}$ and $\vec{Q}/2$ represents the real spin.  $\langle [Q_1,Q_2]\rangle=2i\langle Q_3\rangle\neq0$ breaks this TRS.

Other discrete symmetries, such as the parity $\mathcal{P}$ and the charge conjugation $\mathcal{C}$, if they exist, can be discussed in the same way.

\bibliography{reference}
\end{document}